\documentclass[12pt,english]{article}
\usepackage{times}
\usepackage[T1]{fontenc}
\usepackage{geometry}
\usepackage{array}
\usepackage{rotating}
\usepackage{graphicx}
\usepackage{setspace}
\usepackage{amssymb}

\makeatletter

\providecommand{\tabularnewline}{\\}

\usepackage{bm}

\usepackage{babel}
\makeatother
\begin{document}

\section*{Multi-Scaling Differential Contraction Integral Method for Inverse
Scattering Problems with Inhomogeneous Media}

\noindent ~

\noindent \vfill

\noindent Y. Zhong$^{\left(1\right)\left(2\right)}$, F. Zardi$^{\left(3\right)\left(4\right)}$,
M. Salucci$^{\left(3\right)\left(4\right)}$, G. Oliveri$^{\left(3\right)\left(4\right)}$,
and A. Massa$^{\left(3\right)\left(4\right)\left(5\right)\left(6\right)}$

\noindent \vfill

\noindent {\footnotesize ~}{\footnotesize \par}

\noindent {\footnotesize ~}{\footnotesize \par}

\noindent {\scriptsize $^{(1)}$} \emph{\scriptsize FINIAC Pte. Ltd.}{\scriptsize \par}

\noindent {\scriptsize JTC LaunchPad @ JID, 2 Cleantech Loop, Singapore,
637144}{\scriptsize \par}

\noindent {\scriptsize Email:} \emph{\scriptsize zhongyu@finiac.io}{\scriptsize \par}

\noindent {\scriptsize ~}{\scriptsize \par}

\noindent {\scriptsize $^{(2)}$} \emph{\scriptsize }{\scriptsize Dept.
of Physics and Technology, UiT The Arctic University of Norway}{\scriptsize \par}

\noindent {\scriptsize NO-9037 Tromso, Norway. }{\scriptsize \par}

\noindent {\scriptsize Email:} \emph{\scriptsize yu.zhong@uit.no}{\scriptsize \par}

\noindent {\scriptsize ~}{\scriptsize \par}

\noindent {\scriptsize $^{(3)}$} \emph{\scriptsize ELEDIA Research
Center} {\scriptsize (}\emph{\scriptsize ELEDIA}{\scriptsize @}\emph{\scriptsize UniTN}
{\scriptsize - University of Trento)}{\scriptsize \par}

\noindent {\scriptsize DICAM - Department of Civil, Environmental,
and Mechanical Engineering}{\scriptsize \par}

\noindent {\scriptsize Via Mesiano 77, 38123 Trento - Italy}{\scriptsize \par}

\noindent \textit{\emph{\scriptsize E-mail:}} {\scriptsize \{}\emph{\scriptsize francesco.zardi,
marco.salucci, giacomo.oliveri, andrea.massa}{\scriptsize \}@}\emph{\scriptsize unitn.it}{\scriptsize \par}

\noindent {\scriptsize Website:} \emph{\scriptsize www.eledia.org/eledia-unitn}{\scriptsize \par}

\noindent {\scriptsize ~}{\scriptsize \par}

\noindent {\scriptsize $^{(4)}$} \emph{\scriptsize CNIT - \char`\"{}University
of Trento\char`\"{} ELEDIA Research Unit }{\scriptsize \par}

\noindent {\scriptsize Via Sommarive 9, 38123 Trento - Italy}{\scriptsize \par}

\noindent {\scriptsize Website:} \emph{\scriptsize www.eledia.org/eledia-unitn}{\scriptsize \par}

\noindent {\scriptsize ~}{\scriptsize \par}

\noindent {\scriptsize $^{(5)}$} \emph{\scriptsize ELEDIA Research
Center} {\scriptsize (}\emph{\scriptsize ELEDIA}{\scriptsize @}\emph{\scriptsize UESTC}
{\scriptsize - UESTC)}{\scriptsize \par}

\noindent {\scriptsize School of Electronic Engineering, Chengdu 611731
- China}{\scriptsize \par}

\noindent \textit{\emph{\scriptsize E-mail:}} \emph{\scriptsize andrea.massa@uestc.edu.cn}{\scriptsize \par}

\noindent {\scriptsize Website:} \emph{\scriptsize www.eledia.org/eledia}{\scriptsize -}\emph{\scriptsize uestc}{\scriptsize \par}

\noindent {\scriptsize ~}{\scriptsize \par}

\noindent {\scriptsize $^{(6)}$} \emph{\scriptsize ELEDIA Research
Center} {\scriptsize (}\emph{\scriptsize ELEDIA@TSINGHUA} {\scriptsize -
Tsinghua University)}{\scriptsize \par}

\noindent {\scriptsize 30 Shuangqing Rd, 100084 Haidian, Beijing -
China}{\scriptsize \par}

\noindent {\scriptsize E-mail:} \emph{\scriptsize andrea.massa@tsinghua.edu.cn}{\scriptsize \par}

\noindent {\scriptsize Website:} \emph{\scriptsize www.eledia.org/eledia-tsinghua}{\scriptsize \par}

\noindent \vfill

\noindent \textbf{\emph{This work has been submitted to the IEEE for
possible publication. Copyright may be transferred without notice,
after which this version may no longer be accessible.}}

\noindent \vfill

\newpage
\section*{Multi-Scaling Differential Contraction Integral Method for Inverse
Scattering Problems with Inhomogeneous Media}

~

\noindent ~

\noindent ~

\begin{flushleft}Y. Zhong, F. Zardi, M. Salucci, G. Oliveri, and A.
Massa\end{flushleft}

\noindent \vfill

\begin{abstract}
\noindent Practical applications of microwave imaging often require
the solution of inverse scattering problems with inhomogeneous backgrounds.
Towards this end, a novel inversion strategy, which combines the multi-scaling
(\emph{MS}) regularization scheme and the Difference Contraction Integral
Equation (\emph{DCIE}) formulation, is proposed. Such an integrated
approach mitigates the non-linearity and the ill-posedness of the
problem to obtain reliable high-resolution reconstructions of the
unknown scattering profiles. The arising algorithmic implementation,
denoted as \emph{MS-DCIE}, does not require the computation of the
Green's function of the inhomogeneous background, thus it provides
an efficient and effective way to deal with complex scenarios. The
performance of the \emph{MS-DCIE} are assessed by means of numerical
and experimental tests, in comparison with competitive state-of-the-art
inversion strategies, as well.
\end{abstract}
\noindent \vfill

\noindent \textbf{Key words}: Microwave Imaging, Inverse Scattering,
Multi-Scaling, Difference Contraction Integral Equation, Inhomogeneous
Media.

\newpage
\section{Introduction }

\noindent In engineering, microwave imaging applications are concerned
with both homogeneous \cite{Colton 2013}-\cite{Xu 2018a} and inhomogeneous
\cite{Abubakar 2000}-\cite{Chu 2019} backgrounds. In both cases,
the behavior of time-harmonic electromagnetic fields (\emph{EM}s)
and their interactions with the environment (i.e., scatterers, host
medium, and receivers) can be faithfully described in terms of either
the wave equation, which is a partial differential equation (\emph{PDE}),
or the corresponding integral equation (\emph{IE}). 

\noindent Well-known examples of inverse scattering (\emph{IS}) approaches
formulated within the Lippmann-Schwinger \emph{IE} (\emph{LSIE}) are
the methods based on the Born \cite{Oliveri 2019}\cite{Wang 1989}\cite{Afsari 2019},
the Rytov \cite{Zhang 2009}, and the distorted Born \cite{Chew 1990}\cite{Cui 2001}
approximations or Newton-type techniques \cite{Mojabi 2009}. \emph{PDE}-based
methods (e.g., \cite{Kilic 2015} and \cite{Brown 2019}) have been
extensively studied, as well. In principle, all these methods can
handle \emph{IS} problems (\emph{ISP}s) with inhomogeneous media,
but their efficiency and accuracy might be underwhelming since a repeated
solution of the corresponding forward scattering problem is required
to iteratively update the unknown contrast \cite{vanLeeuwen 2016}.
Additionally, some approaches rely on linear approximations, which
do not hold true for highly-contrasted objects when multiple-scattering
phenomena are non-negligible \cite{Caorsi 2003}.

\noindent To avoid multiple calls of a forward solver and to enable
the application of different regularization techniques \cite{vandenBerg 2003}\cite{Zhong 2009}\cite{Zhong 2011},
by also reducing the non-linearity of the \emph{ISP} at hand thanks
to new theoretical formulations (e.g., the contraction integral equation
(\emph{CIE}) \cite{Zhong 2016}\cite{Xu 2018a}) for dealing with
strong contrasts or electrically-large scatterers, an alternative
class of inversion strategies has been considered. Namely, the modified
gradient method \cite{Kleinman 1992}, the contrast source inversion
(\emph{CSI}) method \cite{Kleinman 1997}, and the subspace-based
optimization method (\emph{SOM}) \cite{Chen 2010}\cite{Zhong 2009}
have been proposed where the equivalent source is now the {}``secondary''
unknown instead of the \emph{EM} field, while the contrast is still
the {}``primary'' quantity to be determined.

\noindent To address the \emph{ISP}s with inhomogeneous scenarios,
the \emph{CSI} \cite{Abubakar 2008} and the \emph{SOM} \cite{Chen 2010b}
methods adopt a \emph{PDE}-based modeling for computing the Green's
function of the inhomogeneous background so that the inversion is
carried out within the \emph{LSIE} framework. Despite their successful
application, these approaches suffer from the heavy computational
burden of numerically determining the inhomogeneous-media Green's
function without using acceleration techniques as for homogeneous
backgrounds (e.g., the Conjugate Gradient Fast Fourier Transform (\emph{CG-FFT})
\cite{Catedra 1995} or the Fast Multipole Method (\emph{FMM}) \cite{Chew 2001}). 

\noindent Otherwise, the Difference \emph{CIE} (\emph{DCIE}) method
\cite{Xu 2018b} exploits an alternative \emph{ISP} formulation that
only needs the closed-form expression of the Green's function for
the homogeneous background. However, the arising inversion process
has to be stabilized \cite{Zhong 2016}\cite{Xu 2018a} by means of
an effective regularization technique such as the multi-scaling (\emph{MS})
strategy, which has proved able to handle both two-dimensional (\emph{2D})
\cite{Caorsi 2003} and three-dimensional (\emph{3D}) \cite{Franceschini 2005}\cite{Donelli 2009}\cite{Ye 2015}
scenarios as well as aspect-limited configurations \cite{Conci 2005}\cite{Salucci 2021}.
Moreover, it has been successfully combined with different optimization
methods \cite{Caorsi 2003}\cite{Donelli 2006} and formulations \cite{Oliveri 2011}\cite{Zhong 2020}.

\noindent Generally speaking, the \emph{MS} strategy is a multi-zoom
meta-level scheme aimed at identifying, at each step, the region of
the investigation domain where the unknown scatterers are supposed
to be most likely located, referred to as Region of Interest (\emph{RoI}).
In this latter, the reconstruction of the objects descriptors, whose
number is kept close to the amount of available information from the
scattering data/measurements, is then performed with a suitable inversion
method. In this way, while mitigating the ill-posedness and the non-linearity
of the successive inversions, a multi-resolution imaging of the scenario
is yielded, the higher spatial resolution being only in the \emph{RoI}
identified at each step.

\noindent In this work, the \emph{DCIE} formulation is extended to
the \emph{MS} scheme for defining a new inversion strategy able to
effectively and efficiently address imaging problems involving unknown
objects in inhomogeneous backgrounds. Such an integration is not straightforward
since the differential formulation takes into account the \emph{EM}
interactions generated by the unknown objects also outside \emph{}the
\emph{RoI}. Indeed, external equivalent currents are generated because
of the coupling between unknown scatterers and the inhomogeneous background
and they cannot be neglected without compromising the validity of
the \emph{IS} model. Therefore, an ad-hoc \emph{RoI} estimation technique
is proposed to derive a novel and customized \emph{ISP} solution approach,
namely the \emph{MS-DCIE}. This method features the advantages of
both the \emph{DCIE} inversion (i.e., computationally fast and weakly
non-linear) and the \emph{MS} scheme (i.e., reduced ill-posedness
and multi-resolution reconstruction).

\noindent The outline of the paper is as follows. The \emph{ISP} is
mathematically formulated within the \emph{DCIE} framework in Sect.
\ref{sec:mathematical-formulation}, while Section \ref{sec:solution_approach}
details the \emph{MS-DCIE} inversion strategy. In Sect. \ref{sec:results},
a representative set of numerical and experimental test cases are
presented and discussed to assess the reliability and the effectiveness
of the proposed method. Finally, some concluding remarks are drawn
(Sect. \ref{sec:Conclusions}).

\section{\emph{ISP} Formulation (\emph{DCIE} Framework)\label{sec:mathematical-formulation}}

\noindent Let $\mathcal{H}$ be a square investigation domain of side
$L_{\mathcal{H}}$ and characterized by a known and inhomogeneous
permittivity distribution $\varepsilon_{H}\left(\mathbf{r}\right)$,
which is univocally described by the contrast function $\tau_{H}\left(\mathbf{r}\right)\triangleq\frac{\varepsilon_{H}\left(\mathbf{r}\right)}{\varepsilon_{B}}-1$,
$\varepsilon_{B}$ and $\tau_{B}\left(\mathbf{r}\right)=0$ being
the permittivity and the contrast of the homogeneous external background
{[}Fig. 1(\emph{a}){]}, respectively. An unknown scatterer of arbitrary
shape and permittivity $\varepsilon_{O}\left(\mathbf{r}\right)$ is
located in a region $\mathcal{O}$ within $\mathcal{H}$ {[}Fig. 1(\emph{b}){]}
so that the contrast function in $\mathcal{H}$ turns out to be $\tau\left(\mathbf{r}\right)\triangleq\frac{\varepsilon\left(\mathbf{r}\right)}{\varepsilon_{B}}-1$
(i.e., $\tau\left(\mathbf{r}\right)=\tau_{O}\left(\mathbf{r}\right)$
if $\mathbf{r}\in\mathcal{O}$, $\tau\left(\mathbf{r}\right)=\tau_{H}\left(\mathbf{r}\right)$
if $\mathbf{r}\in\mathcal{H}-\mathcal{O}$).

\noindent The host domain $\mathcal{H}$ is probed by a transverse
magnetic (\emph{TM}) plane wave from $V$ different directions \{$\phi_{v}$;
$v=1,...,V$\}, $\xi_{inc}^{\left(v\right)}$ being the $z$-th component
of the $v$-th ($v=1,...,V$) incident field, and the field $\xi^{\left(v\right)}$
arising from the \emph{EM} scattering interactions is measured by
$M$ probes located in the observation domain, outside $\mathcal{H}$,
at the positions \{$\mathbf{r}_{m}$; $m=1,...,M$\}. By defining
the scattered field $\xi_{sca}^{\left(v\right)}$ as $\xi_{sca}^{\left(v\right)}\triangleq\xi^{\left(v\right)}-\xi_{inc}^{\left(v\right)}$
($v=1,...,V$), the \emph{ISP} at hand can be then formulated as the
retrieval of the differential contrast $\tau_{\Delta}$ {[}$\tau_{\Delta}\left(\mathbf{r}\right)\triangleq\tau\left(\mathbf{r}\right)-\tau_{H}\left(\mathbf{r}\right)$
- Fig. 1(\emph{c}){]} starting from the knowledge of the measured
samples \{$\xi_{sca}^{\left(v\right)}\left(\mathbf{r}_{m}\right)$;
$m=1,...,M$; $v=1,...,V$\}.

\noindent Mathematically, the relation between data and unknowns,
subject to the knowledge of the (inhomogeneous) host medium (i.e.,
$\tau_{H}\left(\mathbf{r}\right)$, $\mathbf{r}\in\mathcal{H}$),
is described by the following {}``\emph{differential}'' version
of the \emph{LSIE} equations\begin{equation}
\xi_{sca}^{\left(v\right)}\left(\mathbf{r}_{m}\right)=\int_{\mathcal{O}}\mathcal{G}_{H}\left(\mathbf{r}_{m},\,\mathbf{r}^{\prime}\right)\tau_{\Delta}\left(\mathbf{r}^{\prime}\right)\xi^{\left(v\right)}\left(\mathbf{r}^{\prime}\right)d\mathbf{r}^{\prime}\label{eq:LSIE-data differential GH}\end{equation}
\begin{equation}
\xi^{\left(v\right)}\left(\mathbf{r}\right)=\xi_{H}^{\left(v\right)}\left(\mathbf{r}\right)+\int_{\mathcal{O}}\mathcal{G}_{H}\left(\mathbf{r},\,\mathbf{r}^{\prime}\right)\tau_{\Delta}\left(\mathbf{r}^{\prime}\right)\xi^{\left(v\right)}\left(\mathbf{r}^{\prime}\right)d\mathbf{r}^{\prime}\label{eq:LSIE-state differential GH}\end{equation}
where $\mathcal{G}_{H}$ is the Green's function of the inhomogeneous
host medium with permittivity $\varepsilon_{H}\left(\mathbf{r}\right)$,
the differential contrast $\tau_{\Delta}$ is equal to $\tau_{\Delta}\left(\mathbf{r}\right)=\tau_{O}\left(\mathbf{r}\right)-\tau_{H}\left(\mathbf{r}\right)$
if $\mathbf{r}\in\mathcal{O}$, $\tau_{\Delta}\left(\mathbf{r}\right)=0$
otherwise, $\xi_{H}^{\left(v\right)}$ is the $v$-th ($v=1,...,V$)
electric field without the object given by

\noindent \begin{equation}
\xi_{H}^{\left(v\right)}\left(\mathbf{r}\right)=\xi_{inc}^{\left(v\right)}\left(\mathbf{r}\right)+\int_{\mathcal{H}}\mathcal{G}_{B}\left(\mathbf{r},\,\mathbf{r}^{\prime}\right)\tau_{H}\left(\mathbf{r}^{\prime}\right)\xi_{H}^{\left(v\right)}\left(\mathbf{r}^{\prime}\right)d\mathbf{r}^{\prime},\label{eq:LSIE-state}\end{equation}
$\mathcal{G}_{B}$ being the Green's function of the background homogeneous
medium with permittivity $\varepsilon_{B}$. It is worth noticing
that (\ref{eq:LSIE-data differential GH})(\ref{eq:LSIE-state differential GH})
require the evaluation of $\mathcal{G}_{H}$, which can be computationally-demanding.

\noindent Alternatively, it is still possible to deduce a {}``\emph{differential}''
\emph{LSIE} formulation (\emph{DLSIE}) without having available $\mathcal{G}_{H}$
(see Appendix A)\begin{equation}
\xi_{sca,\Delta}^{\left(v\right)}\left(\mathbf{r}_{m}\right)=\int_{\mathcal{H}}\mathcal{G}_{B}\left(\mathbf{r}_{m},\,\mathbf{r}^{\prime}\right)\left\{ \tau_{\Delta}\left(\mathbf{r}^{\prime}\right)\xi^{\left(v\right)}\left(\mathbf{r}^{\prime}\right)+\tau_{H}\left(\mathbf{r}^{\prime}\right)\left[\xi^{\left(v\right)}\left(\mathbf{r}^{\prime}\right)-\xi_{H}^{\left(v\right)}\left(\mathbf{r}^{\prime}\right)\right]\right\} d\mathbf{r}^{\prime}\label{eq:LSIE-data differential}\end{equation}
 \begin{equation}
\xi^{\left(v\right)}\left(\mathbf{r}\right)=\xi_{H}^{\left(v\right)}\left(\mathbf{r}\right)+\int_{\mathcal{H}}\mathcal{G}_{B}\left(\mathbf{r},\,\mathbf{r}^{\prime}\right)\left\{ \tau_{\Delta}\left(\mathbf{r}^{\prime}\right)\xi^{\left(v\right)}\left(\mathbf{r}^{\prime}\right)+\tau_{H}\left(\mathbf{r}^{\prime}\right)\left[\xi^{\left(v\right)}\left(\mathbf{r}^{\prime}\right)-\xi_{H}^{\left(v\right)}\left(\mathbf{r}^{\prime}\right)\right]\right\} d\mathbf{r}^{\prime}\label{eq:LSIE-state differential}\end{equation}

\noindent As a matter of fact, the available information on the host
medium is now fully included in (\ref{eq:LSIE-data differential})(\ref{eq:LSIE-state differential})
through $\tau_{H}$ and $\xi_{H}^{\left(v\right)}$.

\noindent The corresponding differential contrast-source formulation
(\emph{DCSIE}) is obtained by multiplying (\ref{eq:LSIE-state differential})
by the contrast function $\tau$ \cite{Kleinman 1997}\cite{Chen 2010}
and defining the $v$-th ($v=1,...,V$) differential equivalent current
$J_{\Delta}^{\left(v\right)}$ as\begin{equation}
J_{\Delta}^{\left(v\right)}\left(\mathbf{r}\right)\triangleq J^{\left(v\right)}\left(\mathbf{r}\right)-J_{H}^{\left(v\right)}\left(\mathbf{r}\right),\label{eq:equivalent current differential}\end{equation}
$J^{\left(v\right)}$($J_{H}^{\left(v\right)}$) being the equivalent
current induced on the investigation domain $\mathcal{H}$ with contrast
$\tau$ ($\tau_{H}$) and radiating in the free-space background with
$\tau_{B}\left(\mathbf{r}\right)=0$ {[}i.e., $J^{\left(v\right)}\left(\mathbf{r}\right)\triangleq\tau\left(\mathbf{r}\right)\xi^{\left(v\right)}\left(\mathbf{r}\right)$
and $J_{H}^{\left(v\right)}\left(\mathbf{r}\right)\triangleq\tau_{H}\left(\mathbf{r}\right)\xi_{H}^{\left(v\right)}\left(\mathbf{r}\right)${]},
so that, after some simple manipulations, it turns out that\begin{equation}
\xi_{sca,\Delta}^{\left(v\right)}\left(\mathbf{r}_{m}\right)=\int_{\mathcal{H}}\mathcal{G}_{B}\left(\mathbf{r}_{m},\,\mathbf{r}^{\prime}\right)J_{\Delta}^{\left(v\right)}\left(\mathbf{r}^{\prime}\right)d\mathbf{r}^{\prime}\label{eq:CSI-data differential}\end{equation}
\begin{equation}
\begin{array}{c}
J_{\Delta}^{\left(v\right)}\left(\mathbf{r}\right)=\tau_{\Delta}\left(\mathbf{r}\right)\left[\xi_{H}^{\left(v\right)}\left(\mathbf{r}\right)+\int_{\mathcal{H}}\mathcal{G}_{B}\left(\mathbf{r},\,\mathbf{r}^{\prime}\right)J_{\Delta}^{\left(v\right)}\left(\mathbf{r}^{\prime}\right)d\mathbf{r}^{\prime}\right]\\
+\tau_{H}\left(\mathbf{r}\right)\int_{\mathcal{H}}\mathcal{G}_{B}\left(\mathbf{r},\,\mathbf{r}^{\prime}\right)J_{\Delta}^{\left(v\right)}\left(\mathbf{r}^{\prime}\right)d\mathbf{r}^{\prime}.\end{array}\label{eq:CSI-state differential}\end{equation}
By introducing the auxiliary parameter $\beta\left(\mathbf{r}\right)$
\cite{Zhong 2016} and defining the modified contrast function $\chi$
{[}$\chi\left(\mathbf{r}\right)\triangleq\frac{\beta\left(\mathbf{r}\right)\tau\left(\mathbf{r}\right)}{\beta\left(\mathbf{r}\right)\tau\left(\mathbf{r}\right)+1}${]},
(\ref{eq:CSI-state differential}) is then rewritten in the co-called
Differential \emph{CIE} (\emph{DCIE}) form \cite{Xu 2018b}

\noindent \begin{equation}
\begin{array}{r}
\beta\left(\mathbf{r}\right)J_{\Delta}^{\left(v\right)}\left(\mathbf{r}\right)=\chi_{\Delta}\left(\mathbf{r}\right)\left[\xi_{H}^{\left(v\right)}\left(\mathbf{r}\right)+\int_{\mathcal{H}}\mathcal{G}_{B}\left(\mathbf{r},\,\mathbf{r}^{\prime}\right)J_{\Delta}^{\left(v\right)}\left(\mathbf{r}^{\prime}\right)d\mathbf{r}^{\prime}+\beta\left(\mathbf{r}\right)J_{\Delta}^{\left(v\right)}\left(\mathbf{r}^{\prime}\right)+\beta\left(\mathbf{r}\right)J_{H}^{\left(v\right)}\left(\mathbf{r}\right)\right]\\
+\chi_{H}\left(\mathbf{r}\right)\left[\int_{\mathcal{H}}\mathcal{G}_{B}\left(\mathbf{r},\,\mathbf{r}^{\prime}\right)J_{\Delta}^{\left(v\right)}\left(\mathbf{r}^{\prime}\right)d\mathbf{r}^{\prime}+\beta\left(\mathbf{r}\right)J_{\Delta}^{\left(v\right)}\left(\mathbf{r}^{\prime}\right)\right]\end{array}\label{eq:DCIE-state}\end{equation}
where $\chi_{H}$ {[}$\chi_{H}\left(\mathbf{r}\right)\triangleq\frac{\beta\left(\mathbf{r}\right)\tau_{H}\left(\mathbf{r}\right)}{\beta\left(\mathbf{r}\right)\tau_{H}\left(\mathbf{r}\right)+1}${]}
and $\chi_{\Delta}$ {[}$\chi_{\Delta}\left(\mathbf{r}\right)\triangleq\chi\left(\mathbf{r}\right)-\chi_{H}\left(\mathbf{r}\right)${]}
are the host-medium and the differential modified contrast, respectively.

\noindent Within the above \emph{DCIE} formulation, the solution of
the original \emph{ISP} is then recast to that of determining the
distribution of the differential modified contrast $\chi_{\Delta}$
(i.e., the \emph{primary} unknown) and the $V$ differential equivalent
currents, \{$J_{\Delta}^{\left(v\right)}$; $v=1,...,V$\} (i.e.,
the \emph{secondary} unknowns), that fulfil (\ref{eq:CSI-data differential})
and (\ref{eq:DCIE-state}). 

\noindent Towards this end, (\ref{eq:CSI-data differential})(\ref{eq:DCIE-state})
are first discretized by partitioning the investigation domain $\mathcal{H}$
into $N$ square sub-units, \{$\mathcal{H}_{n}$; $n=1,...,N$\} ($\mathcal{H}=\sum_{n=1}^{N}\mathcal{H}_{n}$),
centered at \{$\mathbf{r}_{n}$; $n=1,...,N$\} and using $M$($N$)
Dirac's test functions to sample (\ref{eq:CSI-data differential})(\ref{eq:DCIE-state})
at the $M$($N$) locations of the probes in the observation(investigation)
domain \{$\mathbf{r}_{m}$; $m=1,...,M$\}(\{$\mathbf{r}_{m}$; $n=1,...,N$\}).
The following numerical forms of (\ref{eq:CSI-data differential})(\ref{eq:DCIE-state})
are then derived\begin{equation}
\overline{\xi}_{sca,\Delta}^{\left(v\right)}=\overline{\overline{\mathcal{G}}}_{{\rm ext}}\overline{J}_{\Delta}^{\left(v\right)}\label{eq:DCIE-data-discretized}\end{equation}
\begin{equation}
\begin{array}{r}
\overline{\beta}\otimes\overline{J}_{\Delta}^{\left(v\right)}=\overline{\chi}_{\Delta}\otimes\left[\overline{\xi}_{H}^{\left(v\right)}+\overline{\overline{\mathcal{G}}}_{{\rm int}}\overline{J}_{\Delta}^{\left(v\right)}+\overline{\beta}\otimes\overline{J}_{\Delta}^{\left(v\right)}+\overline{\beta}\otimes\overline{J}_{H}^{\left(v\right)}\right]\\
+\overline{\chi}_{H}\otimes\left[\overline{\overline{\mathcal{G}}}_{{\rm int}}\overline{J}_{\Delta}^{\left(v\right)}+\overline{\beta}\otimes\overline{J}_{\Delta}^{\left(v\right)}\right],\end{array}\label{eq:DCIE-state-discretized}\end{equation}
 where $\overline{\xi}_{sca,\Delta}^{\left(v\right)}$ $=$ \{$\xi_{sca,\Delta}^{\left(v\right)}\left(\mathbf{r}_{m}\right)$;
$m=1,...,M$\}, $\overline{\beta}$ $=$ \{$\beta\left(\mathbf{r}_{n}\right)$;
$n=1,...,N$\}, $\overline{J}_{\Delta}^{\left(v\right)}$ $=$ \{$J_{\Delta}^{\left(v\right)}\left(\mathbf{r}_{n}\right)$;
$n=1,...,N$\}, $\overline{\chi}_{\Delta}$ $=$ \{$\chi_{\Delta}\left(\mathbf{r}_{n}\right)$;
$n=1,...,N$\}, $\overline{\xi}_{H}^{\left(v\right)}=\{\xi_{H}^{\left(v\right)}\left(\mathbf{r}_{n}\right);\, n=1,...,N\}$,
and $\overline{J}_{H}^{\left(v\right)}=\{ J_{H}^{\left(v\right)}\left(\mathbf{r}_{n}\right);\, n=1,...,N\}$,
while $\otimes$ stands for the element-wise multiplication. Moreover,
$\overline{\overline{\mathcal{G}}}_{{\rm int}}$($\overline{\overline{\mathcal{G}}}_{{\rm ext}}$)
is the $N\times N$($M\times N$) internal(external) Green's matrix
whose ($p$, $q$)-th{[}($m$, $n$)-th{]} ($p$, $q$ $=$ $1,...,N$)($m=1,...,M$;
$n=1,...,N$) entry is given by $\mathcal{G}_{{\rm int}}^{pq}=j\frac{k_{B}^{2}}{4}\int_{\mathcal{H}_{q}}H_{0}^{\left(2\right)}\left(k_{B}\left|\mathbf{r}_{p}-\mathbf{r}'\right|\right)d\mathbf{r}'$($\mathcal{G}_{{\rm ext}}^{mn}$
$=$ $j\frac{k_{B}^{2}}{4}\int_{\mathcal{H}_{n}}H_{0}^{\left(2\right)}\left(k_{B}\left|\mathbf{r}_{m}-\mathbf{r}'\right|\right)d\mathbf{r}'$),
$k_{B}$ being the wavenumber ($k_{B}\triangleq\frac{2\pi}{\lambda_{B}}$),
while $H_{0}^{\left(2\right)}\left(\,.\,\right)$ is the $0$-th order
Hankel function of the second kind.

\section{\noindent \emph{ISP} Solution (\emph{MS-DCIE} Inversion Method)\label{sec:solution_approach}}

\noindent The solution of the inverse problem formulated in Sect.
\ref{sec:mathematical-formulation} (i.e., the estimation of the spatial
distribution within $\mathcal{H}$ of both the differential modified
contrast, $\overline{\chi}_{\Delta}$, and the differential equivalent
currents, \{$\overline{J}_{\Delta}^{\left(v\right)}$; $v=1,...,V$\})
is addressed with an approach based on the application of the \emph{MS}
scheme to the \emph{DCIE} formulation of the \emph{ISP}. More specifically,
the scattering-data inversion is carried out by means of an iterative
strategy that performs $S$ successive {}``zooming'' steps. At each
$s$-th ($s=1,...,S$; $s$ being the step index) step, the spatial
distribution of the generic unknown $\overline{\varphi}$ ($\overline{\varphi}\in\left\{ \overline{\chi}_{\Delta},\,\left(\overline{J}_{\Delta}^{\left(v\right)};\, v=1,...,V\right)\right\} $)
in the corresponding \emph{RoI}, $\mathcal{S}_{\varphi}$, which is
the portion of the investigation domain $\mathcal{H}$ where $\varphi\left(\mathbf{r}_{n}\right)\ne0$,
is retrieved by means of an inversion algorithm as applied to (\ref{eq:DCIE-data-discretized})
and (\ref{eq:DCIE-state-discretized}). Such a reconstruction is then
exploited to improve the \emph{RoI} estimate by also enhancing the
spatial resolution of the retrieval. The process is repeated until
a data-matching convergence criterion holds true.

\noindent The implementation of such a multi-level process needs:
(\emph{a}) to define the \emph{RoI} for both the primary, $\mathcal{S}_{\Delta\chi}$,
and the secondary, $\mathcal{S}_{\Delta J}$, unknowns (Sect. {}``\emph{RoI}
\emph{Definition}''); (\emph{b}) to choose an inversion method to
process, at each $s$-step ($s=1,...,S$), the scattering data for
determining the spatial distribution of the generic unknown $\overline{\varphi}$
in the corresponding \emph{RoI}, $\mathcal{S}_{\varphi}^{\left(s\right)}$
(Sect. {}``\emph{SOM Inversion}''); (\emph{c}) to define a suitable
cost function that faithfully links the \emph{ISP} at hand with its
mathematical formulation within the \emph{DCIE} framework so that
the actual \emph{ISP} solution coincides with the global minimum of
the cost function itself (Sect. {}``\emph{Cost Function Definition}'');
(\emph{d}) to customize the meta-level \emph{MS} strategy to both
such a formulation (i.e., problem unknowns and cost function) and
the integration with the optimization level (Sect. {}``\emph{MS Implementation}'').
These items will be detailed or briefly recalled in the following.

\subsubsection*{\noindent \emph{RoI} Definition\label{sub:RoI-Definition}}

\noindent To properly address this issue, let us first recall the
case of the \emph{MS} as applied to the \emph{CIE} formulation for
the \emph{ISP} with homogeneous media. There, the \emph{CIE} unknowns
are the modified contrast function, $\chi$, and the $V$ equivalent
currents, \{$\overline{J}^{\left(v\right)}$; $v=1,...,V$\}, whose
supports, namely $\mathcal{S}_{\chi}$ and $\mathcal{S}_{J}$, coincide
with the extension $\mathcal{S}_{\Delta\tau}$ of the unknown object
$\mathcal{O}$, which is modeled by $\tau_{\Delta}$ ($\mathcal{S}_{\chi}=\mathcal{S}_{J}\equiv\mathcal{S}_{\Delta\tau}$).
Therefore, the \emph{RoI}s of both the primary unknown and the secondary
one, which are identified at each $s$-th ($s=1,...,S$) \emph{MS}
step (i.e., $\mathcal{S}_{\chi}^{\left(s\right)}$ and $\mathcal{S}_{J}^{\left(s\right)}$),
are the same region where the unknown object is most likely to be
present \cite{Zhong 2020} (i.e., $\mathcal{S}_{\chi}^{\left(s\right)}=\mathcal{S}_{J}^{\left(s\right)}\equiv\mathcal{S}_{\Delta\tau}^{\left(s\right)}$).

\noindent Otherwise, the problem unknowns for the \emph{DCIE} formulation
are the differential modified contrast, $\chi_{\Delta}$, and the
$V$ differential equivalent currents, \{$\overline{J}_{\Delta}^{\left(v\right)}$;
$v=1,...,V$\}, $\mathcal{S}_{\Delta\chi}$ and $\mathcal{S}_{\Delta J}$
being the corresponding supports, respectively. While $\mathcal{S}_{\Delta\chi}$
is equal to the area occupied by the unknown scatterer and the standard
\emph{RoI} definition applies ($\mathcal{S}_{\Delta\chi}\equiv\mathcal{S}_{\Delta\tau}$),
$\mathcal{S}_{\Delta J}$ might also span outside the object region
$\mathcal{O}$ (i.e., $\mathcal{S}_{\Delta J}\supseteq\mathcal{S}_{\Delta\tau}$)
since the equivalent currents, \{$\overline{J}^{\left(v\right)}$;
$v=1,...,V$\}, as well as the differential ones, \{$\overline{J}_{\Delta}^{\left(v\right)}$;
$v=1,...,V$\}, are here induced also in the external inhomogeneous
host medium. As detailed in Appendix B, the \emph{RoI} of $\overline{J}_{\Delta}^{\left(v\right)}$
at the $s$-th ($s=1,...,S$) \emph{MS} step turns out to be\begin{equation}
\mathcal{S}_{\Delta J}^{\left(s\right)}=\mathcal{S}_{\Delta\chi}^{\left(s\right)}\cup\mathcal{S}_{\chi_{H}}.\label{eq:current_support}\end{equation}

\subsubsection*{\emph{SOM} Inversion\label{sub:SOM-Inversion}}

\noindent Concerning the inversion method, the \emph{SOM} \cite{Chen 2018}
is adopted for the following reasons. First, the scattering operator
$\overline{\overline{\mathcal{G}}}_{{\rm ext}}$ in (\ref{eq:DCIE-data-discretized})
is compact, thus the \emph{ISP} at hand is ill-posed \cite{Bucci 1997}.
In particular, the \emph{ISP} is not unique because of the (possible)
presence of non-radiating components of the induced equivalent currents,
which do not contribute to the scattered field $\xi_{sca,\Delta}^{\left(v\right)}$
collected in the observation domain external to $\mathcal{H}$. To
recover uniqueness, it is then necessary to take into account these
components during the inversion process as the \emph{SOM} does. Second,
thanks to the properties of the \emph{DCIE} formulation, the \emph{ISP}
nonlinearity is mitigated, thus the use of a deterministic fast inversion
method, instead of computationally-demanding {}``bare'' global optimization
techniques \cite{Goudos 2021}, could be profitable.

\noindent According to the guidelines in \cite{Chen 2010}, the \emph{SOM}
is customized to the \emph{DCIE} formulation, towards the integration
within the \emph{MS} processing scheme, as follows. The $v$-th ($v=1,...,V$)
differential equivalent current $\overline{J}_{\Delta}^{\left(v\right)}$
is decomposed in two parts, namely the deterministic component, $\overline{J}_{\Delta,DP}^{\left(v\right)}$,
and the ambiguous one, $\overline{J}_{\Delta,AP}^{\left(v\right)}$,
which includes the non-radiating terms\begin{equation}
\overline{J}_{\Delta}^{\left(v\right)}=\overline{J}_{\Delta,DP}^{\left(v\right)}+\overline{J}_{\Delta,AP}^{\left(v\right)}.\label{eq:current_decomposition}\end{equation}
The former, $\overline{J}_{\Delta,DP}^{\left(v\right)}$ ($v=1,...,V$),
is computed from (\ref{eq:DCIE-data-discretized}) by applying the
singular value decomposition (\emph{SVD}) to the external Green's
matrix $\overline{\overline{\mathcal{G}}}_{{\rm ext}}$. It turns
out that\begin{equation}
\overline{J}_{\Delta,DP}^{\left(v\right)}=\sum_{n=1}^{N_{th}}\frac{\left(\overline{U}_{n}\right)^{*}\cdot\overline{\xi}_{sca,\Delta}^{\left(v\right)}}{\sigma_{n}}\overline{W}_{n}\label{eq:deterministic_currents}\end{equation}
where $^{*}$ stands for conjugate transposition, while $\cdot$ denotes
the scalar product. Moreover, \{$\sigma_{n}$; $n=1,...,N$\} are
the $N$ singular values of $\overline{\overline{G}}_{ext}$, while
\{$\overline{U}_{n}$; $n=1,...,N$\} and \{$\overline{W}_{n}$; $n=1,...,N$\}
are the $M$-size left-singular vectors and the $N$-size right-singular
vectors, respectively. In (\ref{eq:deterministic_currents}), $N_{th}$
is the \emph{SVD} truncation threshold, which is adaptively set as
follows \cite{Oliveri 2011}\begin{equation}
N_{th}=\arg\min_{N'}\left\{ \left|\frac{\sum_{n=1}^{N'}\sigma_{n}}{\sum_{n=1}^{N}\sigma_{n}}-\alpha\right|\right\} ,\label{eq:adaptive_svd_threshold}\end{equation}
$\alpha$ ($0<\alpha\le1$) being a real user-defined calibration
parameter as detailed in Sect. \ref{sec:results}.

\noindent The ambiguous current component, $\overline{J}_{\Delta,AP}^{\left(v\right)}$
($v=1,...,V$), is related to the smaller singular values of $\overline{\overline{G}}_{ext}$
and it is yielded by the linear combination of the remaining ($N-N_{th}$)
right-singular vectors\begin{equation}
\overline{J}_{\Delta,AP}^{\left(v\right)}=\sum_{n=N_{th}+1}^{N}c_{n-N_{th}}^{\left(v\right)}\overline{W}_{n}\label{eq:ambiguous_current}\end{equation}
where $\overline{c}^{\left(v\right)}$ $=$ \{$c_{n}^{\left(v\right)}$;
$n=1,...,\left(N-N_{th}\right)$\} is the unknown complex algebraic
vector of the weights of the $v$-th ($v=1,...,V$) ambiguous current,
while $\overline{\overline{c}}$ is the corresponding $\left(N-N_{th}\right)\times V$
size matrix ($\overline{\overline{c}}$ $\triangleq$ \{$\overline{c}^{\left(v\right)}$;
$v=1,...,V$\}).

\subsubsection*{\noindent Cost Function Definition\label{sub:Cost-Function-Definition}}

\noindent The cost function $\Psi$ quantifies the error in fulfilling
(\ref{eq:DCIE-data-discretized}) and (\ref{eq:DCIE-state-discretized})
and it is defined as\begin{equation}
\Psi\left(\overline{\overline{c}},\overline{\chi}_{\Delta}\right)=\sum_{v=1}^{V}\left[\frac{\Psi_{{\rm data}}^{\left(v\right)}\left(\overline{c}^{\left(v\right)}\right)}{\left\Vert \overline{\xi}_{sca,\Delta}^{\left(v\right)}\right\Vert ^{2}}+\frac{\Psi_{{\rm state}}^{\left(v\right)}\left(\overline{c}^{\left(v\right)},\overline{\chi}_{\Delta}\right)}{\left\Vert \overline{\xi}_{{\rm inc}}^{\left(v\right)}\right\Vert ^{2}}\right]\label{eq:cost}\end{equation}
where $\left\Vert \cdot\right\Vert $ is the $\ell_{2}$-norm, while
the data equation mismatch, $\Psi_{{\rm data}}^{\left(v\right)}$,
is derived from (\ref{eq:DCIE-data-discretized})

\noindent \begin{equation}
\Psi_{{\rm data}}^{\left(v\right)}\left(\overline{c}^{\left(v\right)}\right)=\left\Vert \overline{\overline{\mathcal{G}}}_{{\rm ext}}\left[\overline{J}_{\Delta,DP}^{\left(v\right)}+\sum_{n=N_{th}+1}^{N}c_{n-N_{th}}^{\left(v\right)}\overline{W}_{n}\right]-\overline{\xi}_{sca,\Delta}^{\left(v\right)}\right\Vert ^{2},\label{eq:cost_data}\end{equation}
and the state equation mismatch $\Psi_{{\rm state}}^{\left(v\right)}$
stems from (\ref{eq:DCIE-state-discretized})\begin{equation}
\Psi_{{\rm state}}^{\left(v\right)}\left(\overline{c}^{\left(v\right)},\overline{\chi}_{\Delta}\right)\triangleq\left\Vert \Psi_{AP}^{\left(v\right)}\left(\overline{c}^{\left(v\right)},\overline{\chi}_{\Delta}\right)-\Psi_{DP}^{\left(v\right)}\left(\overline{\chi}_{\Delta}\right)\right\Vert ^{2},\label{eq:cost_state}\end{equation}
 $\Psi_{AP}^{\left(v\right)}$ and $\Psi_{DP}^{\left(v\right)}$ being
the terms dependent on the \emph{AP} and the \emph{DP} components,
respectively, whose expressions are \begin{equation}
\begin{array}{rl}
\Psi_{AP}^{\left(v\right)}\left(\overline{c}^{\left(v\right)},\overline{\chi}_{\Delta}\right)\triangleq & \overline{\beta}\otimes\overline{J}_{\Delta,AP}^{\left(v\right)}\left(\overline{c}^{\left(v\right)}\right)\\
 & -\left(\overline{\chi}_{\Delta}+\overline{\chi}_{H}\right)\otimes\overline{\overline{\mathcal{G}}}_{{\rm int}}\overline{J}_{\Delta,AP}^{\left(v\right)}\left(\overline{c}^{\left(v\right)}\right)\\
 & -\left(\overline{\chi}_{\Delta}+\overline{\chi}_{H}\right)\otimes\overline{\beta}\otimes\overline{J}_{\Delta,AP}^{\left(v\right)}\left(\overline{c}^{\left(v\right)}\right)\end{array}\label{eq:ELLE}\end{equation}
and\begin{equation}
\begin{array}{rl}
\Psi_{DP}^{\left(v\right)}\left(\overline{\chi}_{\Delta}\right)\triangleq & -\overline{\beta}\otimes\overline{J}_{\Delta,DP}^{\left(v\right)}+\overline{\chi}_{\Delta}\otimes\left(\overline{\xi}_{H}^{\left(v\right)}+\overline{\beta}\otimes\overline{J}_{H}^{\left(v\right)}\right)\\
 & +\left(\overline{\chi}_{\Delta}+\overline{\chi}_{H}\right)\otimes\left(\overline{\overline{\mathcal{G}}}_{{\rm int}}\overline{J}_{\Delta,DP}^{\left(v\right)}+\overline{\beta}\otimes\overline{J}_{\Delta,DP}^{\left(v\right)}\right).\end{array}\label{eq:GAMMA}\end{equation}

\subsubsection*{\emph{MS} Implementation\label{sub:IMSA-Implementation}}

\noindent Within the \emph{DCIE} framework, the algorithmic implementation
of the \emph{MS} scheme as integrated into the \emph{SOM}-based inversion
can be described through the following multi-step iterative ($i$
being the iteration index) process:

\begin{itemize}
\item \noindent \emph{Initialization} - Initialize the \emph{MS} step index
($s=1$) and the \emph{RoI} for both the unknowns to the whole investigation
domain {[}$\mathcal{S}_{\Delta\chi}^{\left(1\right)}=\mathcal{S}_{\Delta J}^{\left(1\right)}=\mathcal{H}$
- Fig. 2(\emph{a}){]};
\item \noindent \emph{MS Loop}

\begin{itemize}
\item \noindent \emph{Unknowns Setup} ($i=0$) - If $s=1$, then reset the
unknowns (i.e., $\left[\overline{\chi}_{\Delta}^{\left(s\right)}\right]_{i}=\overline{0}$
and $\left[\overline{\overline{c}}^{\left(s\right)}\right]_{i}=\overline{\overline{0}}$).
Otherwise (i.e., $s>1$), map the trial solution from the previous
zooming step into the current $s$-th discretization grids of the
\emph{RoI}s \{$\mathcal{S}_{\varphi}^{\left(s\right)}$; $\varphi\in\left\{ \Delta\chi,\,\Delta J\right\} $\}
(i.e., $\left[\overline{\chi}_{\Delta}^{\left(s\right)}\right]_{i}=\Phi_{\Delta\chi}\left\{ \left[\overline{\chi}_{\Delta}^{\left(s-1\right)}\right]_{i};\,\mathcal{S}_{\Delta\chi}^{\left(s\right)}\right\} $
and $\left[\left[\overline{J}_{\Delta}^{\left(v\right)}\right]^{\left(s\right)}\right]_{i}=\Phi_{\Delta J}\left\{ \left[\left[\overline{J}_{\Delta}^{\left(v\right)}\right]^{\left(s-1\right)}\right]_{i};\,\mathcal{S}_{\Delta J}^{\left(s\right)}\right\} $,
$\left[\overline{J}_{\Delta}^{\left(v\right)}\right]^{\left(s-1\right)}$
being equal to $\left[\overline{J}_{\Delta}^{\left(v\right)}\right]^{\left(s-1\right)}=\overline{J}_{\Delta,DP}^{\left(v\right)}+\sum_{n=N_{th}+1}^{N}\left[c_{n-N_{th}}^{\left(v\right)}\right]^{\left(s-1\right)}\overline{W}_{n}$),
$\Phi_{\varphi}$ being the mapping operator from the grid of $\mathcal{S}_{\varphi}^{\left(s-1\right)}$
to the finer one of $\mathcal{S}_{\varphi}^{\left(s\right)}$;
\item \noindent \emph{Scattering-Data Inversion} - Compute the \emph{DCIE}
regularization parameter vector $\overline{\beta}^{\left(s\right)}$
$=$ \{$\beta\left(\mathbf{r}_{n}\right)$; $n=1,...,N$\} by setting
its $N$ entries to the value\begin{equation}
\beta^{\left(s\right)}=\gamma\times\max_{\mathbf{r}_{n}\in\mathcal{S}_{\Delta J}^{(s)}}\left|\int_{\mathcal{S}_{\Delta J}^{(s)}}\mathcal{G}_{{\rm int}}\left(\mathbf{r}_{n},\mathbf{r}'\right)d\mathbf{r}'\right|\label{eq:beta_choice}\end{equation}
 where $\gamma$ is a control parameter \cite{Zhong 2016}. Retrieve
the $s$-th step set of unknowns ($\overline{\chi}_{\Delta}^{\left(s\right)}$,
$\overline{\overline{c}}^{\left(s\right)}$) within the corresponding
\emph{RoI} $\mathcal{S}_{\varphi}^{\left(s\right)}$ by solving the
following optimization problem\begin{equation}
\left(\overline{\chi}_{\Delta}^{\left(s\right)},\overline{\overline{c}}^{\left(s\right)}\right)=\arg\min_{\overline{\chi}_{\Delta},\overline{\overline{c}}}\left\{ \Psi\left(\overline{\chi}_{\Delta},\overline{\overline{c}},\right)\right\} \label{eq:minimization_problem}\end{equation}
 with $I$ iterations of the deterministic Polak-Ribiere version of
the Conjugate Gradient (\emph{CG}) algorithm \cite{Ye 2011} (i.e.,
$\overline{\chi}_{\Delta}^{\left(s\right)}=\left[\overline{\chi}_{\Delta}^{\left(s\right)}\right]_{i=I}$,
$\overline{\overline{c}}^{\left(s\right)}=\left[\overline{\overline{c}}^{\left(s\right)}\right]_{i=I}$)
starting from $\left[\overline{\chi}_{\Delta}^{\left(s\right)}\right]_{i=0}$
and $\left[\overline{\overline{c}}^{\left(s\right)}\right]_{i=0}$.
Update the trial differential current $\left[\overline{J}_{\Delta}^{\left(v\right)}\right]^{\left(s\right)}$
through (\ref{eq:current_decomposition}) by using (\ref{eq:deterministic_currents})
and (\ref{eq:ambiguous_current}) with $c_{n-N_{th}}^{\left(v\right)}$
$\leftarrow$ $\left[c_{n-N_{th}}^{\left(v\right)}\right]^{\left(s\right)}$;
\item \noindent \emph{Step Check} - Halt the \emph{MS} loop if the maximum
number of zooming steps is reached {[}i.e., $s=S$ - Fig. 2(\emph{d}){]},
define the estimated solution by setting $\overline{\chi}_{\Delta}^{opt}=\overline{\chi}_{\Delta}^{\left(S\right)}$
and $\left[\overline{J}_{\Delta}^{\left(v\right)}\right]^{opt}=\left[\overline{J}_{\Delta}^{\left(v\right)}\right]^{\left(S\right)}$as
well as $\beta^{opt}=\beta^{\left(S\right)}$, and go to the {}``\emph{Termination}'';
\item \noindent \emph{RoI} \emph{Update} - Apply the {}``filtering and
clustering'' operations \cite{Caorsi 2003} on $\overline{\chi}_{\Delta}^{\left(s\right)}$
to determine the corresponding new \emph{RoI}, $\mathcal{S}_{\Delta\chi}^{\left(s+1\right)}$
{[}Figs. 2(\emph{b})-2(\emph{c}){]}, by defining its center, $\mathbf{r}_{\Delta\chi}^{\left(s+1\right)}$
{[}$\mathbf{r}_{\Delta\chi}^{\left(s+1\right)}$ $=$ ($x_{\Delta\chi}^{\left(s+1\right)}$,
$y_{\Delta\chi}^{\left(s+1\right)}$){]}, and side, $L_{\Delta\chi}^{\left(s+1\right)}$,
as follows\begin{equation}
\varsigma_{\Delta\chi}^{\left(s+1\right)}=\frac{\sum_{n=1}^{N}\varsigma_{n}^{\left(s\right)}\chi_{\Delta}^{\left(s\right)}\left(\mathbf{r}_{n}\right)}{\sum_{n=1}^{N}\chi_{\Delta}^{\left(s\right)}\left(\mathbf{r}_{n}\right)}\label{eq:doi_barycenter}\end{equation}
($\varsigma\in\left\{ x;\, y\right\} $) and\begin{equation}
L_{\Delta\chi}^{\left(s+1\right)}=2\times\frac{\sum_{n=1}^{N}\left\Vert \mathbf{r}_{n}^{\left(s\right)}-\mathbf{r}_{\Delta\chi}^{\left(s+1\right)}\right\Vert \chi_{\Delta}^{\left(s\right)}\left(\mathbf{r}_{n}\right)}{\sum_{n=1}^{N}\chi_{\Delta}^{\left(s\right)}\left(\mathbf{r}_{n}\right)}.\label{eq:doi_edge}\end{equation}
Identify the new \emph{RoI} for $\overline{J}_{\Delta}^{\left(v\right)}$
(i.e., $\mathcal{S}_{\Delta J}^{\left(s+1\right)}$) through (\ref{eq:current_support});
\end{itemize}
\item \noindent \emph{RoI Check} - Terminate the \emph{MS} loop if the zooming
factor $\eta^{\left(s\right)}$, which is defined as\begin{equation}
\eta^{\left(s\right)}=\frac{\left|L_{\Delta\chi}^{\left(s+1\right)}-L_{\Delta\chi}^{\left(s\right)}\right|}{L_{\Delta\chi}^{\left(s+1\right)}},\label{eq:zooming}\end{equation}
is below a user-defined threshold $\eta_{min}$ (i.e., $\eta^{\left(s\right)}\le\eta_{min}$)
and set the problem solution to the current trial one (i.e., $\overline{\chi}_{\Delta}^{opt}=\overline{\chi}_{\Delta}^{\left(s\right)}$
and $\left[\overline{J}_{\Delta}^{\left(v\right)}\right]^{opt}=\left[\overline{J}_{\Delta}^{\left(v\right)}\right]^{\left(s\right)}$
as well as $\beta^{opt}=\beta^{\left(s\right)}$). Otherwise, update
the \emph{MS} loop index {[}i.e., $s\leftarrow\left(s+1\right)${]}
and restart the {}``\emph{MS Loop}'';
\item \emph{Termination} - Given $\overline{\chi}_{\Delta}^{opt}$, compute
the modified contrast $\overline{\chi}^{opt}$ (i.e., $\overline{\chi}^{opt}=\overline{\chi}_{\Delta}^{opt}+\overline{\chi}_{H}$)
to output the estimated contrast vector $\overline{\tau}^{opt}$ $=$
\{$\tau^{opt}\left(\mathbf{r}_{n}\right)$; $n=1,...,N$\} whose $n$-th
($n=1,...,N$) entry is equal to\begin{equation}
\tau^{opt}\left(\mathbf{r}_{n}\right)=\frac{\chi^{opt}\left(\mathbf{r}_{n}\right)}{\beta^{opt}\left[1-\chi^{opt}\left(\mathbf{r}_{n}\right)\right]}.\label{eq:est_contrast}\end{equation}

\end{itemize}

\section{\noindent Numerical and Experimental Results\label{sec:results}}

\noindent An extensive set of tests has been carried out to assess
the reliability and the performance of the proposed \emph{IS} method.
The most representative ones have been selected and reported in the
following to (\emph{i}) provide some general rules for calibrating
the control parameters of the \emph{MS-DCIE}, (\emph{ii}) show the
effectiveness of such an implementation of the \emph{MS} scheme to
yield high-accuracy reconstructions when dealing with inhomogeneous
scenarios, and (\emph{iii}) provide a comparative analysis with other
state-of-the-art approaches.

\noindent To quantitatively estimate the accuracy of the reconstructions,
let us consider two metrics. The one is a local metric and it is the
local error function, $\mathbb{E}$, defined as\begin{equation}
\mathbb{E}\left(\mathbf{r}\right)\triangleq\frac{\tau\left(\mathbf{r}\right)-\tau^{opt}\left(\mathbf{r}\right)}{\tau\left(\mathbf{r}\right)+1},\label{eq:error_map}\end{equation}
$\tau$ and $\tau^{opt}$ being the actual and the reconstructed contrast,
respectively. The other metric gives a global/integral information
on the imaging process and it is given by \begin{equation}
\Xi\triangleq\frac{1}{\mathcal{A}}\sum_{r_{n}\in\mathcal{A}}\left|\mathbb{E}\left(\mathbf{r}_{n}\right)\right|,\label{eq:error_definition}\end{equation}
where $\mathcal{A}$ is the area of the region where the error figure
is evaluated. More specifically, $\Xi=\Xi_{tot}$, $\Xi=\Xi_{int}$,
and $\Xi=\Xi_{ext}$ if $\mathcal{A=H}$, $\mathcal{A=S}_{\Delta\tau}$,
and $\left\{ \mathcal{H}\smallsetminus\mathcal{S}_{\Delta\tau}\right\} $,
respectively.

\noindent Unless otherwise stated, the side of the square investigation
domain $\mathcal{H}$ has been set to $L_{\mathcal{H}}=3\times\lambda_{B}$,
$\lambda_{B}$ being the wavelength in the external background. Such
a benchmark scenario has been probed by $V=27$ directions, while
the scattering data have been collected in $M=27$ uniformly-spaced
locations on a circle of radius $\rho=2.2\times\lambda_{B}$ and centered
in the host medium region $\mathcal{H}$. In all numerical tests,
the scattered field samples in (\ref{eq:DCIE-data-discretized}) have
been synthetically-generated by solving the forward (\emph{FW}) scattering
problem with a Method of Moments (\emph{MoM}) solver and discretizing
the investigation domain with $N_{FW}=80\times80$ square cells. Moreover,
the data samples have been blurred with an additive Gaussian noise,
characterized by a signal-to-noise ratio (\emph{SNR}), to model realistic
measurement conditions. As for the data inversion, a coarser discretization
grid has been assumed (i.e., $N=30\times30$) to represent the contrast
profile at each $s$-th ($s=1,...,S$) \emph{MS} step. The maximum
number of \emph{MS} steps has been set to $S=6$, while $I=200$ \emph{SOM}
iterations have been executed at each $s$-th \emph{MS} step according
to the guidelines in \cite{Zhong 2020}. Lastly, the value of $\eta_{min}=0.2$
has been selected for the \emph{MS} zooming threshold according to
\cite{Salucci 2021}.

\subsection{\noindent Parameters Calibration and Sensitivity Analysis \label{sub:Parameters-Calibration-and}}

\noindent The behavior of the \emph{MS-DCIE} inversion method mainly
depends on the values of two control parameters: $\alpha$, which
determines the \emph{SVD} truncation threshold $N_{th}$, and $\gamma$,
which controls the \emph{DCIE} regularization at each \emph{MS} step.
To determine their optimal trade-off setup, a sensitivity analysis
has been performed and the representative results from the benchmark
scenario in Fig. 1 are reported hereinafter for illustration purposes.
More in detail, an off-centered square scatterer $0.7\lambda_{B}$-sided
with unknown differential contrast $\tau_{\Delta}=2$ {[}Fig. 1(\emph{c}){]},
located at $\left(x_{O},y_{O}\right)=\left(0.3\,\lambda_{B},\,0.15\,\lambda_{B}\right)$
{[}Fig. 1(\emph{b}){]}, and surrounded by a known host domain composed
by a centered square ring of thickness $0.3\lambda_{B}$ with $\tau_{H}^{\left(1\right)}=0.5$
{[}Fig. 1(\emph{b}){]} and a remaining area of contrast $\tau_{H}^{\left(2\right)}=0.0$,
has been retrieved by applying the \emph{MS-DCIE} strategy with different
values of the control parameters and for different \emph{SNR}s.

\noindent The analysis outcomes are summarized in Fig. 3 where the
behavior of the total reconstruction error $\Xi_{tot}$ versus $\alpha$
{[}Fig. 3(\emph{a}){]} and $\gamma$ {[}Fig. 3(\emph{b}){]} is shown
in correspondence with different noise level on the scattering data.
It turns out that the \emph{$\alpha$} value has a low impact on the
\emph{MS-DCIE} performance, $\Xi_{tot}$ being almost flat or with
a limited range of variations for a given value of \emph{SNR}. On
the contrary, the $\gamma$ parameter has a more notable effect on
the reconstruction accuracy since it {}``controls'' the non-linearity
of the inverse problem at hand \cite{Zhong 2016}. The optimal setup
for $\alpha$ and $\gamma$ has then been chosen according to the
following rule\begin{equation}
\zeta^{*}=\frac{{\displaystyle \int_{SNR}\arg\min_{\zeta}\left\{ \left.\Xi_{tot}\right\rfloor _{SNR}^{\zeta}\right\} dSNR}}{{\displaystyle \int_{SNR}dSNR}}\end{equation}
($\zeta=\left\{ \alpha;\gamma\right\} $) and the result has been
$\left(\alpha^{*},\gamma^{*}\right)=\left(0.4,1.4\right)$.

\subsection{\noindent Numerical Assessment\label{sub:Numerical-Assessment}}

\noindent Once calibrated, the performance of the \emph{MS-DCIE} have
been assessed in comparison with competitive state-of-the-art inversion
approaches. First, the comparison with the single-resolution \emph{DCIE}%
\footnote{\noindent According to the guidelines in \cite{Zhong 2016}, the regularization
parameter has been set to $\beta=2$, while the investigation domain
$\mathcal{H}$ has been partitioned in $N_{DCIE}=46\times46$ square
cells.%
} has been carried out. The test case refers to the {}``Square''
scatterer in Sect. \ref{sub:Parameters-Calibration-and} and the scattering
data have been blurred with $SNR=20$ {[}dB{]}. Figure 4 shows the
evolution of the cost function, $\Psi$, the total error, $\Xi_{tot}$,
and the \emph{MS} zooming factor, $\eta^{(s)}$, during the \emph{MS}
process. Moreover, the color-maps of the retrieved differential contrast,
$\tau_{\Delta}$, are reported in Fig. 5, as well. As it can be observed,
the estimated $\mathcal{S}_{\Delta\tau}$\emph{-RoI} shrinks around
the actual object position {[}Figs. 5(\emph{a})-5(\emph{c}){]} until
there is an \emph{}almost perfect match with the scatterer support
{[}Fig. 5(\emph{d}){]}. This occurs at $s=4$ when the value of $\eta^{(s)}$
falls below the threshold $\eta_{min}$ (Fig. 4) and the inversion
process is stopped. It is worth noticing that during the \emph{MS}
procedure both the cost function and the reconstruction error decrease
(Fig. 4) by pointing out the enhancement of the inversion accuracy
in correspondence with a better fitting with the scattering data.
The contrast profile $\tau^{opt}$ derived by the \emph{MS-DCIE} {[}Fig.
5(\emph{e}){]} is then compared with that retrieved with the \emph{DCIE}
method {[}Fig. 5(\emph{f}){]}. Both pictorially and quantitatively,
it is evident that the \emph{MS}-based inversion significantly improves
the reconstruction inside(outside) the object support. Indeed, the
internal/external error reduces of about $\Theta\Xi_{int}=77$ \%($\Theta\Xi_{ext}=52$
\%) ($\Theta\Xi\triangleq\frac{\Xi^{DCIE}-\Xi^{MS-DCIE}}{\Xi^{DCIE}}$)
with respect to the single-resolution approach.

\noindent The comparative assessment has been extended next to the
strategies based on the \emph{DLSIE} formulation \cite{Xu 2018b}\cite{Zhong 2020},
which are referred in the following as \emph{MS-DLSIE} and \emph{DLSIE},
respectively, and the results from the analysis on scenario {}``Circular-Ring''
object, shown in the inset of Fig. 6, are reported hereinafter. Figure
6 plots the values of $\Xi_{tot}$ versus the \emph{SNR} for the different
inversion methods. One can observe that the \emph{DCIE} formulation
turns out to be better than the \emph{DLSIE} one as visually confirmed
by the error maps at $SNR=20$ {[}dB{]} reported in Fig. 7 {[}e.g.,
Fig. 7(\emph{a}) vs. Fig. 7(\emph{c}) and Fig. 7(\emph{b}) vs. Fig.
7(\emph{d}){]}. Moreover, as expected, the multi-resolution implementations
yield better reconstructions with lower errors than their single-resolution
counterparts for any \emph{SNR} (Fig. 6 - $\Xi_{tot}^{MS-DCIE}<\Xi_{tot}^{DCIE}$
and $\Xi_{tot}^{MS-DLSIE}<\Xi_{tot}^{DLSIE}$ ) by reducing the artifacts
at the scatterer edges and outside the object within the host medium
{[}e.g., Fig. 7(\emph{a}) vs. Fig. 7(\emph{b}) and Fig. 7(\emph{c})
vs. Fig. 7(\emph{d}){]}.

\noindent The performance of the \emph{MS-DCIE} have been assessed
also against the scatterer permittivity, thus verifying its accuracy
against a higher non-linearity of the \emph{ISP}. Towards this end,
the actual contrast of the circular ring object has been varied in
the range from $\tau_{\Delta}=2$ up to $\tau_{\Delta}=4$, while
keeping the noise level to $SNR=20$ {[}dB{]}. The plots of the reconstruction
indexes in Fig. 8 point out that the inversion becomes more and more
difficult when the differential contrast becomes stronger and stronger
since all errors get larger, but they also point out the effectiveness
of the \emph{MS} in dealing with stronger scatterers with an average
improvement of the total(external) error, $\Xi_{tot}$($\Xi_{ext}$),
of about 45 \%(48 \%) with respect to the \emph{DCIE} single-resolution
strategy. Moreover, the higher the contrast is the greater is the
advantage of using the \emph{MS} for imaging the unknown object support
since the internal error gap $\Theta\Xi_{int}$ grows as $\tau_{\Delta}$
tends to $\tau_{\Delta}=4$ (i.e., $\left.\Theta\Xi_{int}\right\rfloor _{\tau_{\Delta}=2}=12$
\%, $\left.\Theta\Xi_{int}\right\rfloor _{\tau_{\Delta}=3}=37$ \%,
and $\left.\Theta\Xi_{int}\right\rfloor _{\tau_{\Delta}=4}=51$ \%).
These outcomes are highlighted by the error maps related to $\tau_{\Delta}=2$
{[}Figs. 7(\emph{a})-7(\emph{b}){]}, $\tau_{\Delta}=3$ {[}Fig. 9(\emph{a})
and Fig. 9(\emph{c}){]}, and $\tau_{\Delta}=4$ {[}Fig. 9(\emph{b})
and Fig. 9(\emph{d}){]}.

\noindent Another numerical experiment has been devoted to assess
the \emph{MS-DCIE} when dealing with lossy scatterers \emph{}(i.e.,
$\sigma_{\Delta}\ne0$ {[}S/m{]}). By keeping the ring target of Fig.
6, its differential contrast has been set to $\textrm{Re}\left\{ \tau_{\Delta}\right\} =2$,
while the actual scatterer conductivity has been varied within the
range $10^{-3}$ {[}S/m{]} $\le$ $\sigma_{\Delta}$ $\le$ $1$ {[}S/m{]}
( $f=300$ {[}MHz{]}). Unlike the dependence on the scatterer permittivity
in Fig. 8, the accuracy improvement of the \emph{MS} strategy reduces
as the object conductivity increases (e.g., $\left.\Theta\Xi_{tot}\right\rfloor _{\sigma_{\Delta}=10^{-3}\,[S/m]}=56$
\%, $\left.\Theta\Xi_{tot}\right\rfloor _{\sigma_{\Delta}=10^{-2}\,[S/m]}=51$
\%, $\left.\Theta\Xi_{tot}\right\rfloor _{\sigma_{\Delta}=10^{-1}\,[S/m]}=44$
\%, and $\left.\Theta\Xi_{tot}\right\rfloor _{\sigma_{\Delta}=10^{0}\,[S/m]}=24$
\% - Fig. 10). However, it is worth noticing that locally the \emph{MS-DCIE}
still significantly outperforms the \emph{DCIE} in retrieving the
imaginary part of the contrast distribution by better detailing the
contours and the support of the unknown circular ring as highlighted
by the error maps in Fig. 11 ($\sigma_{\Delta}=10^{-1}\,[S/m]$).

\noindent The last numerical experiment is aimed at evaluating how
an inaccurate knowledge of the host medium contrast, $\tau_{H}\left(\mathbf{r}\right)$,
affects the \emph{MS-DCIE} reconstruction. Towards this end, the \emph{a-priori}
information on the host medium has been supposed to be affected by
an uncertainty proportional to the constant $\delta$\begin{equation}
\widehat{\tau}_{H}\left(\mathbf{r}\right)=\tau_{H}\left(\mathbf{r}\right)\times\left(1+\delta\right)\label{eq: Masciulla}\end{equation}

\noindent and the analysis has been carried out on the {}``Square
object'' scenario with $\tau_{\Delta}=2$ and $SNR=20$ {[}dB{]}.
By varying the value of $\delta$ from $0$ \% up to $100$ \%, it
is not surprising that all error indexes get worse when the \emph{a-priori}
information on the host medium is more and more imprecise (Fig. 12).
However, the \emph{MS-DCIE} always yields better reconstructions as
quantitatively confirmed by the total error gap $\Theta\Xi_{tot}$
values (i.e., $\Theta\Xi_{tot}\ge32$ \%) even though the improvements
with respect to the \emph{DCIE} diminish as the host medium knowledge
is getting more inaccurate (i.e., $\left.\Theta\Xi_{tot}\right\rfloor _{\delta=0.05}=71$
\% - Fig. 13(\emph{a}) vs. Fig. 13(\emph{b}), $\left.\Theta\Xi_{tot}\right\rfloor _{\delta=0.2}=60$
\% - Fig. 13(\emph{c}) vs. Fig. 13(\emph{d}), and $\left.\Theta\Xi_{tot}\right\rfloor _{\delta=0.8}=45$
\% - Fig. 13(\emph{e}) vs. Fig. 13(\emph{f}){]}.

\subsection{\noindent Experimental Assessment\label{sub:Experimental-Assessment}}

\noindent To complete the validation of the \emph{MS-DCIE} with real
data, the results of this section refer to the experimental measurements
from the Istitut Fresnel for the {}``FoamDielIntTM'' object \cite{Geffrin 2005}.
In this test case, the square investigation domain of side $L_{\mathcal{H}}=0.4$
{[}m{]} has been illuminated by a ridged-horn antenna from $V=8$
different angular directions and the electric-field data samples have
been collected by another horn antenna moved in $M=241$ uniformly-distributed
locations on a circle of radius $\rho=1.67$ {[}m{]}. Within the investigation
domain, there is an inner cylinder with a diameter of $D_{1}=31$
{[}mm{]} and relative permittivity $\varepsilon_{1}=3$, which is
centered at $(x_{O},y_{O})=(-5\textrm{ [mm]},\,0\textrm{ [mm]})$
and it is surrounded by a larger centered cylinder of diameter $D_{2}=80$
{[}mm{]} having a relative permittivity equal to $\varepsilon_{2}=1.45$
{[}Fig. 14(\emph{a}){]}.

\noindent In order to test the differential formulation of the proposed
\emph{IS} method for inhomogeneous media, the outer cylinder has been
assumed to be part of the known host medium with contrast $\tau_{H}\left(\mathbf{r}\right)$
{[}Fig. 14(\emph{b}){]} so that the differential contrast $\tau_{\Delta}\left(\mathbf{r}\right)$
of the inner cylinder is the unknown to be retrieved {[}Fig. 14(\emph{c}){]}.
As for the \emph{a-priori} knowledge on the host medium, it is worthwhile
to note that the permittivity of the second/larger cylinder ($\varepsilon_{2}=1.45$)
is known with a precision of about $\pm0.15$ \cite{Geffrin 2005},
which corresponds to an uncertainty in (\ref{eq: Masciulla}) of approximately
$\delta=\pm10$ \%.

\noindent For comparison purposes, such real data have been processed
with the \emph{IP} methods based on both \emph{DCIE} and \emph{DLSIE}
formulations as well as single- and multi-resolution strategies. The
data inversion outcomes in terms of the total reconstruction error,
$\Xi_{tot}$, versus the dataset frequency, $f$, are summarized in
Tab. I. Generally, \emph{MS}-based implementations yield lower reconstruction
errors when compared to their single-step counterparts. The same holds
true for the \emph{DCIE} formulation versus the \emph{DLSIE}-based
one. Always, the \emph{MS-DCIE} outperforms the other alternative
techniques by confirming the conclusions drawn from the numerical
analyses in Sect. \ref{sub:Numerical-Assessment}.

\noindent As for the local accuracy of the reconstructions, let us
consider the error maps for some representative cases. For instance,
Figures 15-16 show the local error distribution, $\mathbb{E}\left(\mathbf{r}\right)$,
yielded when processing the datasets at $f=7$ {[}GHz{]} (Fig. 15)
and $f=8$ {[}GHz{]} (Fig. 16). As one can observe, the \emph{DLSIE-}based
methods fail at correctly estimating the contrast of the unknown scatterer
as highlighted by the higher values of the error index in Figs. 15(\emph{c})-15(\emph{d})
and Figs. 16(\emph{c})-16(\emph{d}). On the contrary, the single-resolution
\emph{DCIE} reconstructions provide a more reliable estimation of
the unknown cylinder permittivity {[}Fig. 15(\emph{b}) and Fig. 16(\emph{b}){]},
even though the retrieved distributions still present several artifacts
outside the scatterer support along with non-negligible errors on
the scatterer edges. As expected, the \emph{MS-DCIE} provides the
smoothest error distribution in the whole investigation domain {[}Fig.
15(\emph{a}) and Fig. 16(\emph{a}){]} and it obtains a careful representation
of the target edges. It is also worthwhile to point out that \emph{}the
average reconstruction error of the \emph{MS-DCIE} within the support
of the outer/larger known cylinder is $\left.\Xi_{2}\right\rfloor _{f=7\,[GHz]}^{MR-DCIE}=14.8$
\% and $\left.\Xi_{2}\right\rfloor _{f=8\,[GHz]}^{MR-DCIE}=13.5$
\%, respectively, those values are very close to the tolerance on
the \emph{a-priori} knowledge of the outer cylinder permittivity ($\delta=\pm10$
\%). Such results further assess the remarkable reconstruction capabilities
of the \emph{MS-DCIE} despite the approximate knowledge of the inhomogeneous
host medium.

\section{\noindent Conclusions\label{sec:Conclusions}}

\noindent A novel inversion method, named \emph{MS-DCIE}, has been
developed to address the \emph{ISP}s with inhomogeneous media. The
proposed approach combines the \emph{DCIE} formulation with the \emph{MS}
inversion strategy and it has proved to be reliable and effective
in a wide range of scenarios and under different conditions. As a
matter of fact, the effectiveness of the developed \emph{IS} technique
has been tested against both numerical and experimental scattering
data by considering lossless and lossy profiles as well as varying
the object contrast. The effects of some uncertainty on the \emph{a-priori}
knowledge of the host medium have been evaluated, as well.

\noindent Future works, beyond the scope of the current manuscript,
will be aimed at extending the formulation to three-dimensional (\emph{3D})
geometries as well as at customizing the proposed implementation to
biomedical scenarios of great applicative interest.

\section*{\noindent Appendix A}

\noindent With reference to Fig. 1(\emph{a}) (Scenario {}``\emph{with}
\emph{object}'') and according to the \emph{LSIE} theory, the \emph{EM}
phenomena are described by the {}``\emph{data equation}'',\begin{equation}
\xi_{sca}^{\left(v\right)}\left(\mathbf{r}_{m}\right)=\int_{\mathcal{H}}\mathcal{G}_{B}\left(\mathbf{r}_{m},\,\mathbf{r}^{\prime}\right)\tau\left(\mathbf{r}^{\prime}\right)\xi^{\left(v\right)}\left(\mathbf{r}^{\prime}\right)d\mathbf{r}^{\prime},\label{eq:LSIE-data}\end{equation}
and the {}``\emph{state equation}'',

\noindent \begin{equation}
\xi^{\left(v\right)}\left(\mathbf{r}\right)=\xi_{inc}^{\left(v\right)}\left(\mathbf{r}\right)+\int_{\mathcal{H}}\mathcal{G}_{B}\left(\mathbf{r},\,\mathbf{r}^{\prime}\right)\tau\left(\mathbf{r}^{\prime}\right)\xi^{\left(v\right)}\left(\mathbf{r}^{\prime}\right)d\mathbf{r}^{\prime}.\label{eq:LSIE-state}\end{equation}
Analogously, the \emph{LSIE} equations for the the Scenario {}``\emph{without
the object}'' {[}Fig. 1(\emph{a}){]} turn out to be\begin{equation}
\xi_{sca,H}^{\left(v\right)}\left(\mathbf{r}_{m}\right)=\int_{\mathcal{H}}\mathcal{G}_{B}\left(\mathbf{r}_{m},\,\mathbf{r}^{\prime}\right)\tau_{H}\left(\mathbf{r}^{\prime}\right)\xi_{H}^{\left(v\right)}\left(\mathbf{r}^{\prime}\right)d\mathbf{r}^{\prime},\label{eq:LSIE-data H}\end{equation}
\begin{equation}
\xi_{H}^{\left(v\right)}\left(\mathbf{r}\right)=\xi_{inc}^{\left(v\right)}\left(\mathbf{r}\right)+\int_{\mathcal{H}}\mathcal{G}_{B}\left(\mathbf{r},\,\mathbf{r}^{\prime}\right)\tau_{H}\left(\mathbf{r}^{\prime}\right)\xi_{H}^{\left(v\right)}\left(\mathbf{r}^{\prime}\right)d\mathbf{r}^{\prime}.\label{eq:LSIE-state H}\end{equation}
By subtracting (\ref{eq:LSIE-data H}) from (\ref{eq:LSIE-data})
and (\ref{eq:LSIE-state H}) from (\ref{eq:LSIE-state}), it turns
out that\begin{equation}
\xi_{sca,\Delta}^{\left(v\right)}\left(\mathbf{r}_{m}\right)=\int_{\mathcal{H}}\mathcal{G}_{B}\left(\mathbf{r}_{m},\,\mathbf{r}^{\prime}\right)\left[\tau\left(\mathbf{r}^{\prime}\right)\xi^{\left(v\right)}\left(\mathbf{r}^{\prime}\right)-\tau_{H}\left(\mathbf{r}^{\prime}\right)\xi_{H}^{\left(v\right)}\left(\mathbf{r}^{\prime}\right)\right]d\mathbf{r}^{\prime},\label{eq:W Rsciugni}\end{equation}
\begin{equation}
\xi^{\left(v\right)}\left(\mathbf{r}\right)=\xi_{H}^{\left(v\right)}\left(\mathbf{r}\right)+\int_{\mathcal{H}}\mathcal{G}_{B}\left(\mathbf{r},\,\mathbf{r}^{\prime}\right)\left[\tau\left(\mathbf{r}^{\prime}\right)\xi^{\left(v\right)}\left(\mathbf{r}^{\prime}\right)-\tau_{H}\left(\mathbf{r}^{\prime}\right)\xi_{H}^{\left(v\right)}\left(\mathbf{r}^{\prime}\right)\right]d\mathbf{r}^{\prime}.\label{eq:W Masciulla}\end{equation}
where $\xi_{sca,\Delta}^{\left(v\right)}$ is the differential scattered
field given by the difference between the scattered field with $\xi_{sca}^{\left(v\right)}$
and without $\xi_{sca,H}^{\left(v\right)}$ the unknown object {[}$\xi_{sca,\Delta}^{\left(v\right)}\left(\mathbf{r}_{m}\right)\triangleq\xi_{sca}^{\left(v\right)}\left(\mathbf{r}_{m}\right)-\xi_{sca,H}^{\left(v\right)}\left(\mathbf{r}_{m}\right)$;
$m=1,...,M$; $v=1,...,V${]}.

\noindent Finally, since $\tau\left(\mathbf{r}\right)=\tau_{\Delta}\left(\mathbf{r}\right)-\tau_{H}\left(\mathbf{r}\right)$
and after simple manipulations, (\ref{eq:LSIE-data differential})
and (\ref{eq:LSIE-state differential}) are yielded.

\section*{\noindent Appendix B}

The support $\mathcal{S}_{\varphi}$ of a function $\varphi$, referred
here as \emph{RoI} of $\varphi$, is defined as the region of the
investigation domain $\mathcal{H}$ where the value of the function
is non-null\begin{equation}
\mathcal{S}_{\varphi}\triangleq\left\{ \mathbf{r}:\,\varphi\left(\mathbf{r}\right)\neq0\right\} .\label{eq:support}\end{equation}
Since the difference equivalent current, $J_{\Delta}^{\left(v\right)}$,
is given by (\ref{eq:equivalent current differential}), then its
value is non-zero only if either $J^{\left(v\right)}$ or $J_{H}^{\left(v\right)}$
are non-zero, vice-versa $J_{\Delta}^{\left(v\right)}\left(\mathbf{r}\right)=0$
if $J^{\left(v\right)}\left(\mathbf{r}\right)=0$ and $J_{H}^{\left(v\right)}\left(\mathbf{r}\right)=0$.
Therefore, the following relation on the support $\mathcal{S}_{\Delta J}$
of $J_{\Delta}^{\left(v\right)}$ holds true

\begin{equation}
\mathcal{S}_{\Delta J}\subseteq\left(\mathcal{S}_{J}\cup\mathcal{S}_{J_{H}}\right).\label{eq:result1}\end{equation}
Moreover, the $v$-th ($v=1,...,V$) equivalent current, $J^{\left(v\right)}\left(\mathbf{r}\right)$,
is zero if the contrast $\tau\left(\mathbf{r}\right)$ is zero being
$J^{\left(v\right)}\left(\mathbf{r}\right)\triangleq\tau\left(\mathbf{r}\right)\xi^{\left(v\right)}\left(\mathbf{r}\right)$,
thus $\mathcal{S}_{J}\subseteq\mathcal{S}_{\tau}$. Analogously, $\mathcal{S}_{J_{H}}\subseteq\mathcal{S}_{\tau_{H}}$
since $J_{H}^{\left(v\right)}\left(\mathbf{r}\right)\triangleq\tau_{H}\left(\mathbf{r}\right)\xi_{H}^{\left(v\right)}\left(\mathbf{r}\right)$.
Accordingly, (\ref{eq:result1}) can be rewritten as follows\begin{equation}
\mathcal{S}_{\Delta J}\subseteq\left(\mathcal{S}_{\tau}\cup\mathcal{S}_{\tau_{H}}\right).\label{eq:result2}\end{equation}
Moving to the modified contrasts, given by $\chi\left(\mathbf{r}\right)\triangleq\frac{\beta\left(\mathbf{r}\right)\tau\left(\mathbf{r}\right)}{\beta\left(\mathbf{r}\right)\tau\left(\mathbf{r}\right)+1}$
and $\chi_{H}\left(\mathbf{r}\right)\triangleq\frac{\beta\left(\mathbf{r}\right)\tau_{H}\left(\mathbf{r}\right)}{\beta\left(\mathbf{r}\right)\tau_{H}\left(\mathbf{r}\right)+1}$
subject to $\beta\left(\mathbf{r}\right)\neq0$, it turns out that
$\mathcal{S}_{\chi}\equiv\mathcal{S}_{\tau}$ and $\mathcal{S}_{\chi_{H}}\equiv\mathcal{S}_{\tau_{H}}$.
By combining these latter conclusions with (\ref{eq:result2}), one
yields that\begin{equation}
\mathcal{S}_{\Delta J}\subseteq\left(\mathcal{S}_{\chi}\cup\mathcal{S}_{\chi_{H}}\right),\label{eq:result3}\end{equation}
that assumes the form (\ref{eq:current_support}) by observing that
$S_{\chi}\equiv\left(\mathcal{S}_{\Delta\chi}\cup\mathcal{S}_{\chi_{H}}\right)$
from the definition of $\chi_{\Delta}\left(\mathbf{r}\right)$ {[}i.e.,
$\chi_{\Delta}\left(\mathbf{r}\right)\triangleq\chi\left(\mathbf{r}\right)-\chi_{H}\left(\mathbf{r}\right)${]}.

\section*{Acknowledgements}

\noindent This work benefited from the networking activities carried
out within the Project \char`\"{}MITIGO - Mitigazione dei rischi naturali
per la sicurezza e la mobilita' nelle aree montane del Mezzogiorno\char`\"{}
(Grant no. ARS01\_00964) funded by the Italian Ministry of Education,
University, and Research within the PON R\&I 2014-2020 Program (CUP:
B64I20000450005), the Project \char`\"{}EMvisioning - Cyber-Physical
Electromagnetic Vision: Context-Aware Electromagnetic Sensing and
Smart Reaction\char`\"{} (Grant no. 2017HZJXSZ) funded by the Italian
Ministry of Education, University, and Research within the PRIN2017
Program (CUP: E64I19002530001), and Project \char`\"{}SPEED\char`\"{}
(Grant No. 6721001) funded by National Science Foundation of China
under the Chang-Jiang Visiting Professorship Program. A. Massa wishes
to thank E. Vico for her never-ending inspiration, support, guidance,
and help.

~

\newpage
\section*{FIGURE CAPTIONS}

\begin{itemize}
\item \textbf{Figure 1.} \emph{Problem Scenario} - Contrast distribution
(\emph{a}) without and (\emph{b}) with the scatterer as well as (\emph{c})
differential profile.
\item \textbf{Figure 2.} \emph{Illustrative} \emph{Example} ($S=4$) - Evolution
of the \emph{RoI}s ($\mathcal{S}_{\Delta\chi}^{(s)}$, $\mathcal{S}_{\Delta J}^{(s)}$)
at different \emph{MS} steps: (\emph{a}) $s=1$, (\emph{b}) $s=2$,
(\emph{c}) $s=3$, and (\emph{d}) $s=S$.
\item \textbf{Figure 3.} \emph{Control-Parameters Calibration} ({}``Square''
object {[}$\tau_{\Delta}=2${]}, $M=V=27$, \emph{MS-DCIE}) - Plots
of the total reconstruction error, $\Xi_{tot}$, as a function of
(\emph{a}) $\alpha$ ($\gamma=1.4$) and (\emph{b}) $\gamma$ ($\alpha=0.4$).
\item \textbf{Figure 4.} \emph{Numerical Assessment} ({}``Square'' object
{[}$\tau_{\Delta}=2${]}, $M=V=27$, $SNR=20$ {[}dB{]}, \emph{MS-DCIE})
- Plots of the cost function, $\Psi$, the total reconstruction error,
$\Xi_{tot}$, and the \emph{MS} zooming factor, $\eta^{(s)}$, versus
the \emph{MS} step index ($s=1,...,S$).
\item \textbf{Figure 5.} \emph{Numerical Assessment} ({}``Square'' object
{[}$\tau_{\Delta}=2${]}, $M=V=27$, $SNR=20$ {[}dB{]}) - Maps of
(\emph{a})-(\emph{d}) the retrieved differential contrast, $\tau_{\Delta}$,
at different steps of the \emph{MS-DCIE} {[}(\emph{a}) $s=1$, (\emph{b})
$s=2$, (\emph{c}) $s=3$, and (\emph{d}) $s=4${]} and (\emph{e})(\emph{f})
the contrast profile, $\tau^{opt}$, reconstructed by (\emph{e}) the
\emph{MS-DCIE} and (\emph{f}) the \emph{DCIE}.
\item \textbf{Figure 6.} \emph{Numerical Assessment} ({}``Circular Ring''
object {[}$\tau_{\Delta}=2${]}, $M=V=27$) - Plots of the total reconstruction
error, $\Xi_{tot}$, versus \emph{SNR}.
\item \textbf{Figure 7.} \emph{Numerical Assessment} ({}``Circular Ring''
object {[}$\tau_{\Delta}=2${]}, $M=V=27$, $SNR=20$ {[}dB{]}) -
Maps of the local error, $\mathbb{E}$, yielded by (\emph{a}) the
\emph{MS-DCIE}, (\emph{b}) the \emph{DCIE}, (\emph{c}) the \emph{MS-DLSIE},
and (\emph{d}) the \emph{DLSIE}.
\item \textbf{Figure 8.} \emph{Numerical Assessment} ({}``Circular Ring''
object, $M=V=27$, $SNR=20$ {[}dB{]}) - Plots of the global reconstruction
errors as a function of the actual differential contrast, $\tau_{\Delta}$\emph{.}
\item \textbf{Figure 9.} \emph{Numerical Assessment} ({}``Circular Ring''
object, $M=V=27$, $SNR=20$ {[}dB{]}) - Maps of the local error,
$\mathbb{E}$, yielded by (\emph{a})(\emph{b}) the \emph{MS-DCIE}
and (\emph{c})(\emph{d}) the \emph{DCIE} when (\emph{a})(\emph{c})
$\tau_{\Delta}=3$ and (\emph{b})(\emph{d}) $\tau_{\Delta}=4$.
\item \textbf{Figure 10.} \emph{Numerical Assessment} ({}``Circular Ring''
object {[}$\textrm{Re}\left\{ \tau_{\Delta}\right\} =2${]}, $M=V=27$,
$SNR=20$ {[}dB{]}) - Plots of the global reconstruction errors as
a function of the actual differential conductivity, $\sigma_{\Delta}$\emph{.}
\item \textbf{Figure 11.} \emph{Numerical Assessment} ({}``Circular Ring''
object {[}$\textrm{Re}\left\{ \tau_{\Delta}\right\} =2$, $\sigma_{\Delta}=10^{-1}$
{[}S/m{]}{]}, $M=V=27$, $SNR=20$ {[}dB{]}) - Maps of the (\emph{a})(\emph{c})
real and the (\emph{b})(\emph{d}) imaginary part of the local error,
$\mathbb{E}$, yielded by (\emph{a})(\emph{b}) the \emph{MS-DCIE}
and (\emph{c})(\emph{d}) the \emph{DCIE}.
\item \textbf{Figure 12.} \emph{Numerical Assessment} ({}``Square'' object
{[}$\tau_{\Delta}=2${]}, $M=V=27$, $SNR=20$ {[}dB{]}) - Plots of
the global reconstruction errors versus the \emph{a-priori} knowledge
uncertainty, $\delta$\emph{.}
\item \textbf{Figure 13.} \emph{Numerical Assessment} ({}``Square'' object
{[}$\tau_{\Delta}=2${]}, $M=V=27$, $SNR=20$ {[}dB{]}) - Maps of
the local error, $\mathbb{E}$, yielded by (\emph{a})(\emph{c})(\emph{e})
the \emph{MS-DCIE} and (\emph{b})(\emph{d})(\emph{f}) the \emph{DCIE}
when (\emph{a})(\emph{b}) $\delta=5$ \%, (\emph{c})(\emph{d}) $\delta=20$
\%, and (\emph{e})(\emph{f}) $\delta=80$ \%.
\item \textbf{Figure 14.} \emph{Experimental Assessment} ({}``FoamDielIntTM''
object {[}$\tau_{\Delta}=2$, $\tau_{2}=0.45${]}, $V=8$, $M=241$)
- Maps of (\emph{a}) the actual contrast, $\tau$, (\emph{b}) the
host medium, $\tau_{H}$, and (\emph{c}) the actual differential contrast,
$\tau_{\Delta}$.
\item \textbf{Figure 15.} \emph{Experimental Assessment} ($f=7$ {[}GHz{]},
{}``FoamDielIntTM'' object {[}$\tau_{\Delta}=2$, $\tau_{2}=0.45${]},
$V=8$, $M=241$) - Maps of the local error, $\mathbb{E}$, yielded
by (\emph{a}) the \emph{MS-DCIE}, (\emph{b}) the \emph{DCIE}, (\emph{c})
the \emph{MS-DLSIE}, and (\emph{d}) the \emph{DLSIE}.
\item \textbf{Figure 16.} \emph{Experimental Assessment} ($f=8$ {[}GHz{]},
{}``FoamDielIntTM'' object {[}$\tau_{\Delta}=2$, $\tau_{2}=0.45${]},
$V=8$, $M=241$) - Maps of the local error, $\mathbb{E}$, yielded
by (\emph{a}) the \emph{MS-DCIE}, (\emph{b}) the \emph{DCIE}, (\emph{c})
the \emph{MS-DLSIE}, and (\emph{d}) the \emph{DLSIE}.
\end{itemize}

\section*{TABLE CAPTIONS}

\begin{itemize}
\item \textbf{Table I.} \emph{Experimental Assessment} ({}``FoamDielIntTM''
object {[}$\tau_{\Delta}=2$, $\tau_{2}=0.45${]}, $V=8$, $M=241$)
- Total reconstruction error, $\Xi_{tot}$.
\end{itemize}
\newpage
\begin{center}~\vfill\end{center}

\begin{center}\begin{tabular}{cc}
\includegraphics[%
  width=0.48\textwidth]{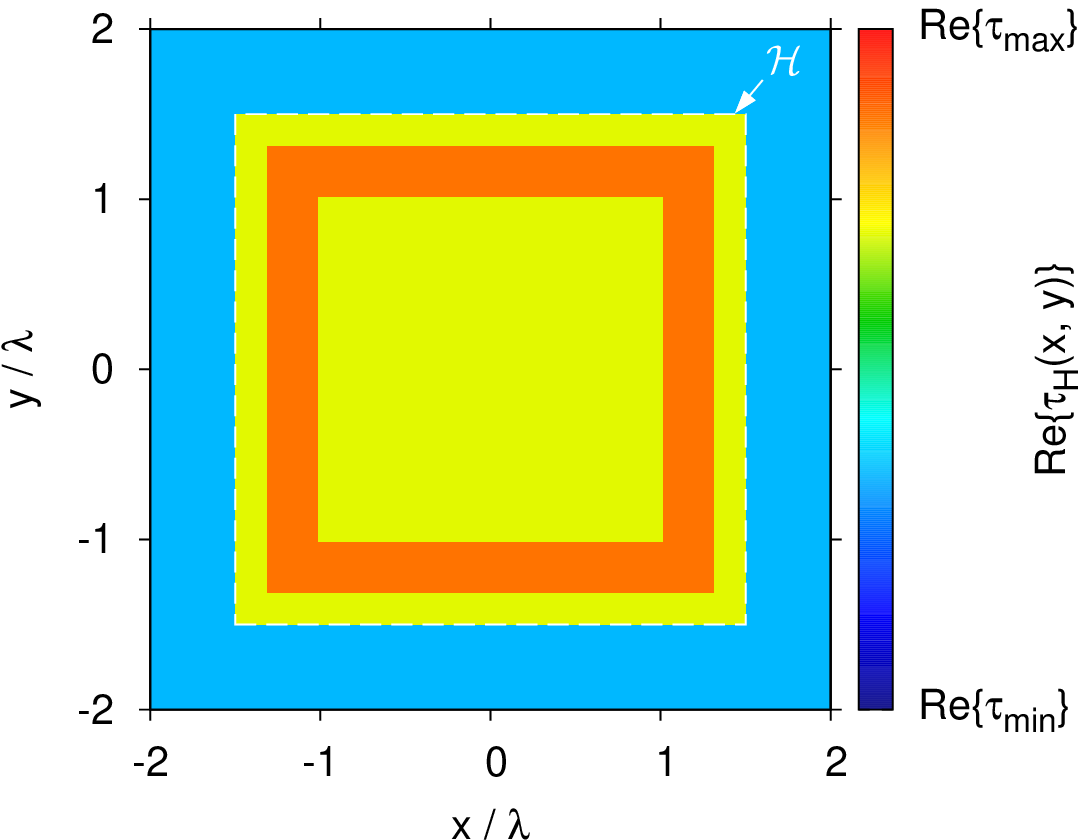}&
\includegraphics[%
  width=0.48\textwidth]{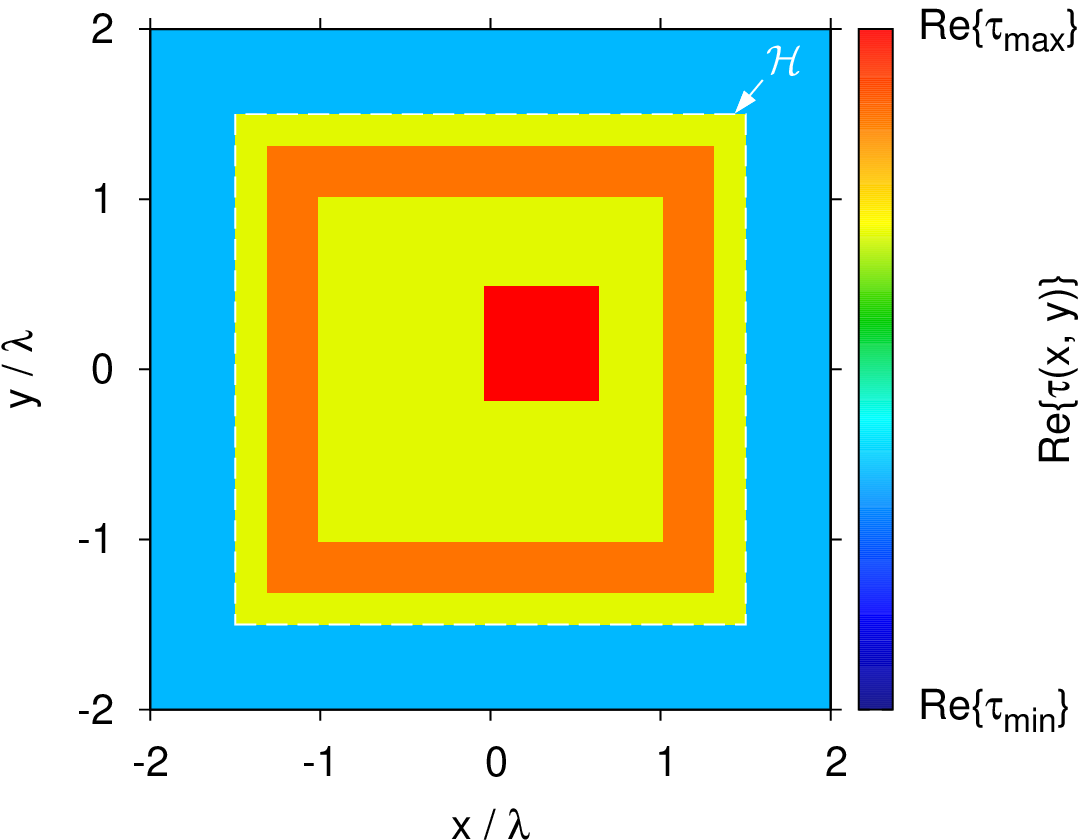}\tabularnewline
(\emph{a})~~~&
(\emph{b})~~~~\tabularnewline
\multicolumn{2}{c}{\includegraphics[%
  width=0.48\textwidth]{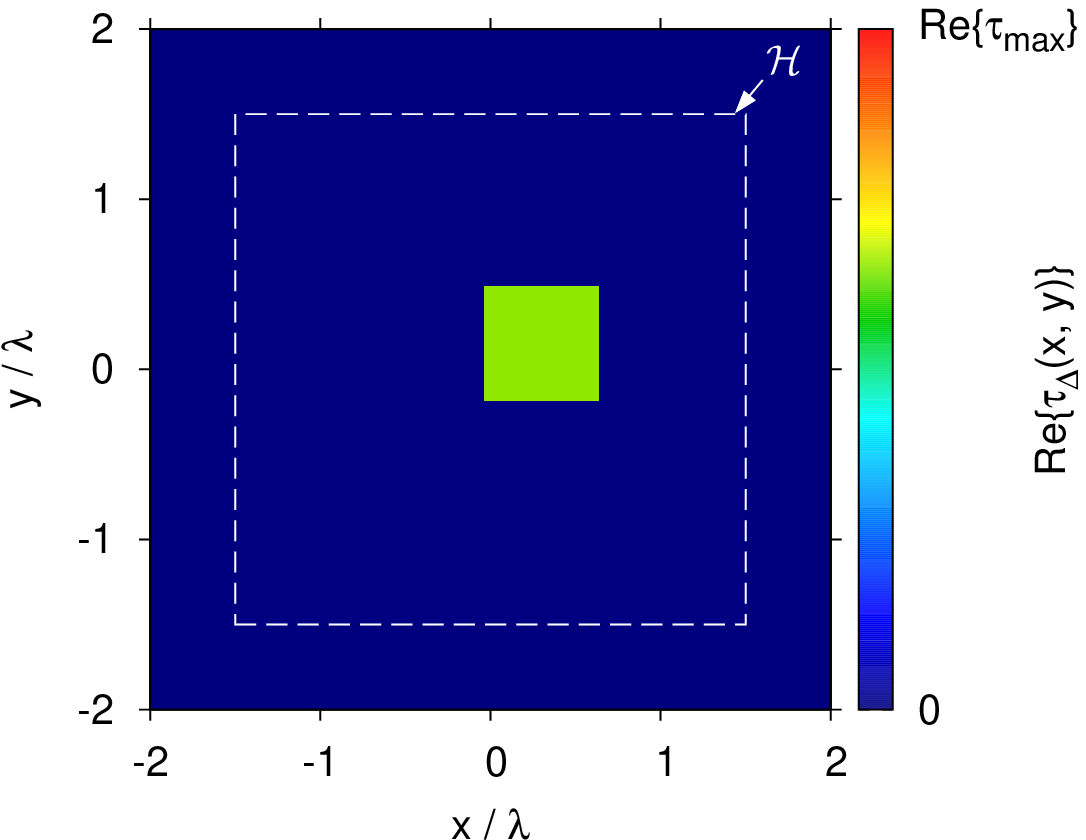}}\tabularnewline
\multicolumn{2}{c}{(\emph{c})~~~}\tabularnewline
\end{tabular}\end{center}

\begin{center}~\vfill\end{center}

\begin{center}\textbf{Fig. 1 - Zhong et} \textbf{\emph{al.}}\textbf{,}
\textbf{\emph{{}``}}Multi-Scaling Differential Contraction ...''\end{center}

\newpage
\begin{center}~\vfill\end{center}

\begin{center}\begin{tabular}{cc}
\includegraphics[%
  width=0.45\textwidth,
  keepaspectratio]{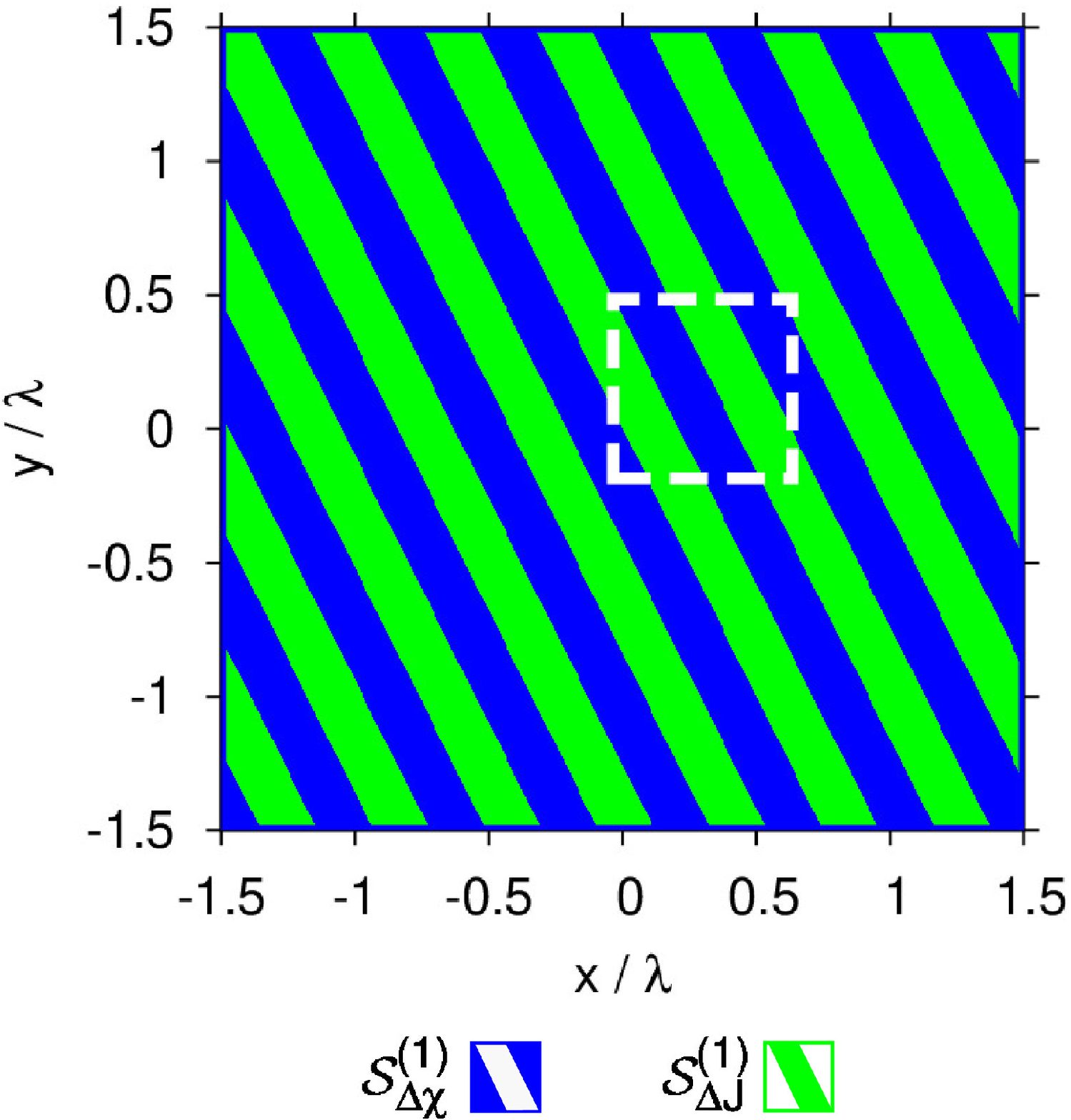}&
\includegraphics[%
  width=0.45\textwidth,
  keepaspectratio]{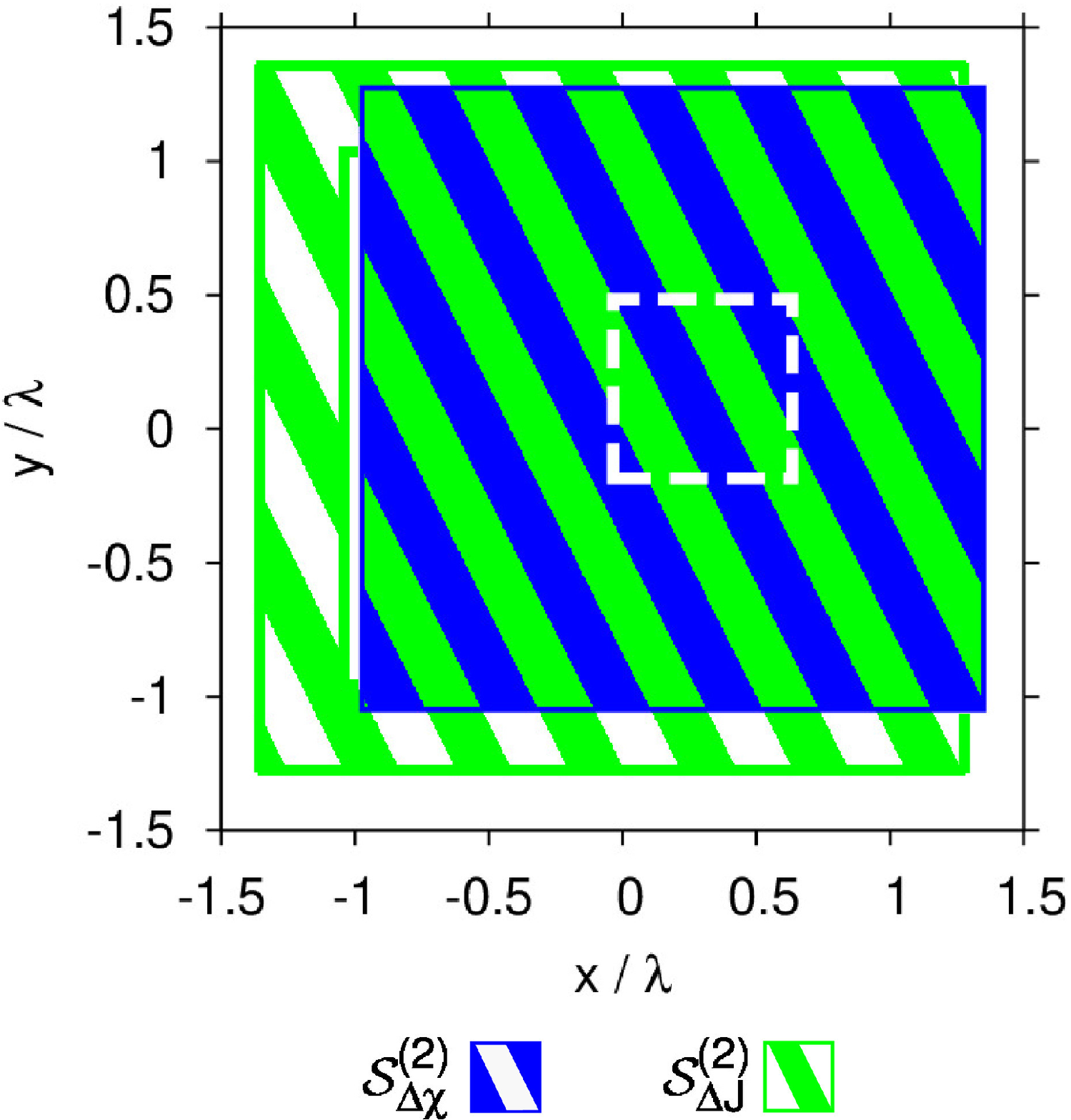}\tabularnewline
~~~~~~~~~~~(\emph{a})&
~~~~~~~~~~~(\emph{b})\tabularnewline
\includegraphics[%
  width=0.45\textwidth,
  keepaspectratio]{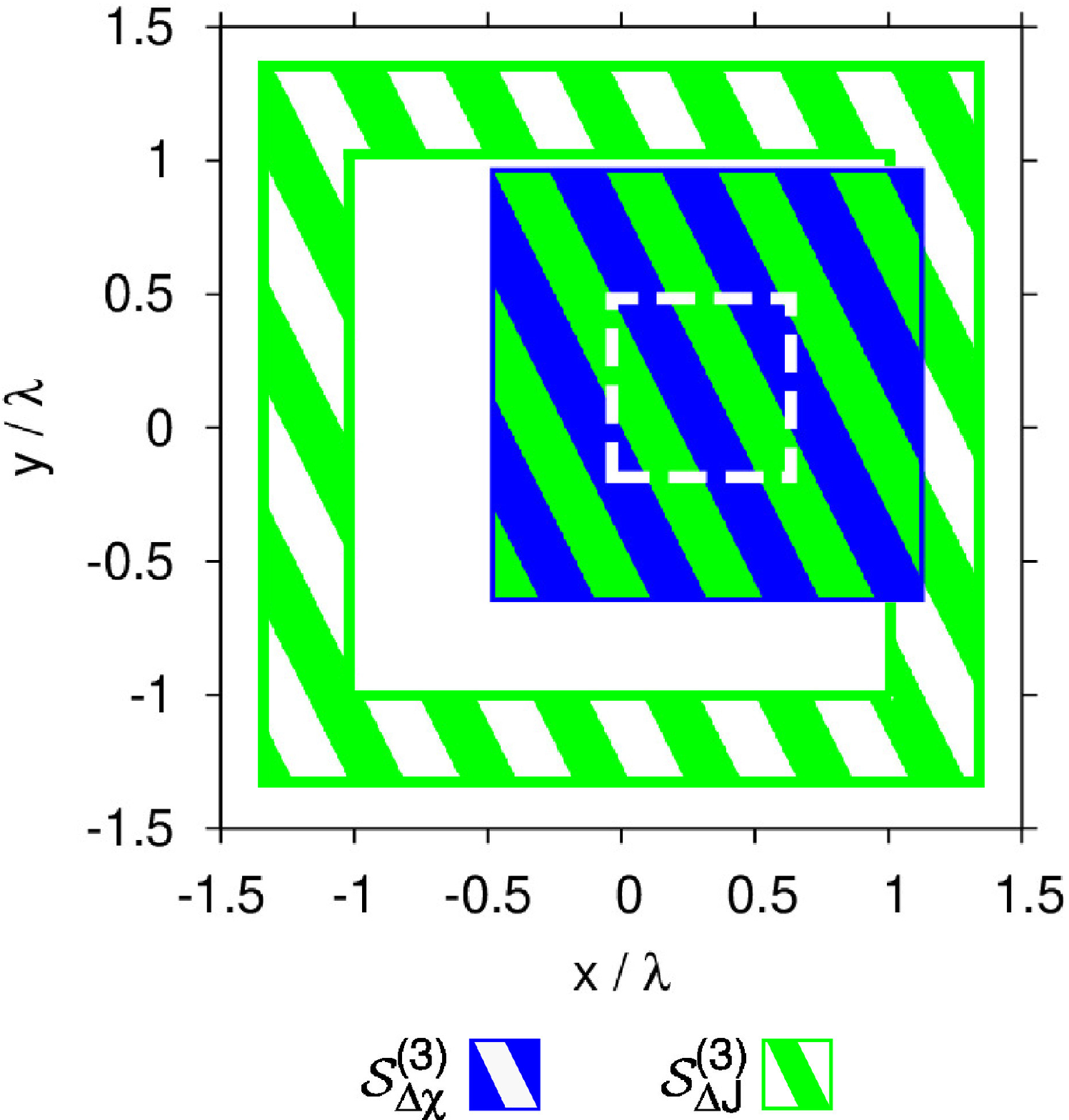}&
\includegraphics[%
  width=0.45\textwidth,
  keepaspectratio]{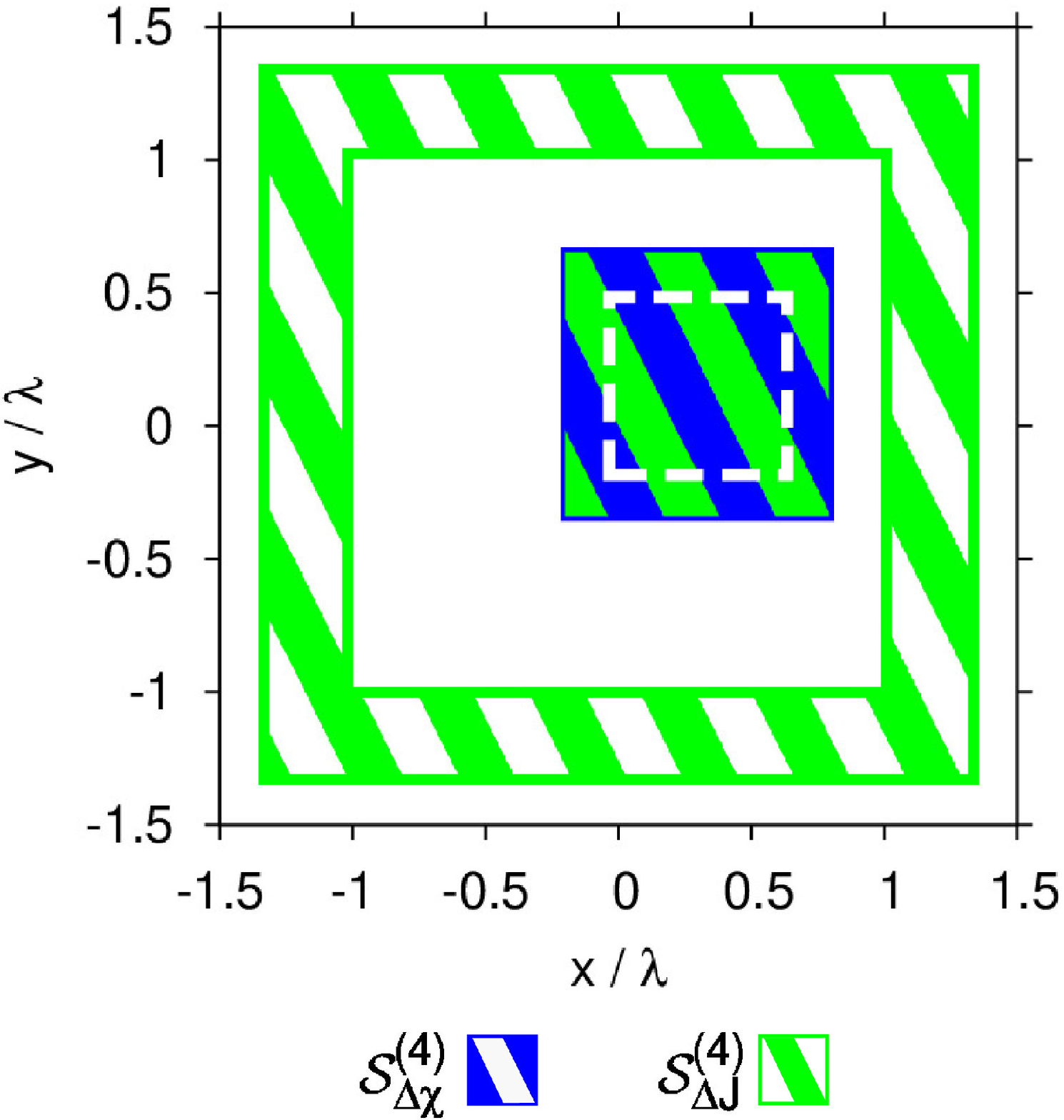}\tabularnewline
~~~~~~~~~~~(\emph{c})&
~~~~~~~~~~~(\emph{d})\tabularnewline
\end{tabular}\end{center}

\begin{center}~\vfill\end{center}

\begin{center}\textbf{Fig. 2 - Zhong et} \textbf{\emph{al.}}\textbf{,}
\textbf{\emph{{}``}}Multi-Scaling Differential Contraction ...''\end{center}

\newpage
\begin{center}~\vfill\end{center}

\begin{center}\begin{tabular}{c}
\includegraphics[%
  width=0.80\textwidth,
  keepaspectratio]{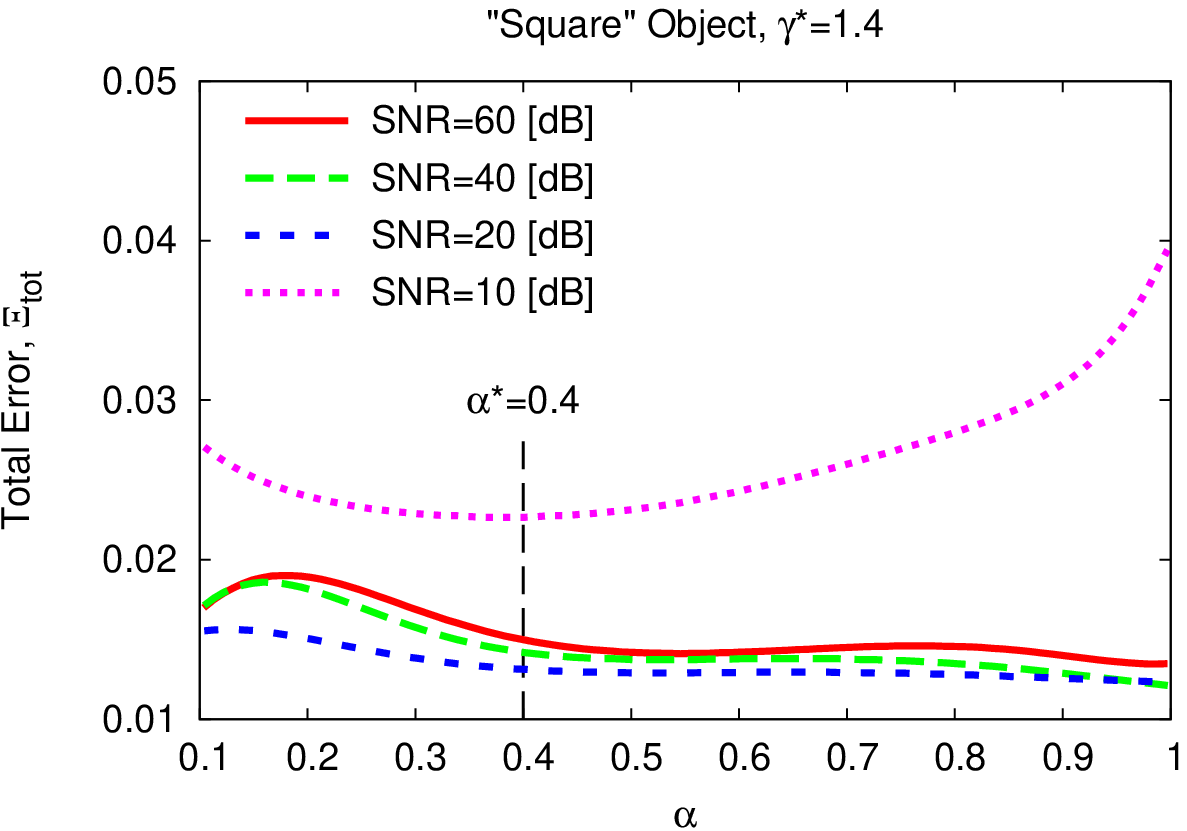}\tabularnewline
~~~~~~~~~~~~~~~~~~~(\emph{a})\tabularnewline
\includegraphics[%
  width=0.80\textwidth,
  keepaspectratio]{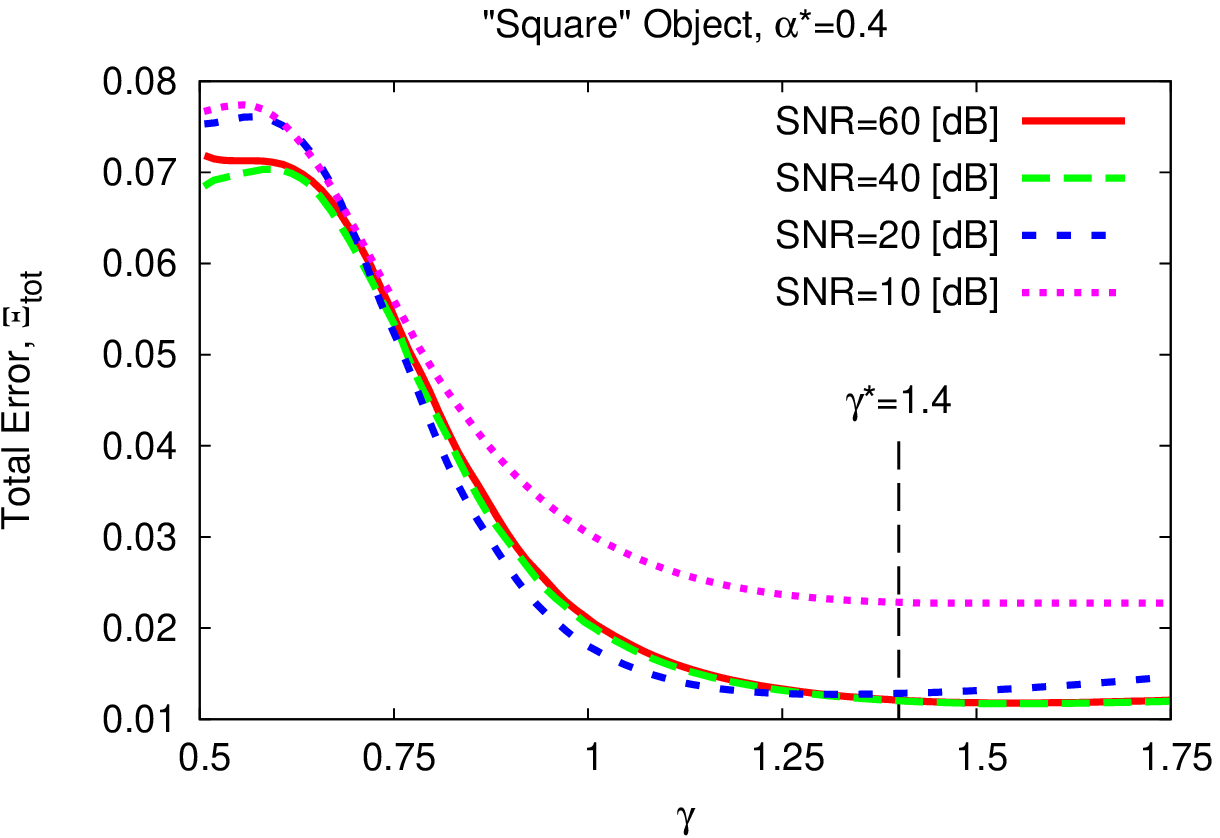}\tabularnewline
~~~~~~~~~~~~~~~~~~(\emph{b})\tabularnewline
\end{tabular}\end{center}

\begin{center}~\vfill\end{center}

\begin{center}\textbf{Fig. 3 - Zhong et} \textbf{\emph{al.}}\textbf{,}
\textbf{\emph{{}``}}Multi-Scaling Differential Contraction ...''\end{center}

\newpage
\begin{center}~\vfill\end{center}

\begin{center}\begin{tabular}{c}
\includegraphics[%
  width=0.80\textwidth]{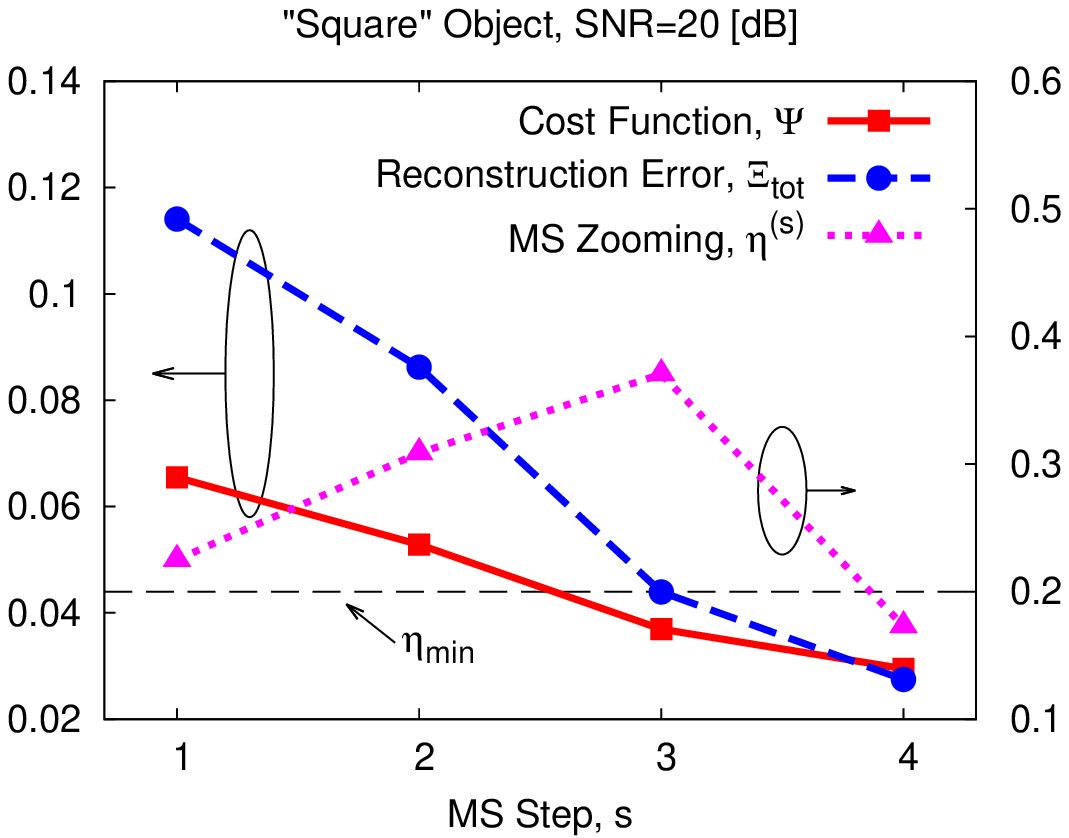}\tabularnewline
\end{tabular}\end{center}

\begin{center}~\vfill\end{center}

\begin{center}\textbf{Fig. 4 - Zhong et} \textbf{\emph{al.}}\textbf{,}
\textbf{\emph{{}``}}Multi-Scaling Differential Contraction ...''\end{center}

\newpage
\begin{center}~\vfill\end{center}

\begin{center}\begin{tabular}{cc}
$s=1$&
$s=2$\tabularnewline
\includegraphics[%
  width=0.35\textwidth,
  keepaspectratio]{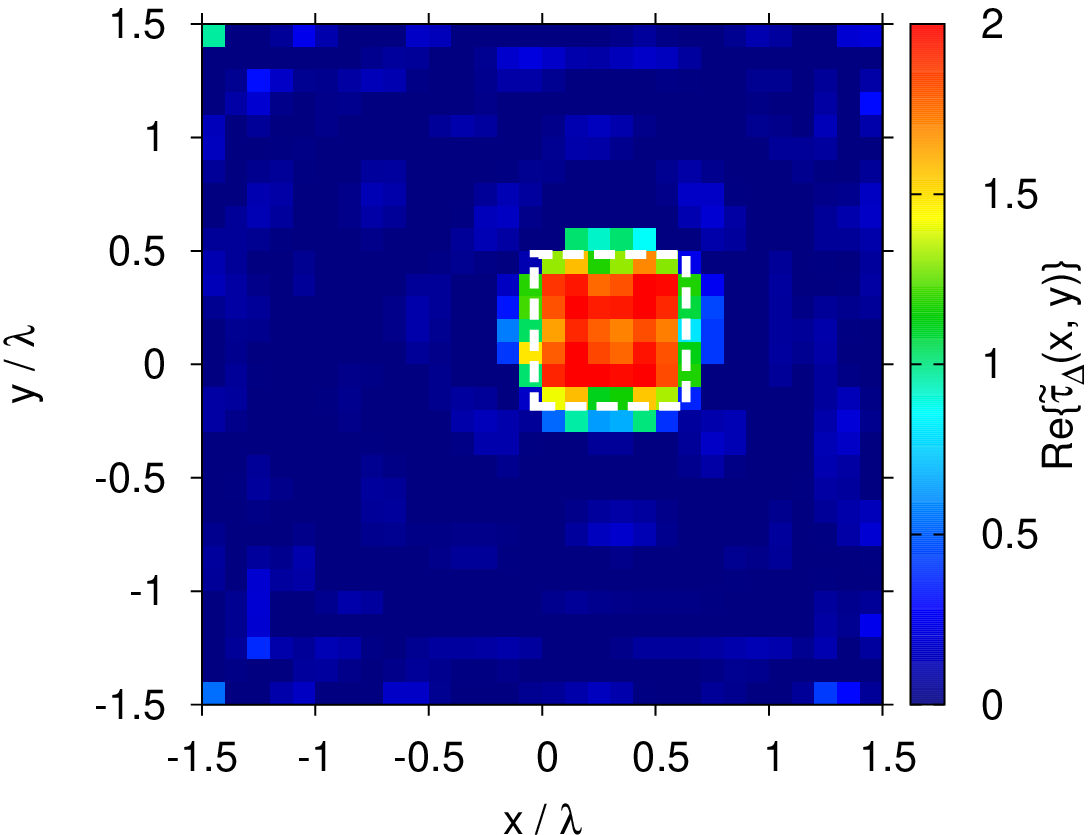}&
\includegraphics[%
  width=0.35\textwidth,
  keepaspectratio]{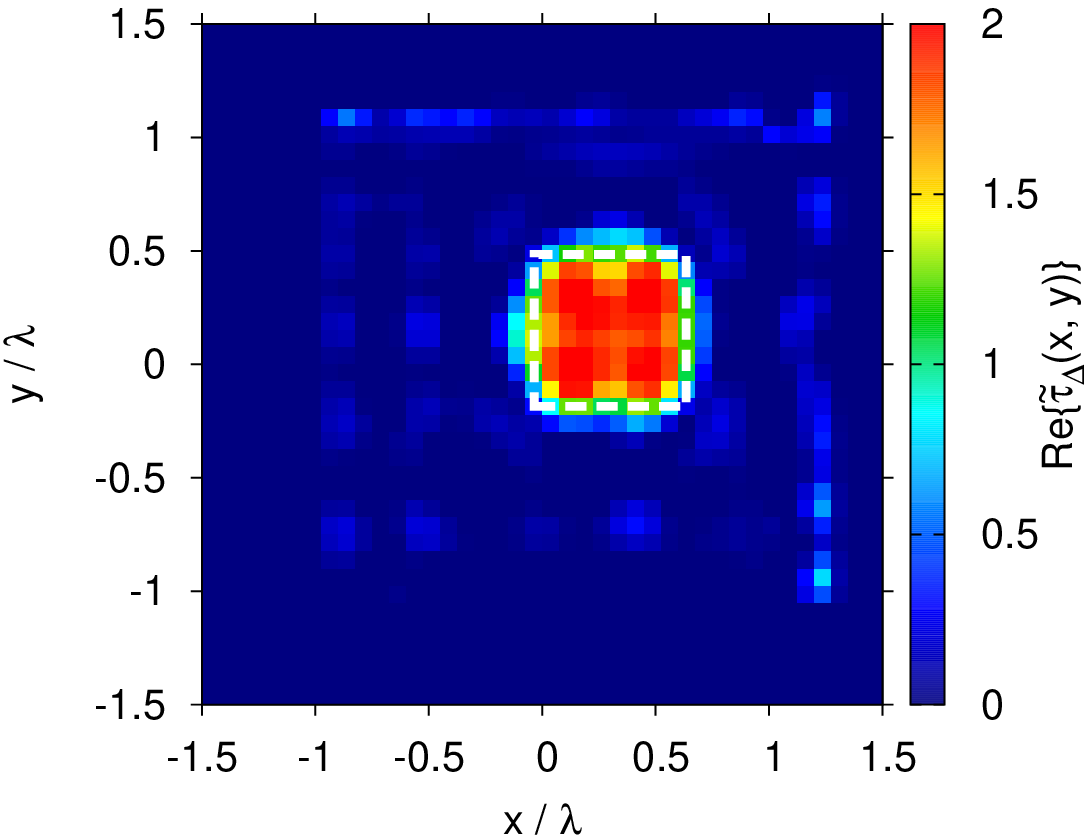}\tabularnewline
(\emph{a})&
(\emph{b})\tabularnewline
$s=3$&
$s=4$\tabularnewline
\includegraphics[%
  width=0.35\textwidth,
  keepaspectratio]{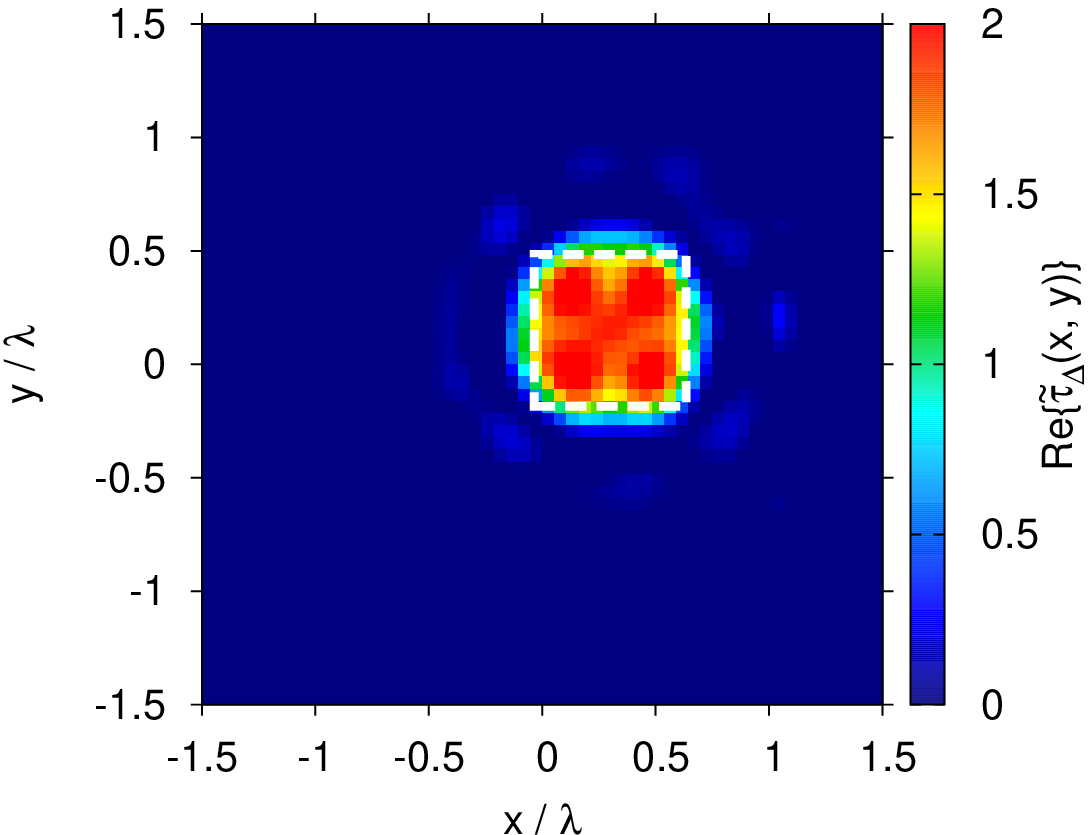}&
\includegraphics[%
  width=0.35\textwidth,
  keepaspectratio]{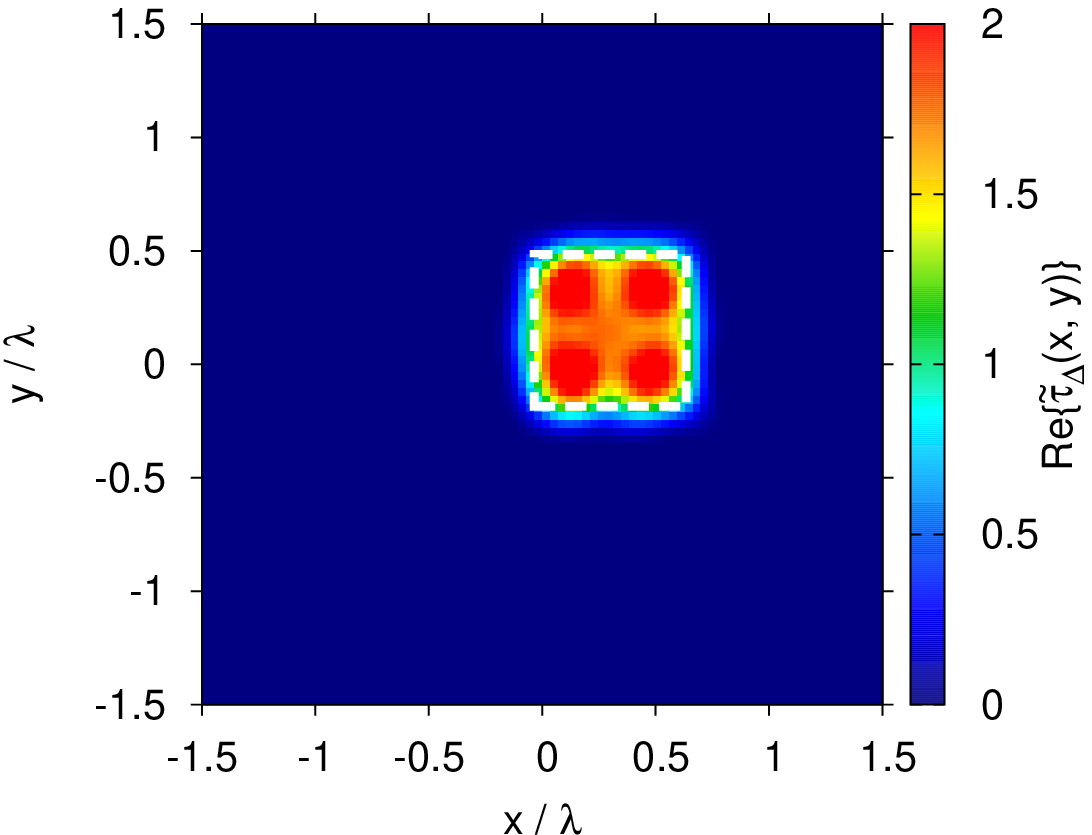}\tabularnewline
(\emph{c})&
(\emph{d})\tabularnewline
\emph{MS-DCIE}&
\emph{DCIE}\tabularnewline
\includegraphics[%
  width=0.35\textwidth,
  keepaspectratio]{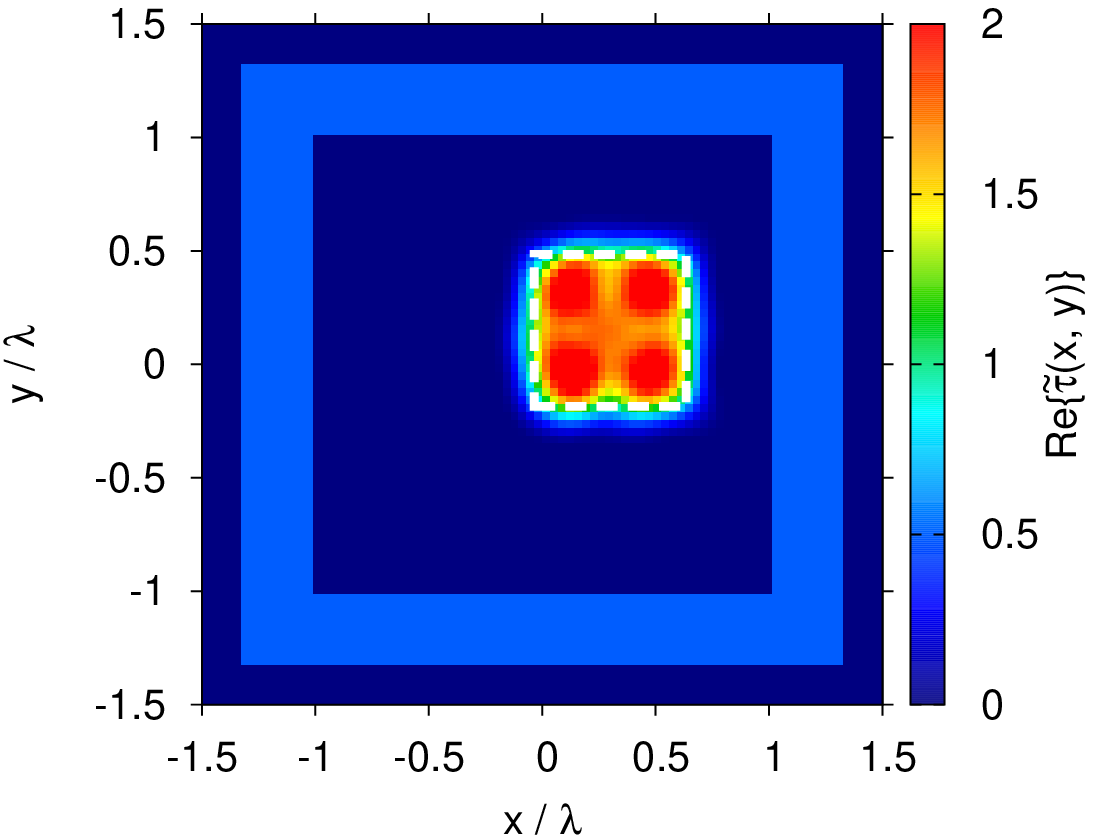}&
\includegraphics[%
  width=0.35\textwidth,
  keepaspectratio]{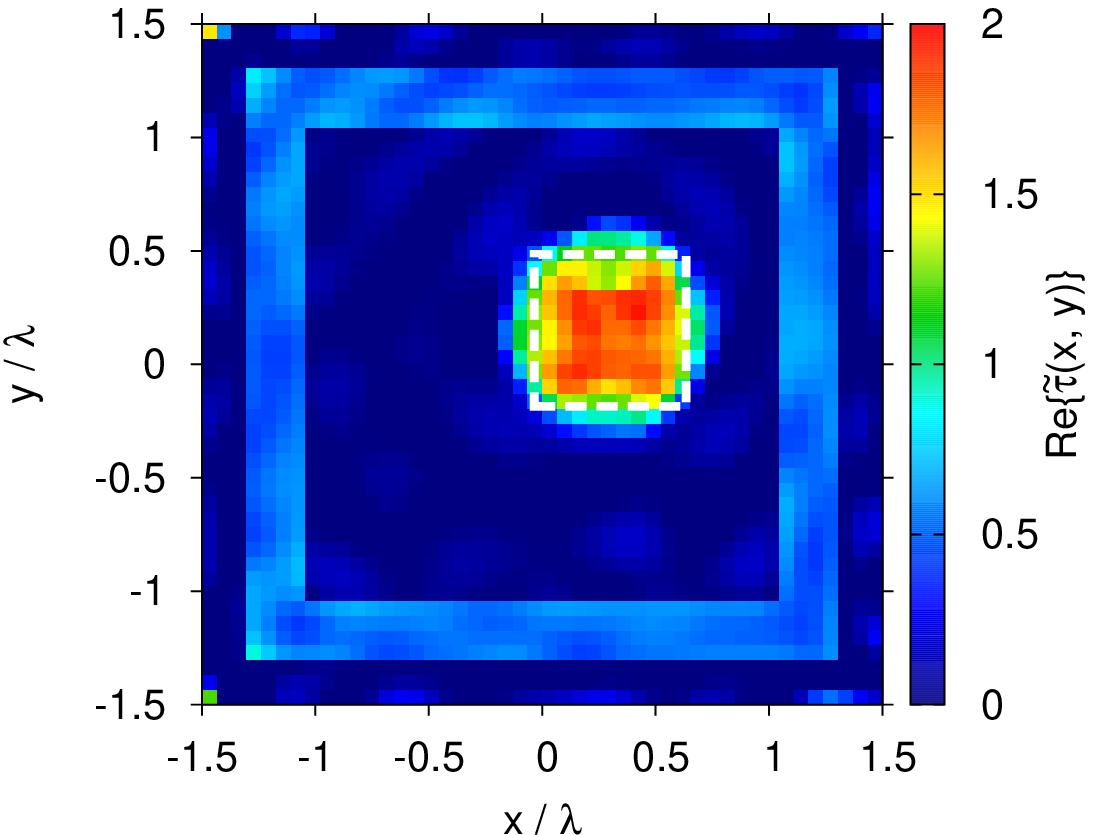}\tabularnewline
(\emph{e})&
(\emph{f})\tabularnewline
\end{tabular}\end{center}

\begin{center}~\vfill\end{center}

\begin{center}\textbf{Fig. 5 - Zhong et} \textbf{\emph{al.}}\textbf{,}
\textbf{\emph{{}``}}Multi-Scaling Differential Contraction ...''\end{center}

\newpage
\begin{center}~\vfill\end{center}

\begin{center}\begin{tabular}{c}
\includegraphics[%
  width=0.80\textwidth]{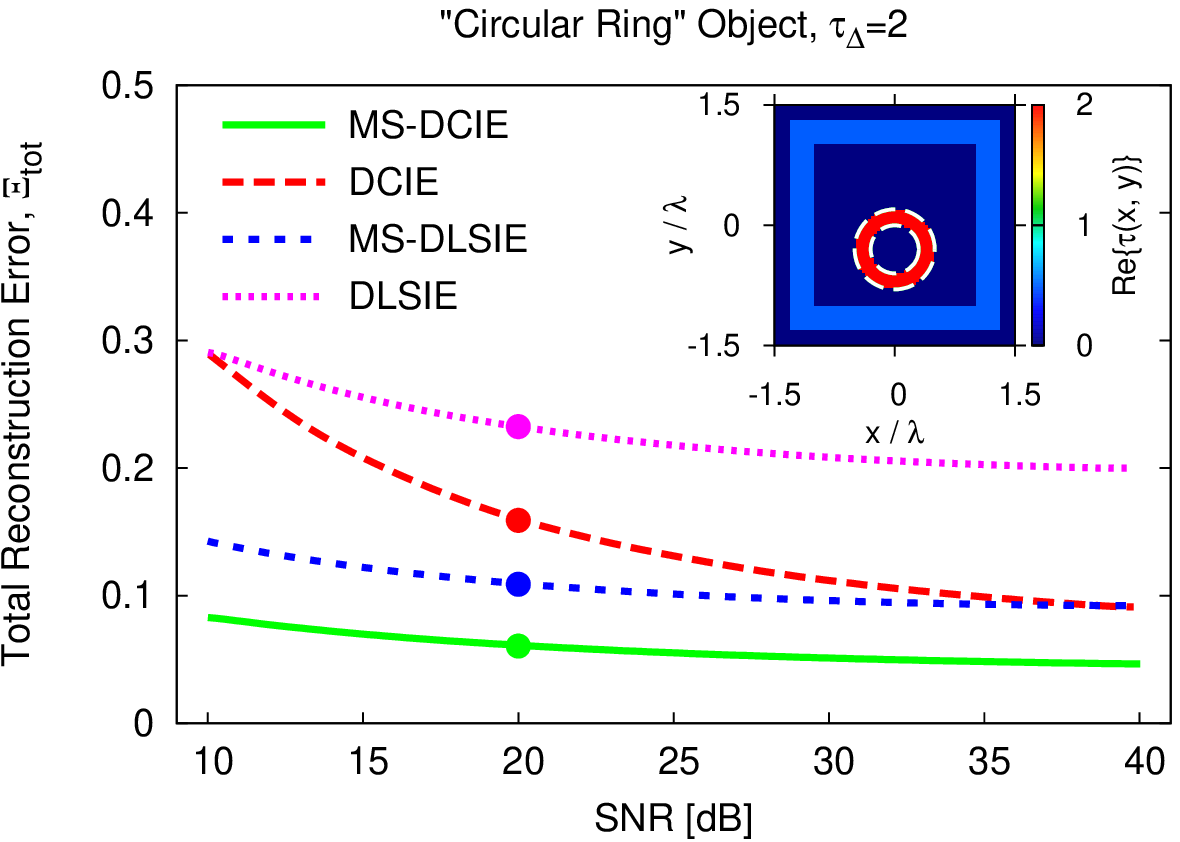}\tabularnewline
\end{tabular}\end{center}

\begin{center}~\vfill\end{center}

\begin{center}\textbf{Fig. 6 - Zhong et} \textbf{\emph{al.}}\textbf{,}
\textbf{\emph{{}``}}Multi-Scaling Differential Contraction ...''\end{center}

\newpage
\begin{center}~\vfill\end{center}

\begin{center}\begin{tabular}{cc}
\emph{MS-DCIE}&
\emph{DCIE}\tabularnewline
\includegraphics[%
  width=0.48\textwidth]{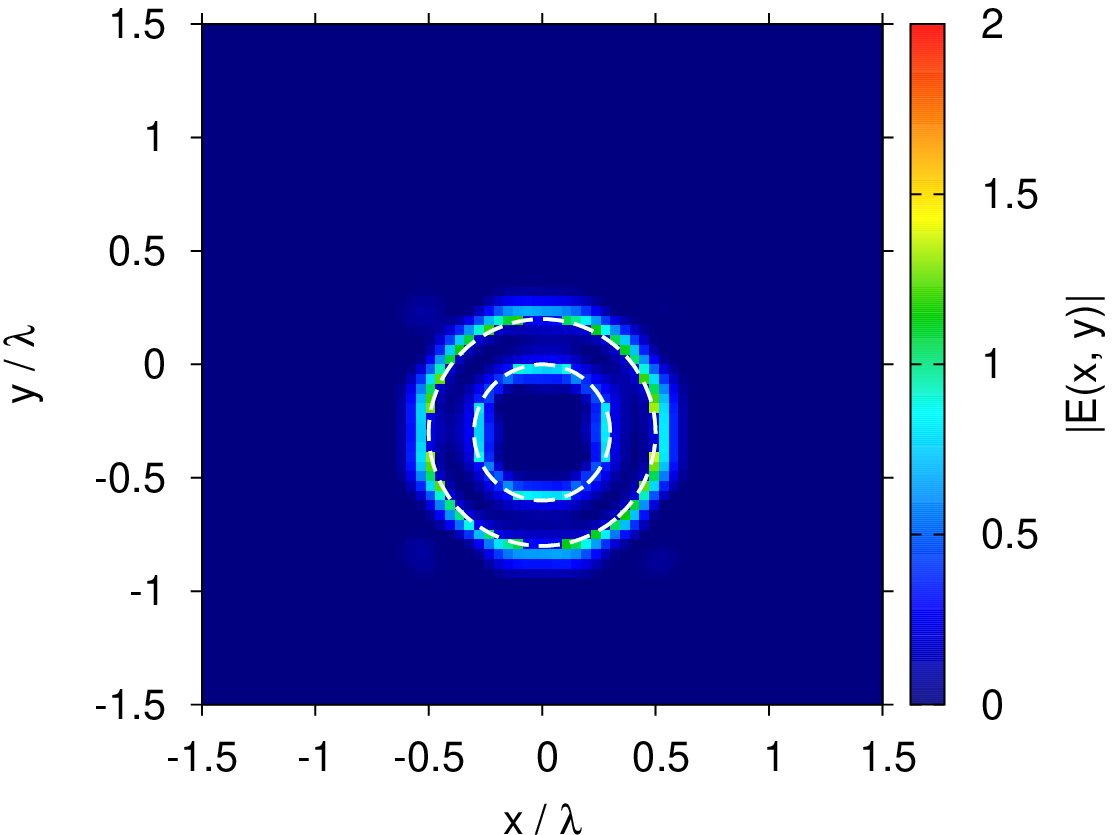}&
\includegraphics[%
  width=0.48\textwidth]{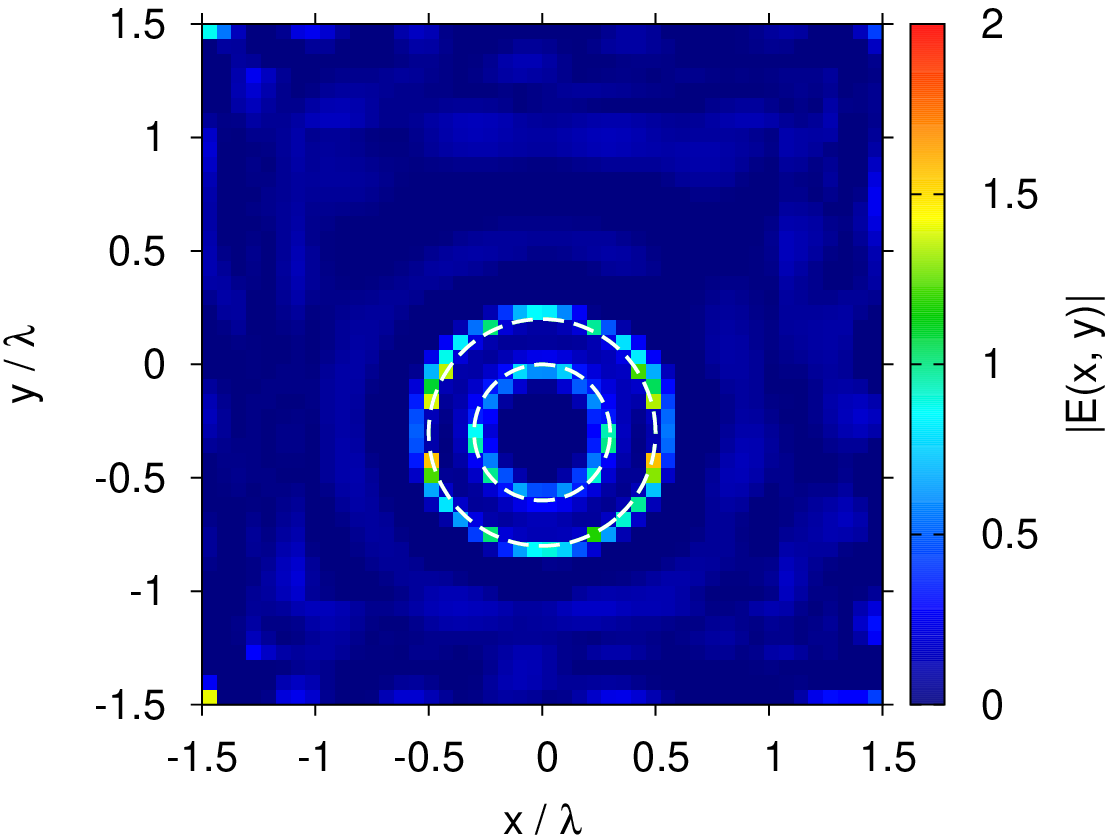}\tabularnewline
~(\emph{a})&
~(\emph{b})\tabularnewline
\emph{MS-DLSIE}&
\emph{DLSIE}\tabularnewline
\includegraphics[%
  width=0.48\textwidth]{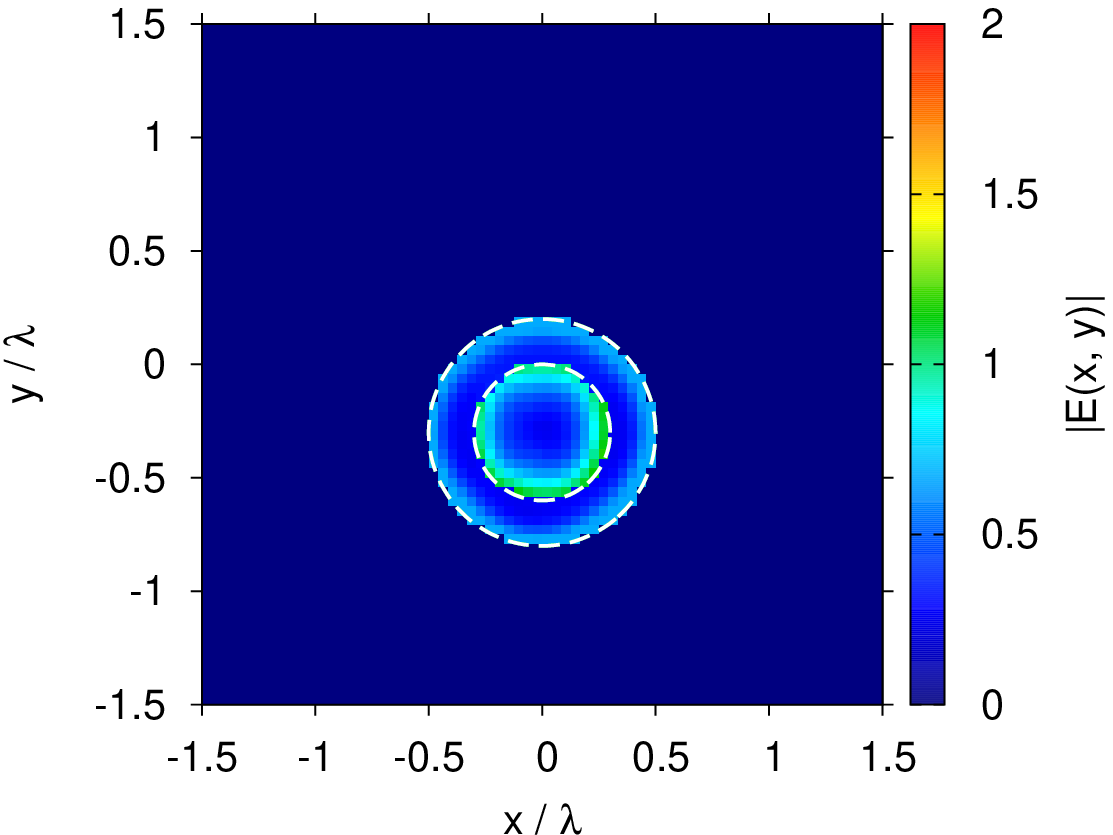}&
\includegraphics[%
  width=0.48\textwidth]{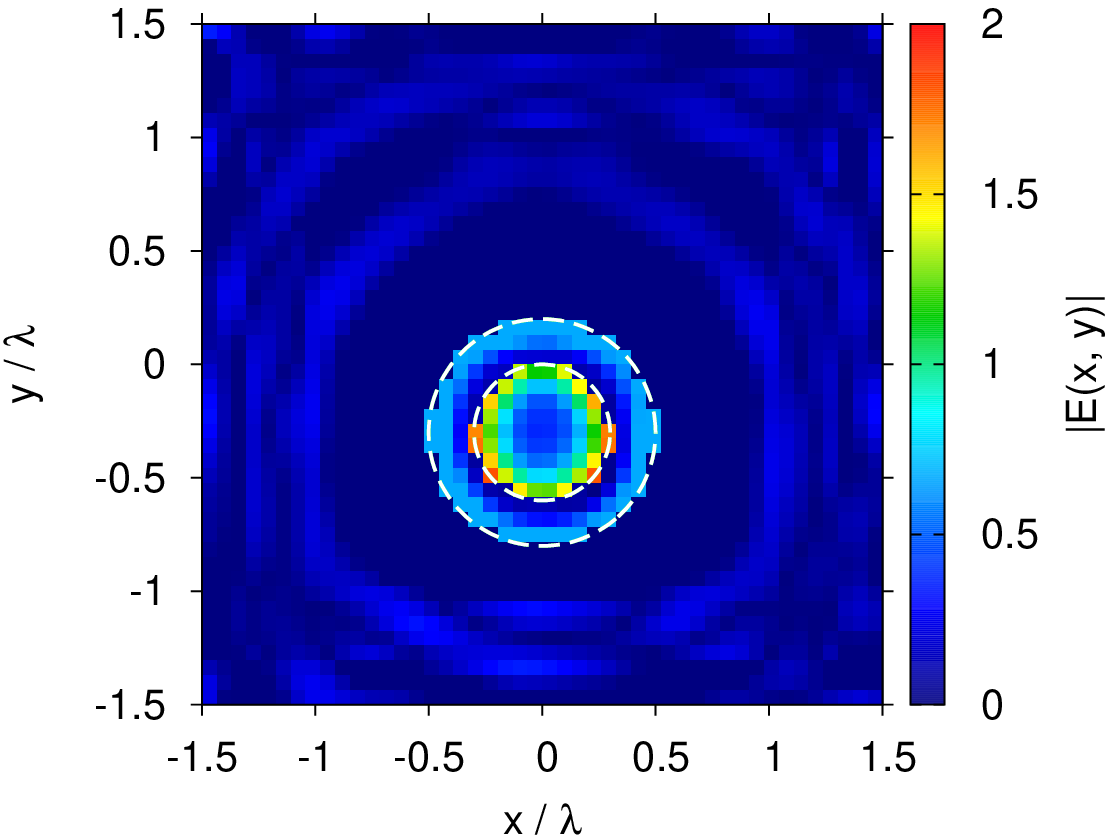}\tabularnewline
~(\emph{c})&
~(\emph{d})\tabularnewline
\end{tabular}\end{center}

\begin{center}~\vfill\end{center}

\begin{center}\textbf{Fig. 7 - Zhong et} \textbf{\emph{al.}}\textbf{,}
\textbf{\emph{{}``}}Multi-Scaling Differential Contraction ...''\end{center}

\newpage
\begin{center}~\vfill\end{center}

\begin{center}\begin{tabular}{c}
\includegraphics[%
  width=0.80\textwidth]{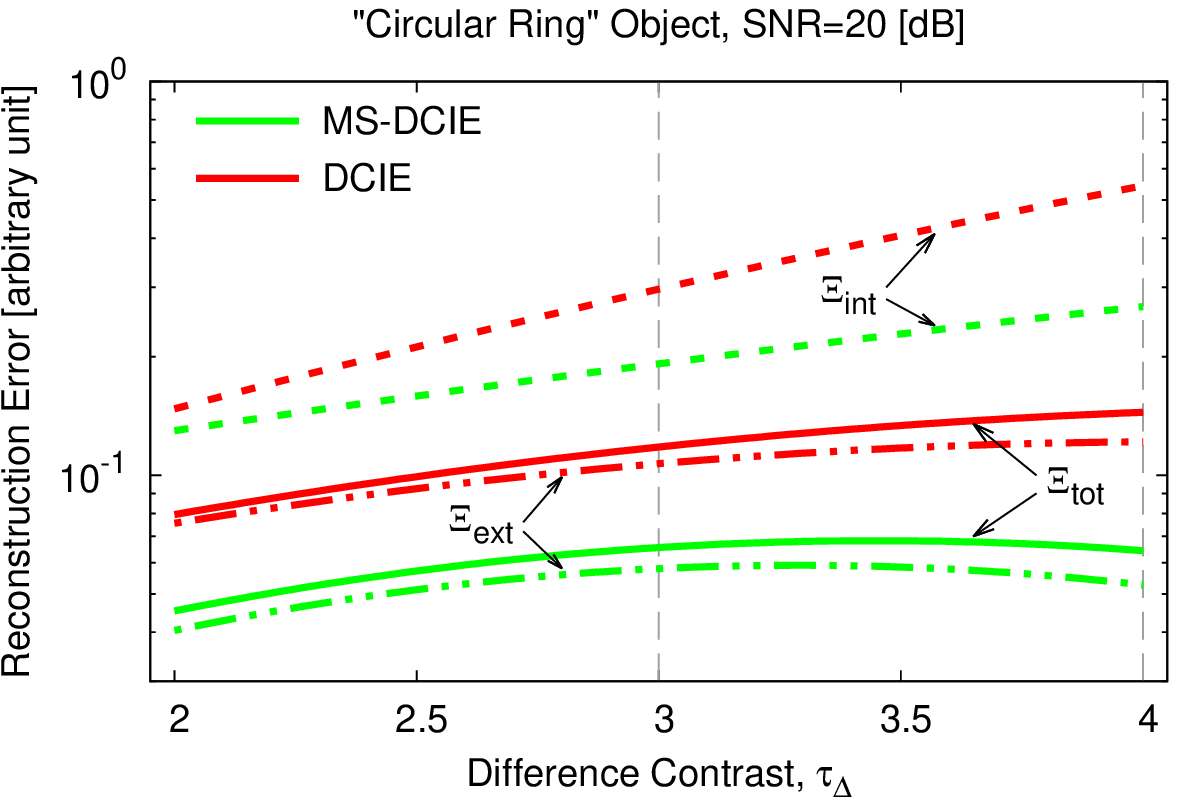}\tabularnewline
\end{tabular}\end{center}

\begin{center}~\vfill\end{center}

\begin{center}\textbf{Fig. 8 - Zhong et} \textbf{\emph{al.}}\textbf{,}
\textbf{\emph{{}``}}Multi-Scaling Differential Contraction ...''\end{center}

\newpage
\begin{center}~\vfill\end{center}

\begin{center}\begin{tabular}{ccc}
&
\emph{~~}$\tau_{\Delta}=3$&
\emph{~~}$\tau_{\Delta}=4$\tabularnewline
\begin{sideways}
\emph{~~~~~~~~~~~}~~~~~~~~~~\emph{MS-DCIE}%
\end{sideways}&
\includegraphics[%
  width=0.40\columnwidth,
  keepaspectratio]{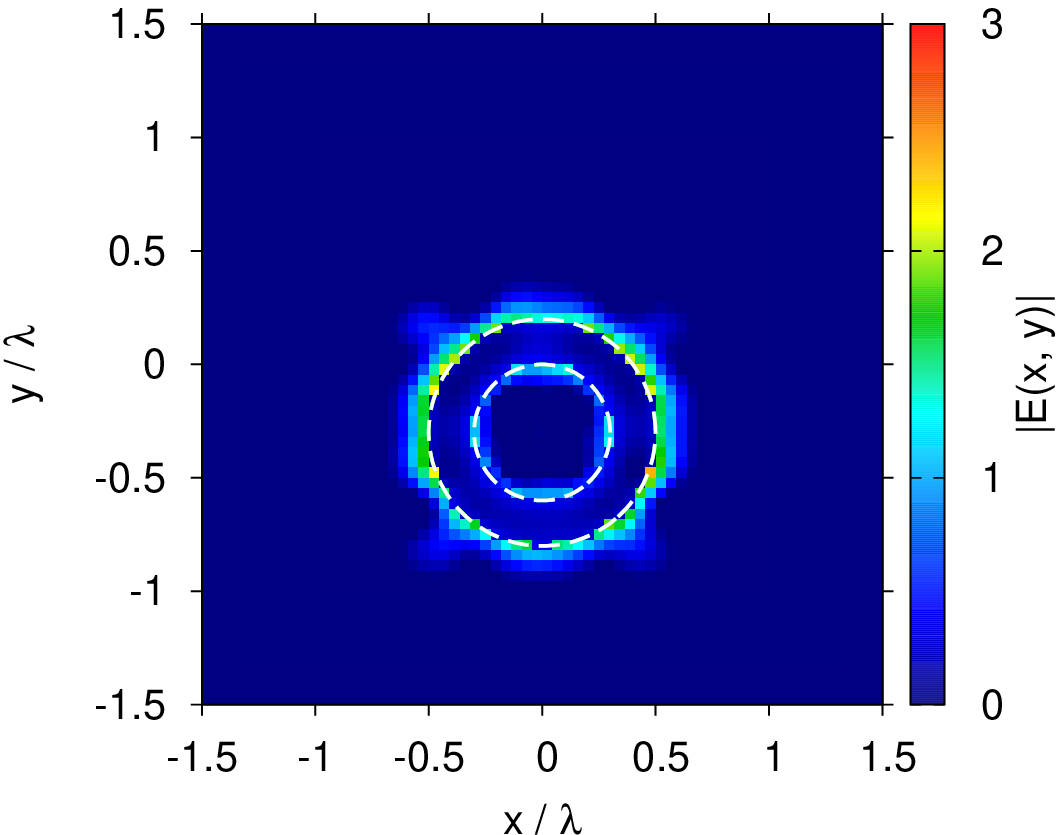}&
\includegraphics[%
  width=0.40\columnwidth,
  keepaspectratio]{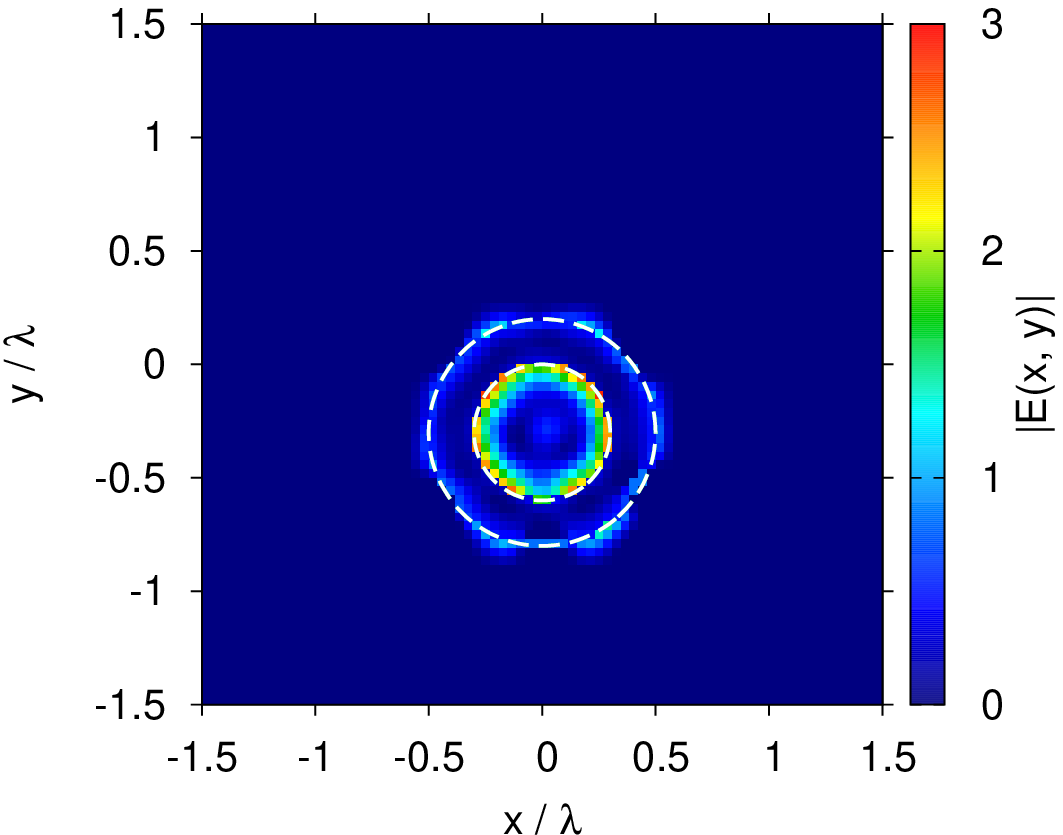}\tabularnewline
&
\emph{~~}(\emph{a})&
\emph{~~}(\emph{b})\tabularnewline
\begin{sideways}
\emph{~~~~~~~~~~~~~~~~}~~~~~\emph{DCIE}%
\end{sideways}&
\includegraphics[%
  width=0.40\columnwidth,
  keepaspectratio]{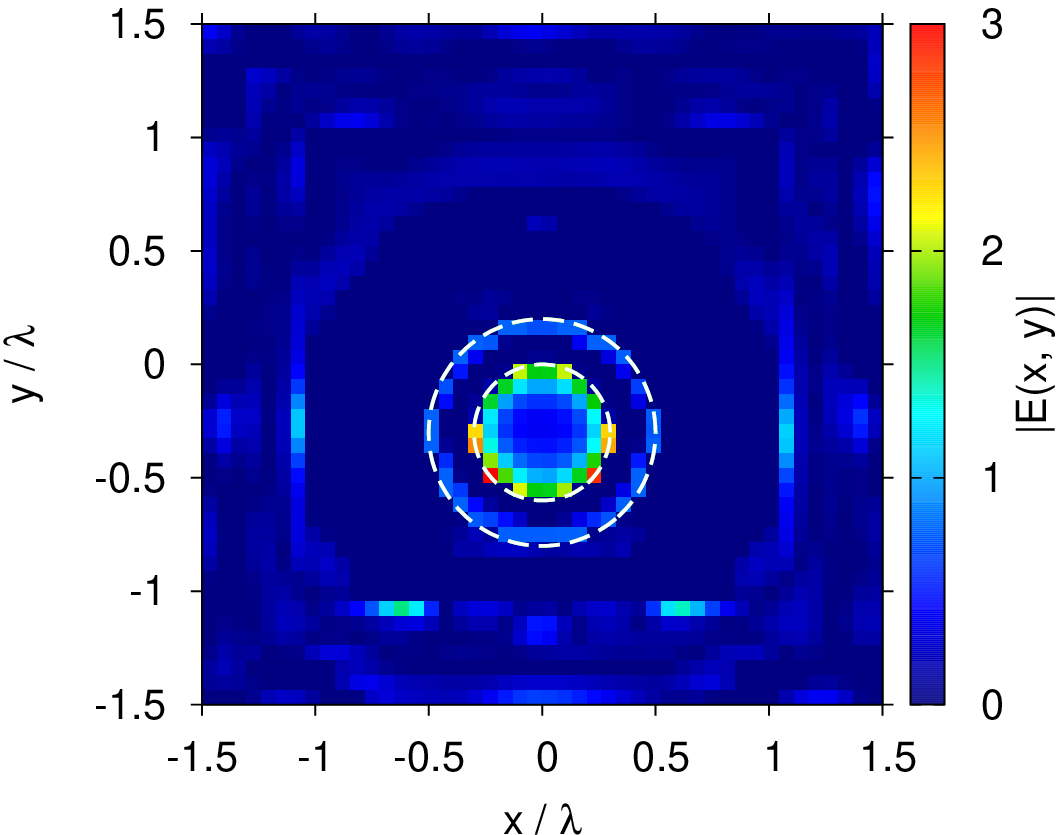}&
\includegraphics[%
  width=0.40\columnwidth,
  keepaspectratio]{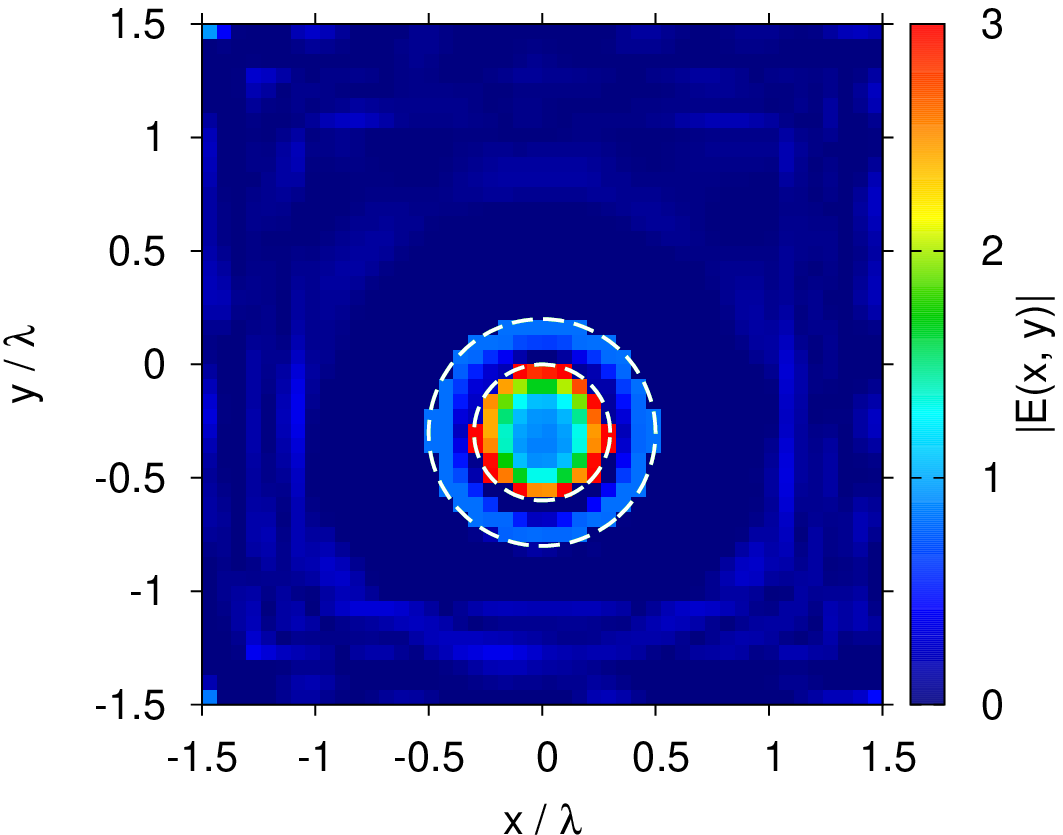}\tabularnewline
&
\emph{~~}(\emph{c})&
\emph{~~}(\emph{d})\tabularnewline
\end{tabular}\end{center}

\begin{center}~\vfill\end{center}

\begin{center}\textbf{Fig. 9 - Zhong et} \textbf{\emph{al.}}\textbf{,}
\textbf{\emph{{}``}}Multi-Scaling Differential Contraction ...''\end{center}

\newpage
\begin{center}~\vfill\end{center}

\begin{center}\begin{tabular}{c}
\includegraphics[%
  width=0.80\textwidth]{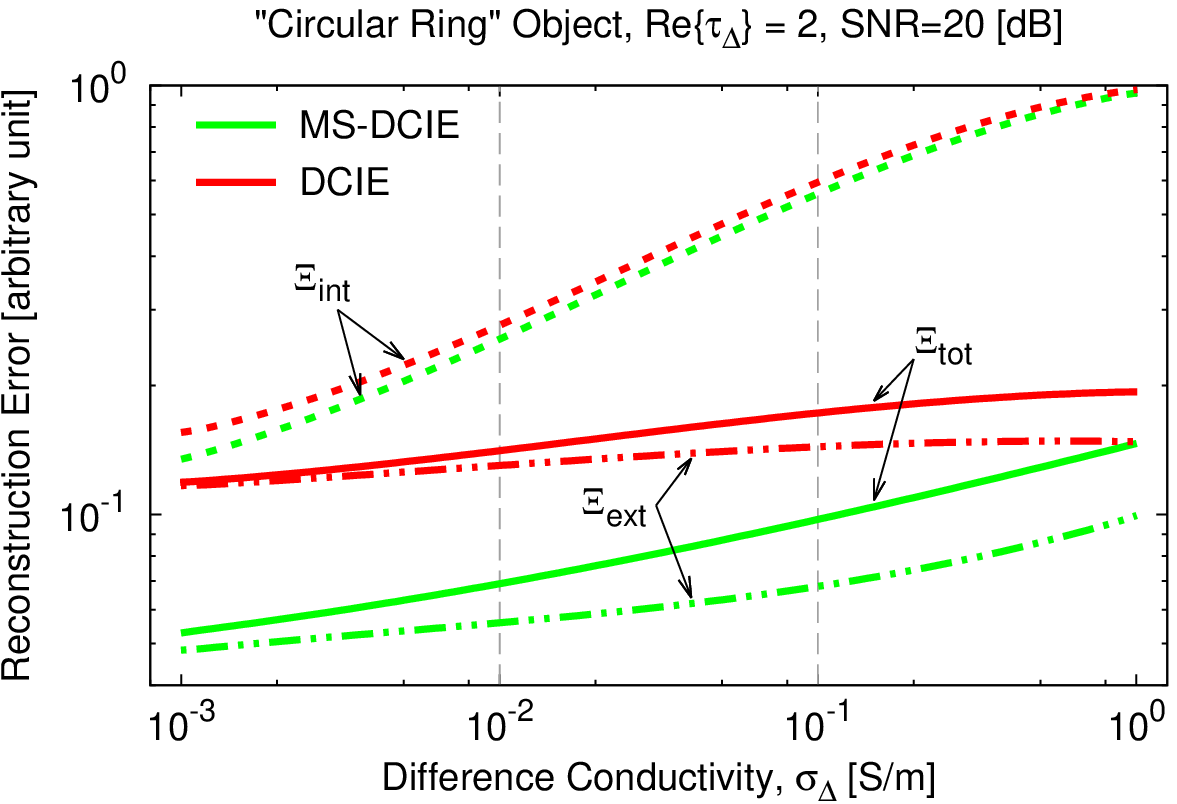}\tabularnewline
\end{tabular}\end{center}

\begin{center}~\vfill\end{center}

\begin{center}\textbf{Fig. 10 - Zhong et} \textbf{\emph{al.}}\textbf{,}
\textbf{\emph{{}``}}Multi-Scaling Differential Contraction ...''\end{center}

\newpage
\begin{center}~\vfill\end{center}

\begin{center}\begin{tabular}{ccc}
&
\emph{Real part}&
\emph{Imaginary part}\tabularnewline
\begin{sideways}
\emph{~~~~~~~}~~~~~~~~~~~\emph{MS-DCIE}%
\end{sideways}&
\includegraphics[%
  width=0.40\columnwidth,
  keepaspectratio]{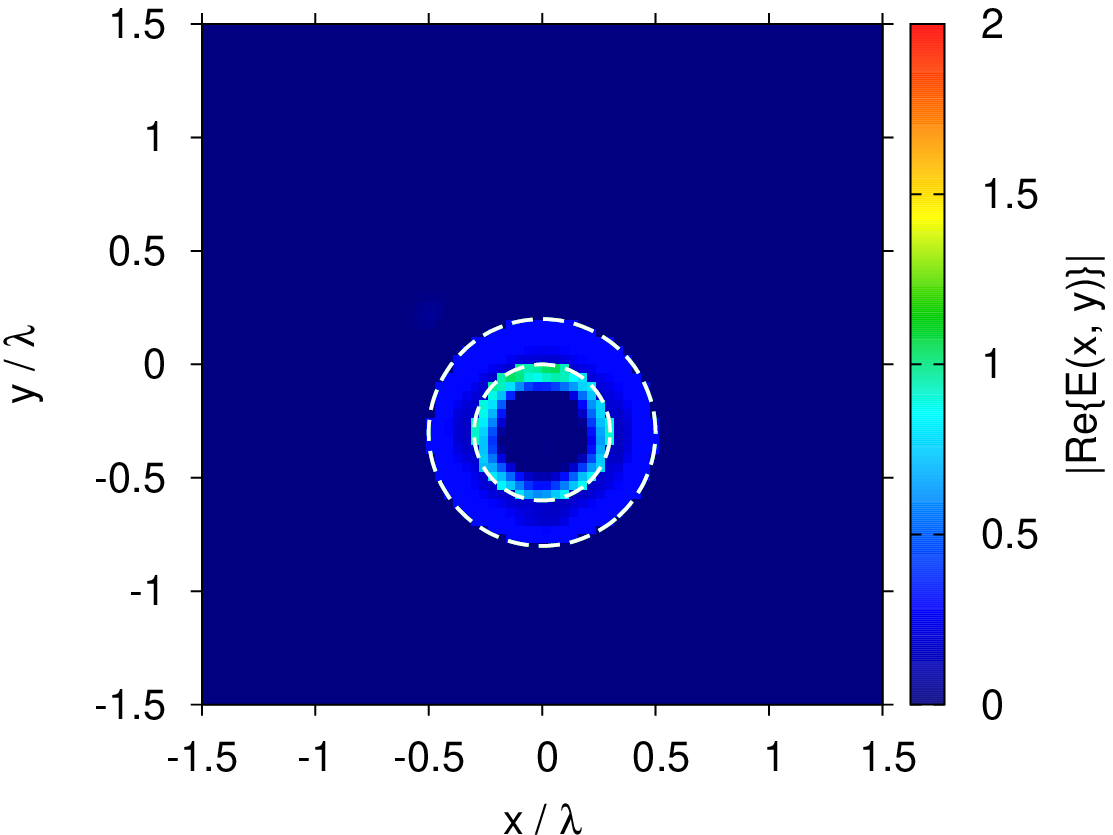}&
\includegraphics[%
  width=0.40\columnwidth,
  keepaspectratio]{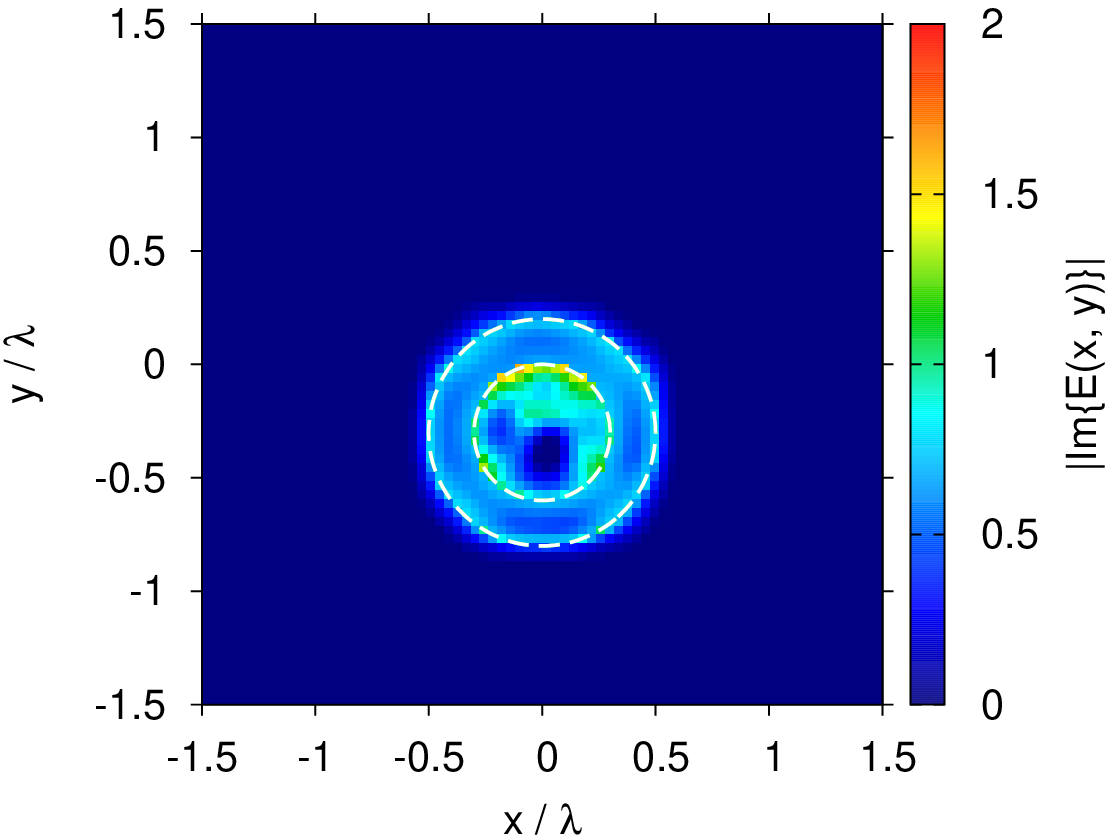}\tabularnewline
&
(\emph{a})&
(\emph{b})\tabularnewline
\begin{sideways}
\emph{~~~~~~~~~~}~~~~~~~~\emph{DCIE}%
\end{sideways}&
\includegraphics[%
  width=0.40\columnwidth,
  keepaspectratio]{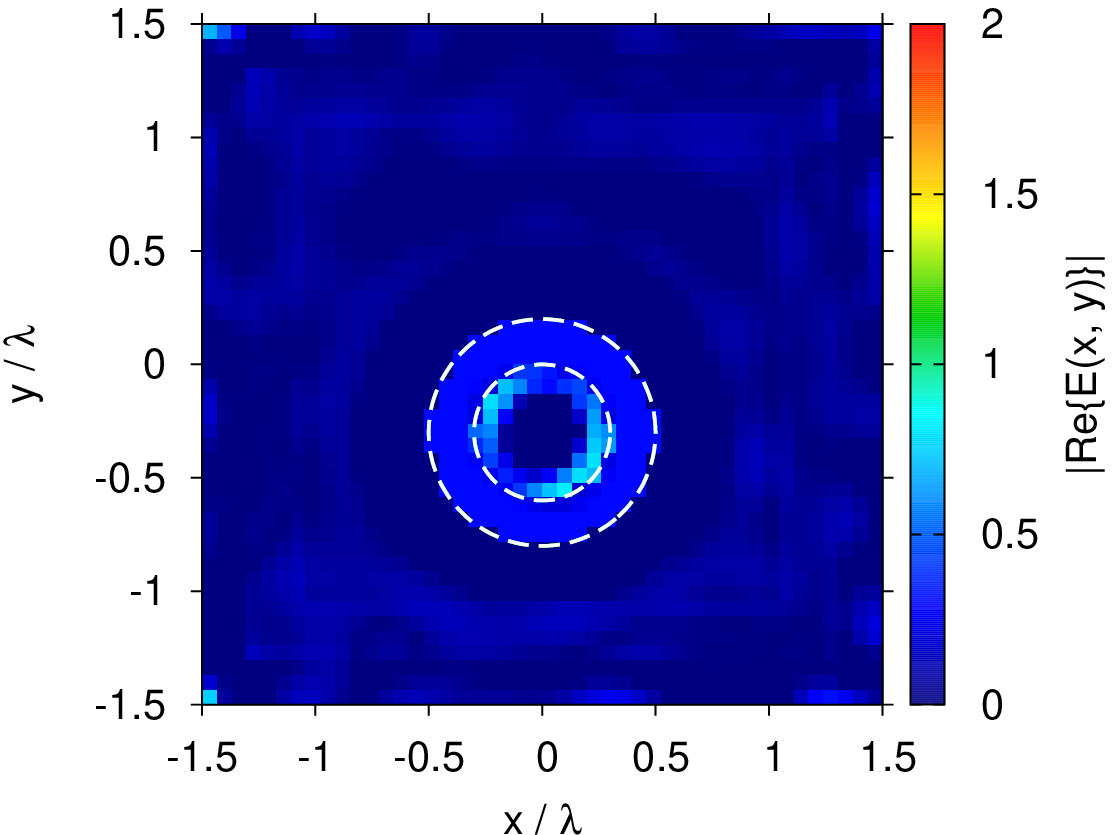}&
\includegraphics[%
  width=0.40\columnwidth,
  keepaspectratio]{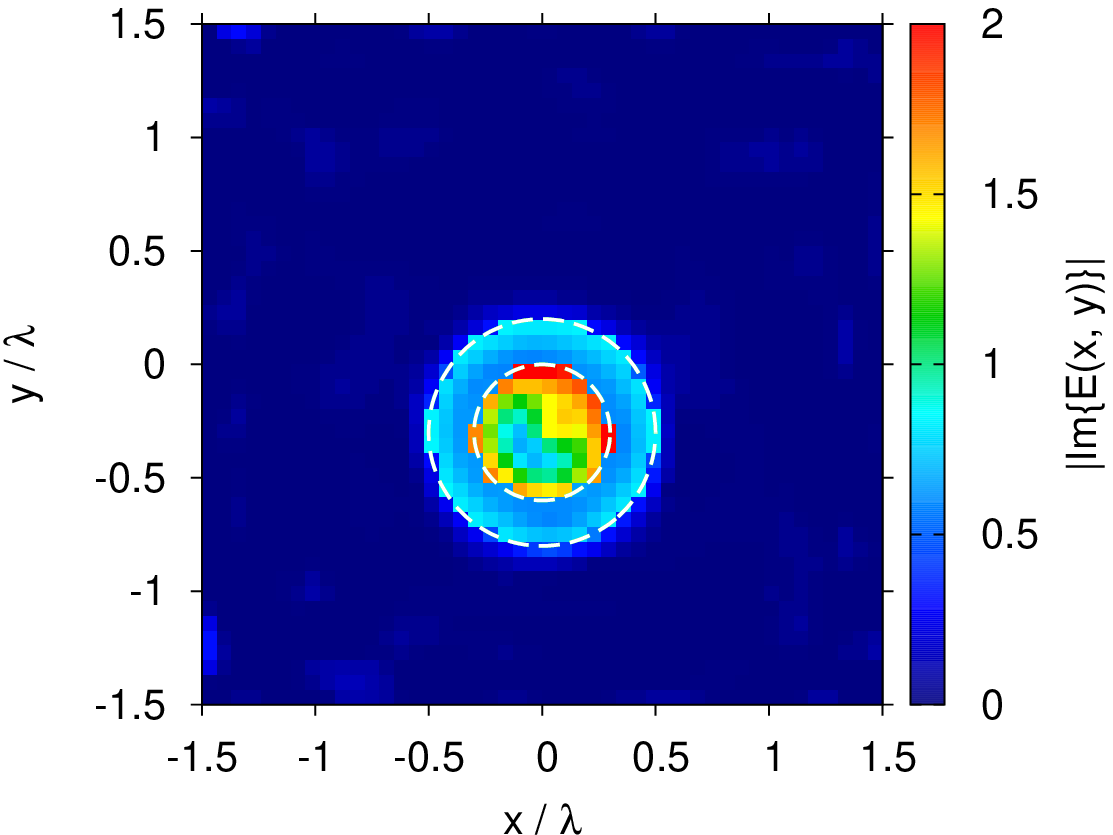}\tabularnewline
&
(\emph{c})&
(\emph{d})\tabularnewline
\end{tabular}\end{center}

\begin{center}~\vfill\end{center}

\begin{center}\textbf{Fig. 11 - Zhong et} \textbf{\emph{al.}}\textbf{,}
\textbf{\emph{{}``}}Multi-Scaling Differential Contraction ...''\end{center}

\newpage
\begin{center}~\vfill\end{center}

\begin{center}\begin{tabular}{c}
\includegraphics[%
  width=0.80\textwidth]{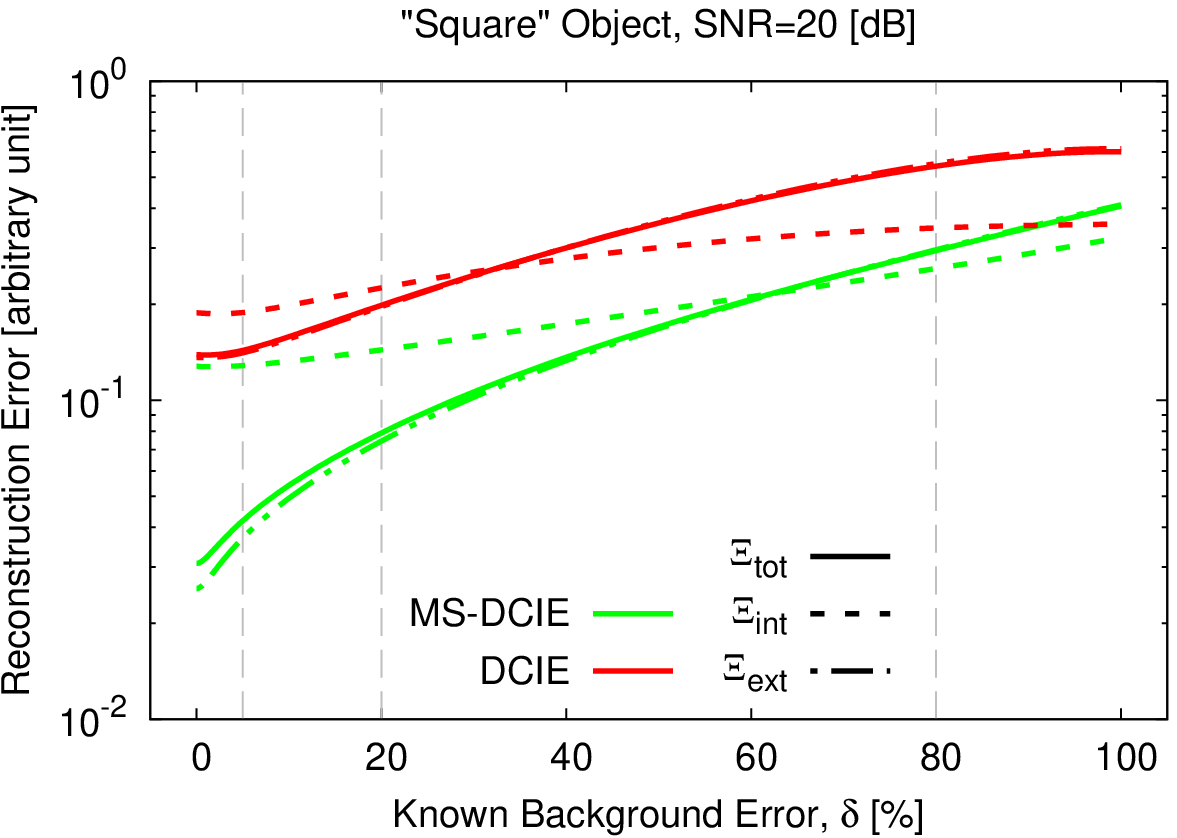}\tabularnewline
\end{tabular}\end{center}

\begin{center}~\vfill\end{center}

\begin{center}\textbf{Fig. 12 - Zhong et} \textbf{\emph{al.}}\textbf{,}
\textbf{\emph{{}``}}Multi-Scaling Differential Contraction ...''\end{center}

\newpage
\begin{center}~\vfill\end{center}

\begin{center}\begin{tabular}{ccc}
&
\emph{MS-DCIE~~}&
\emph{DCIE~~}\tabularnewline
\begin{sideways}
~~~~~~~~~~~~~~~~~$\delta=5$ \%%
\end{sideways}&
\includegraphics[%
  width=0.35\textwidth,
  keepaspectratio]{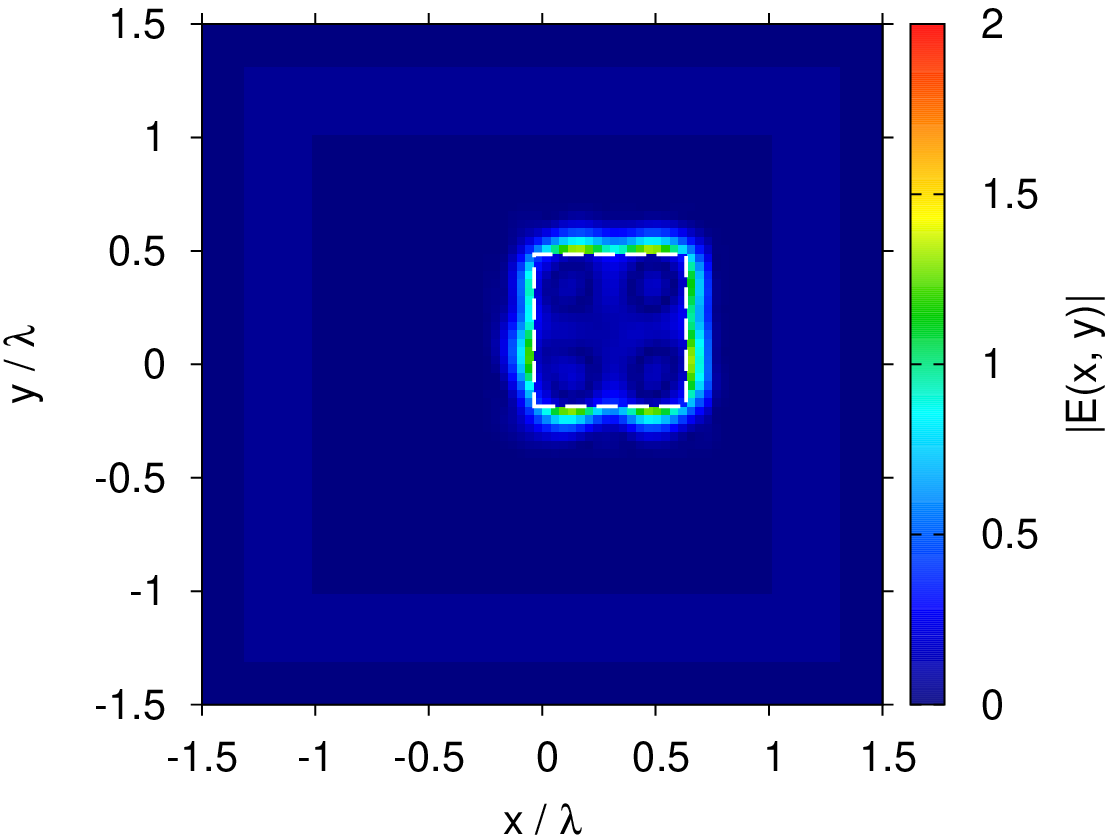}&
\includegraphics[%
  width=0.35\textwidth,
  keepaspectratio]{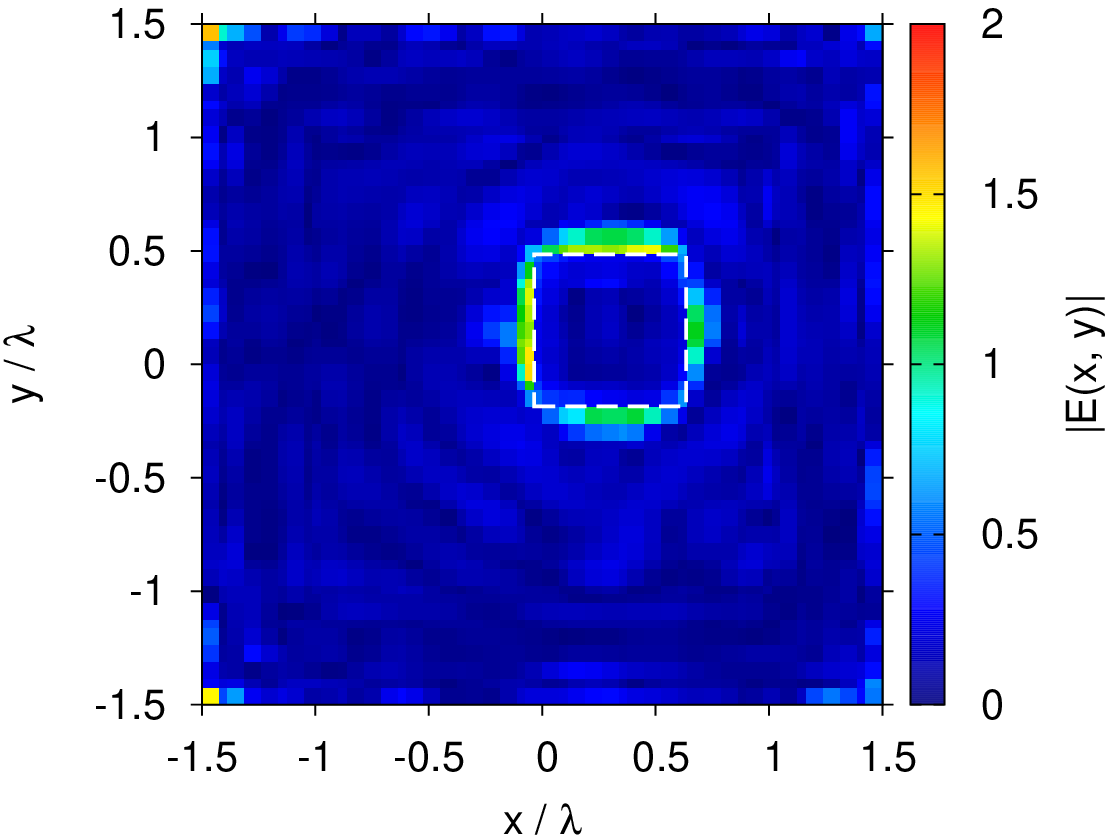}\tabularnewline
&
(\emph{a})&
(\emph{b})\tabularnewline
\begin{sideways}
~~~~~~~~~~~~~~~~~$\delta=20$ \%%
\end{sideways}&
\includegraphics[%
  width=0.35\textwidth,
  keepaspectratio]{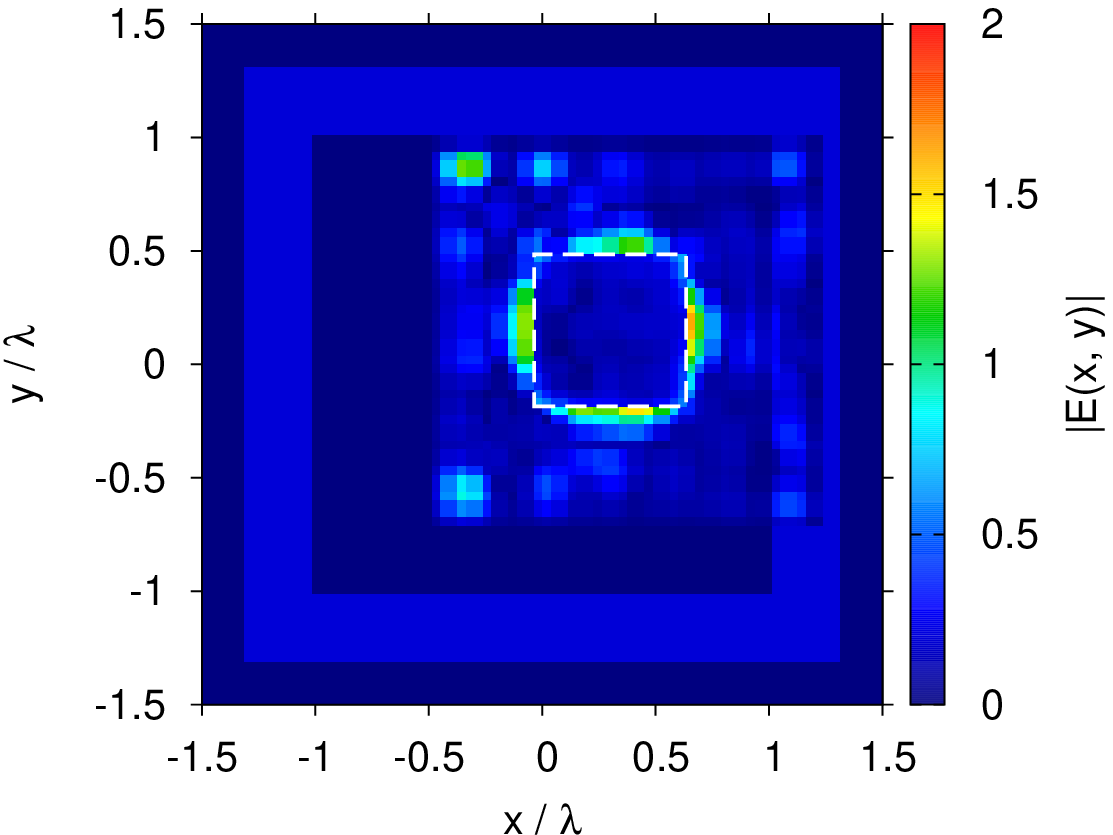}&
\includegraphics[%
  width=0.35\textwidth,
  keepaspectratio]{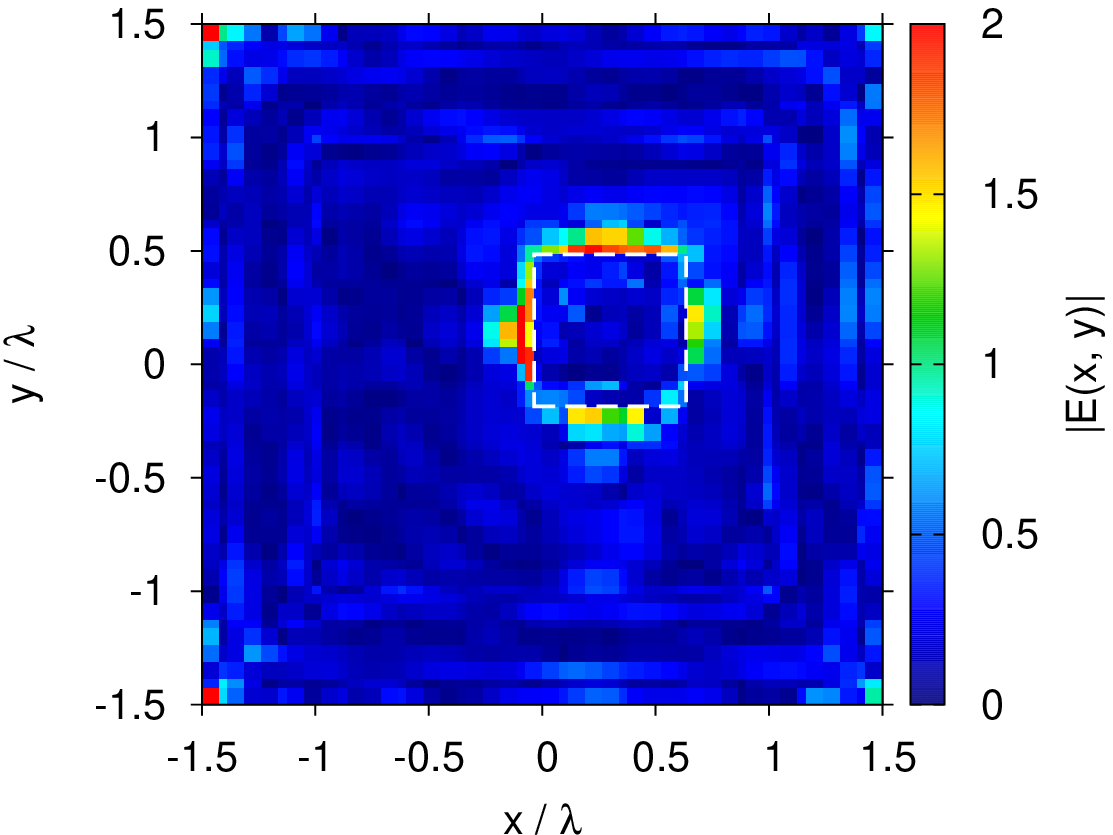}\tabularnewline
&
(\emph{c})&
(\emph{d})\tabularnewline
\begin{sideways}
~~~~~~~~~~~~~~~~~$\delta=80$ \%%
\end{sideways}&
\includegraphics[%
  width=0.35\textwidth,
  keepaspectratio]{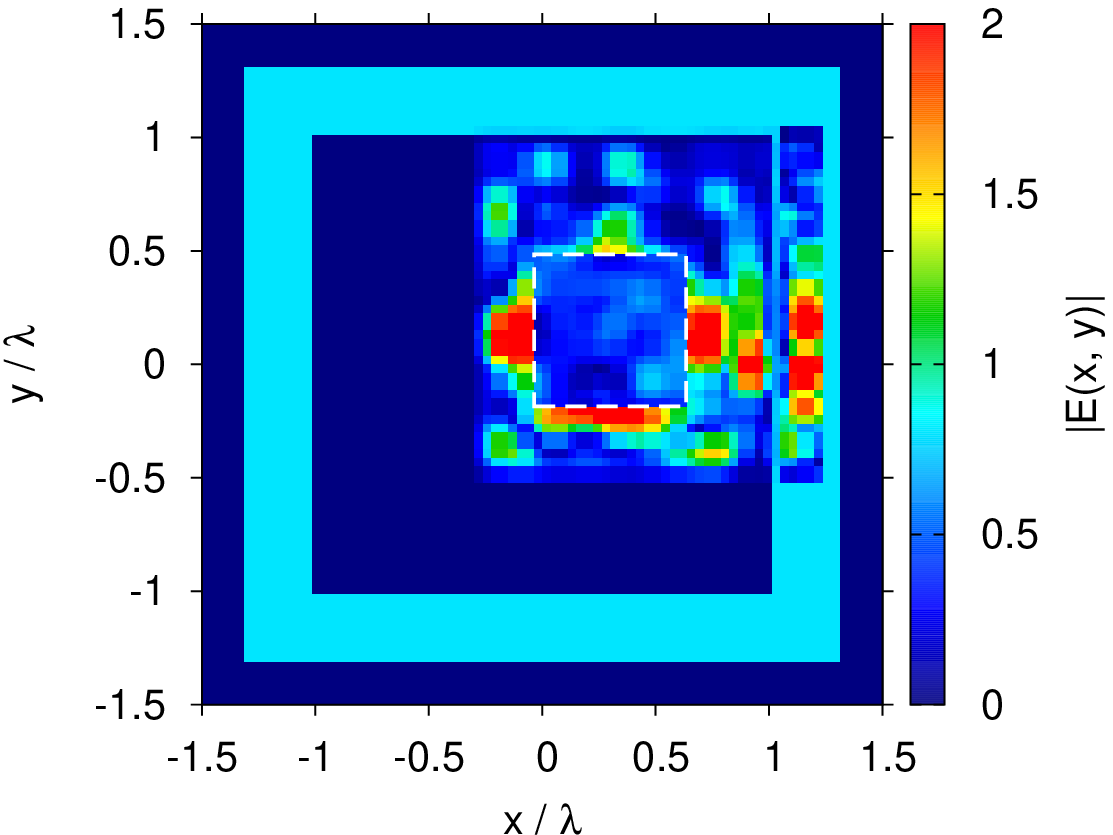}&
\includegraphics[%
  width=0.35\textwidth,
  keepaspectratio]{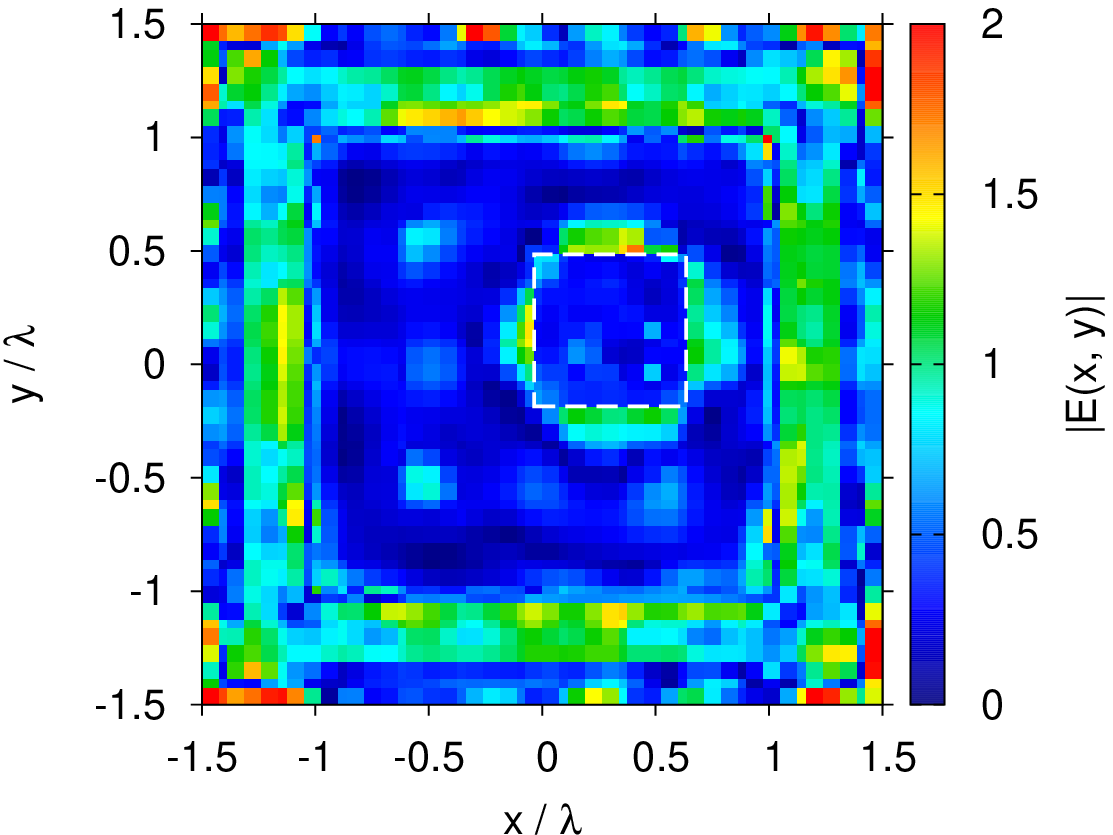}\tabularnewline
&
(\emph{e})&
(\emph{f})\tabularnewline
\end{tabular}\end{center}

\begin{center}~\vfill\end{center}

\begin{center}\textbf{Fig. 13 - Zhong et} \textbf{\emph{al.}}\textbf{,}
\textbf{\emph{{}``}}Multi-Scaling Differential Contraction ...''\end{center}

\newpage
\begin{center}~\vfill\end{center}

\begin{center}\begin{tabular}{cc}
\multicolumn{2}{c}{\includegraphics[%
  width=0.48\textwidth]{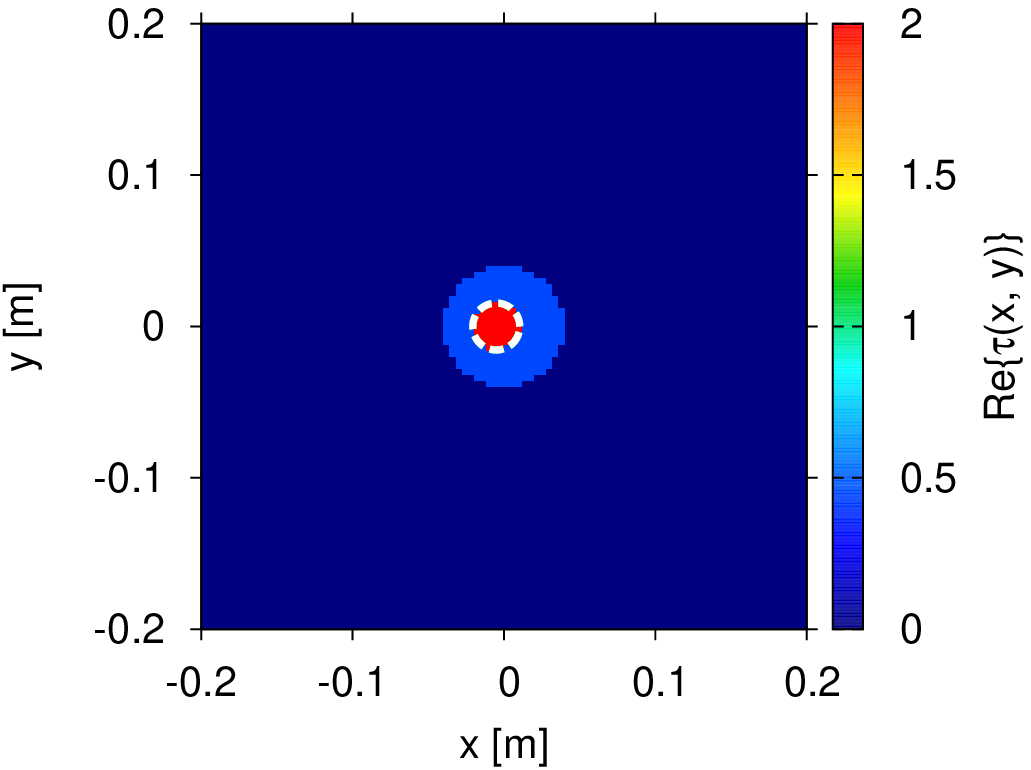}}\tabularnewline
\multicolumn{2}{c}{(\emph{a})~~~~}\tabularnewline
\includegraphics[%
  width=0.48\textwidth]{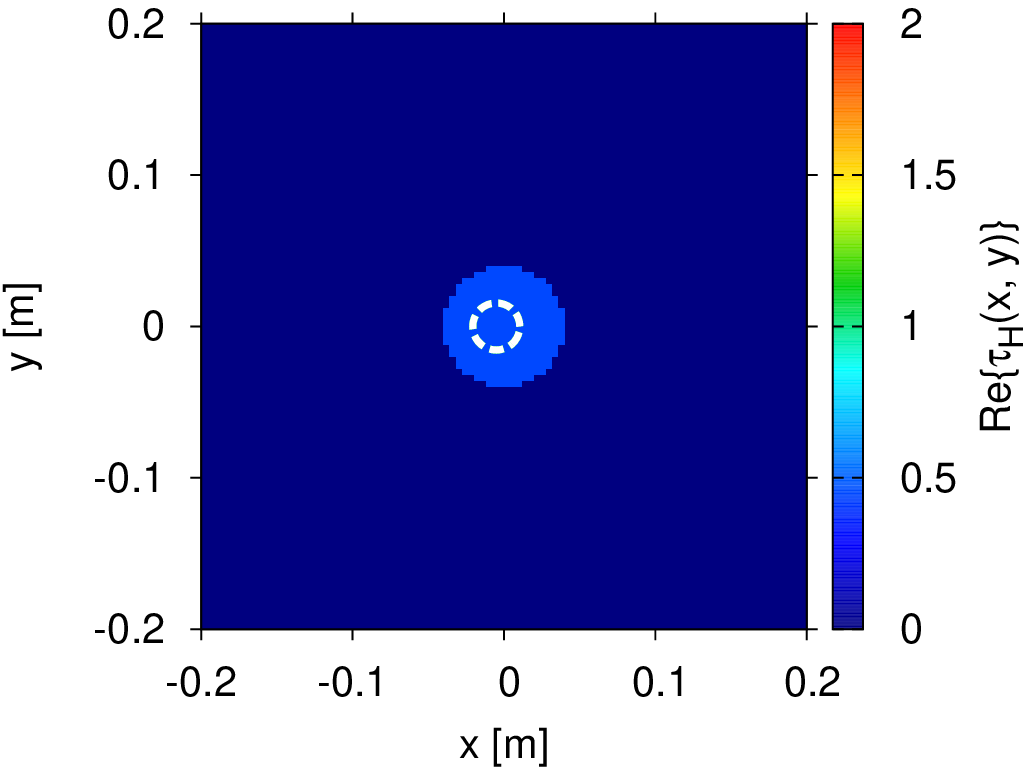}&
\includegraphics[%
  width=0.48\textwidth]{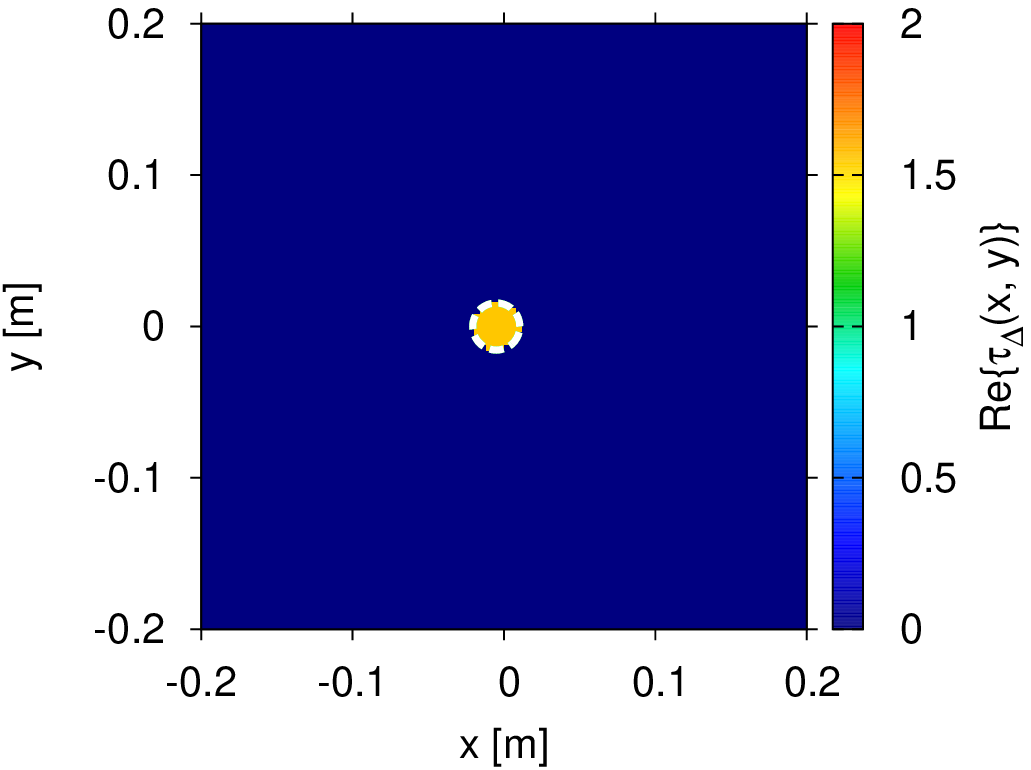}\tabularnewline
(\emph{b})~~~&
(\emph{c})~~~\tabularnewline
\end{tabular}\end{center}

\begin{center}~\vfill\end{center}

\begin{center}\textbf{Fig. 14 - Zhong et} \textbf{\emph{al.}}\textbf{,}
\textbf{\emph{{}``}}Multi-Scaling Differential Contraction ...''\end{center}

\newpage
\begin{center}~\vfill\end{center}

\begin{center}\begin{tabular}{cc}
\emph{MS-DCIE}&
\emph{DCIE}\tabularnewline
\includegraphics[%
  width=0.48\textwidth]{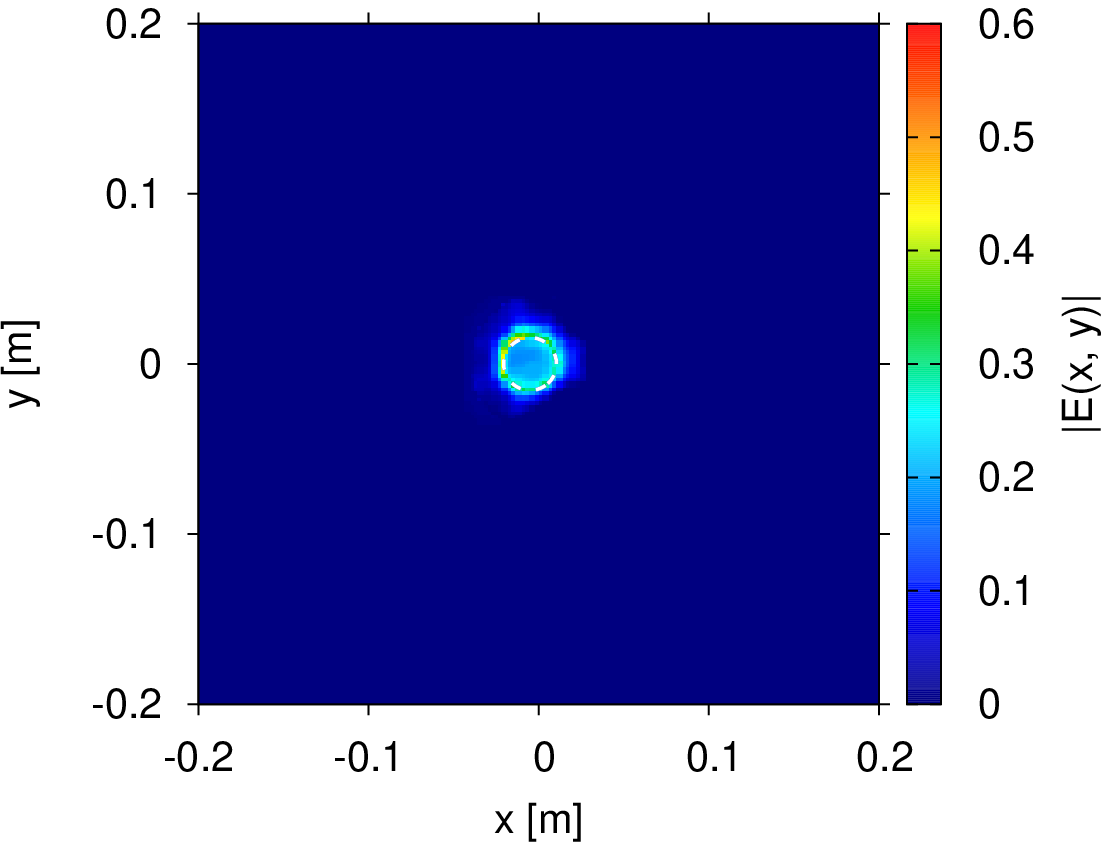}&
\includegraphics[%
  width=0.48\textwidth]{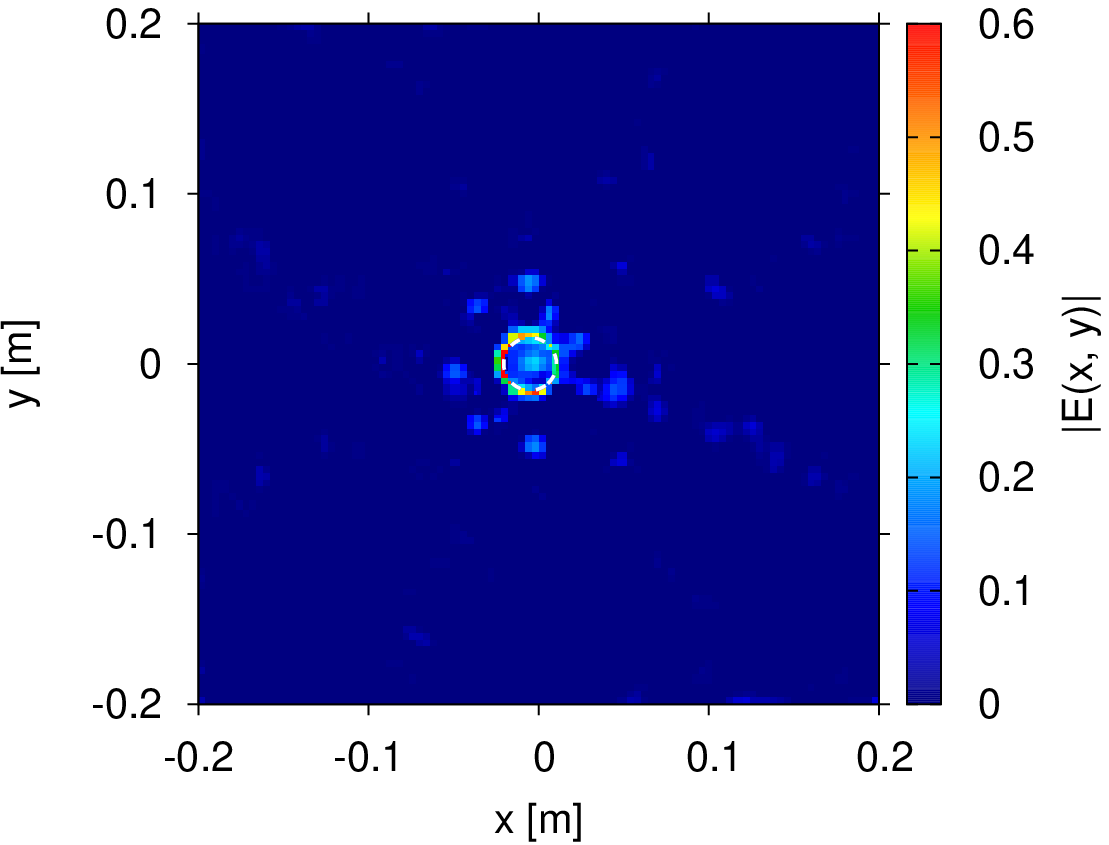}\tabularnewline
(\emph{a})&
(\emph{b})\tabularnewline
\emph{MS-DLSIE}&
\emph{DLSIE}\tabularnewline
\includegraphics[%
  width=0.48\textwidth]{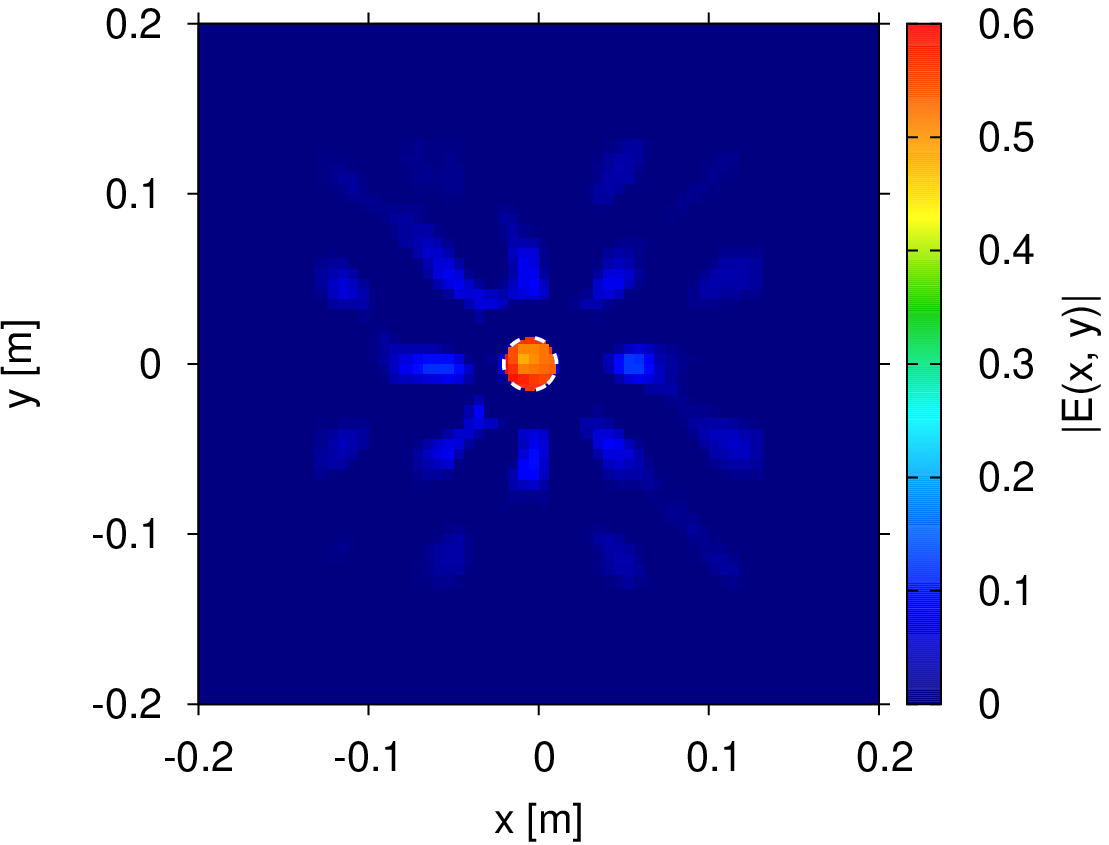}&
\includegraphics[%
  width=0.48\textwidth]{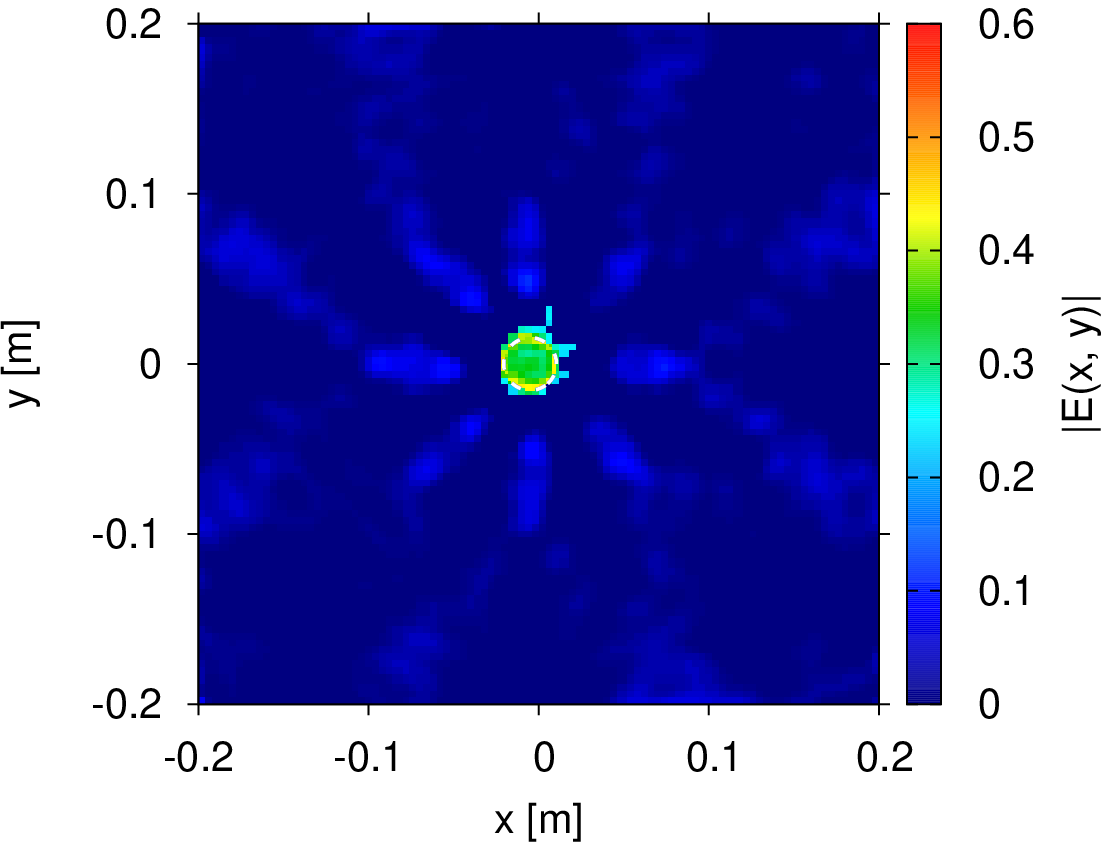}\tabularnewline
(\emph{c})&
(\emph{d})\tabularnewline
\end{tabular}\end{center}

\begin{center}~\vfill\end{center}

\begin{center}\textbf{Fig. 15 - Zhong et} \textbf{\emph{al.}}\textbf{,}
\textbf{\emph{{}``}}Multi-Scaling Differential Contraction ...''\end{center}

\newpage
\begin{center}~\vfill\end{center}

\begin{center}\begin{tabular}{cc}
\emph{MS-DCIE}&
\emph{DCIE}\tabularnewline
\includegraphics[%
  width=0.48\textwidth]{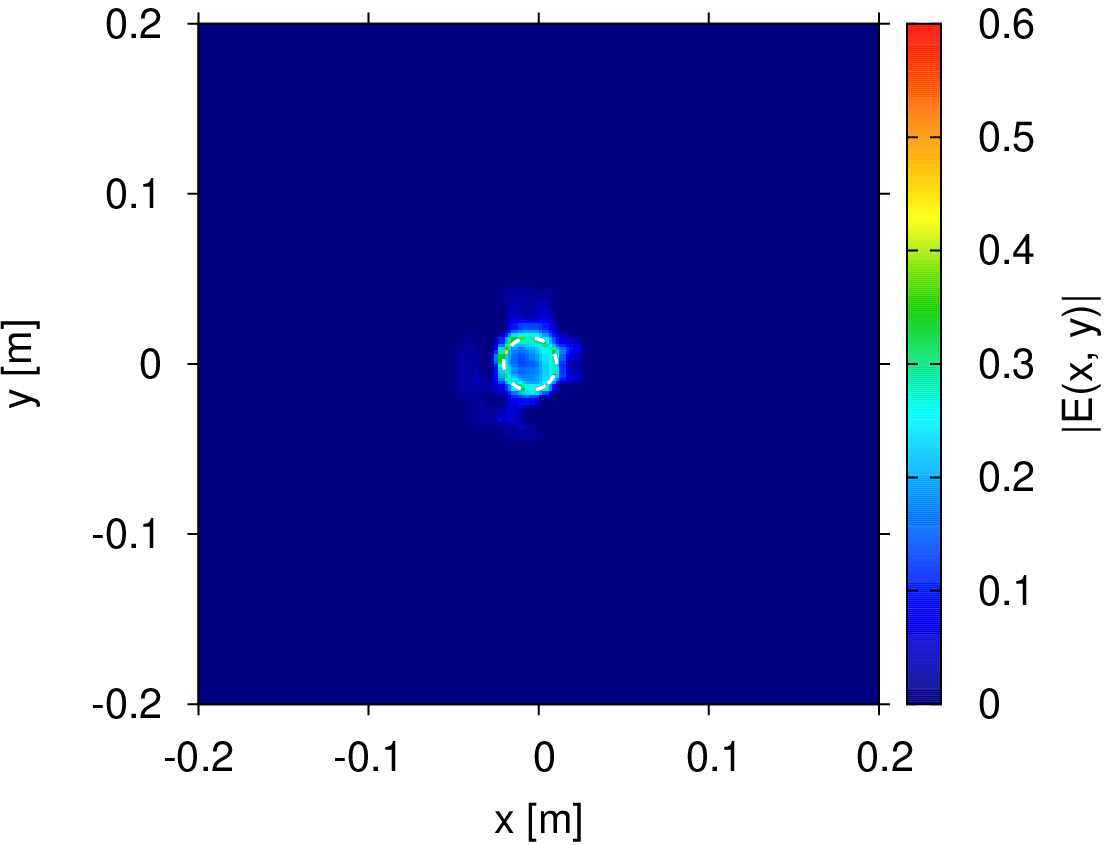}&
\includegraphics[%
  width=0.48\textwidth]{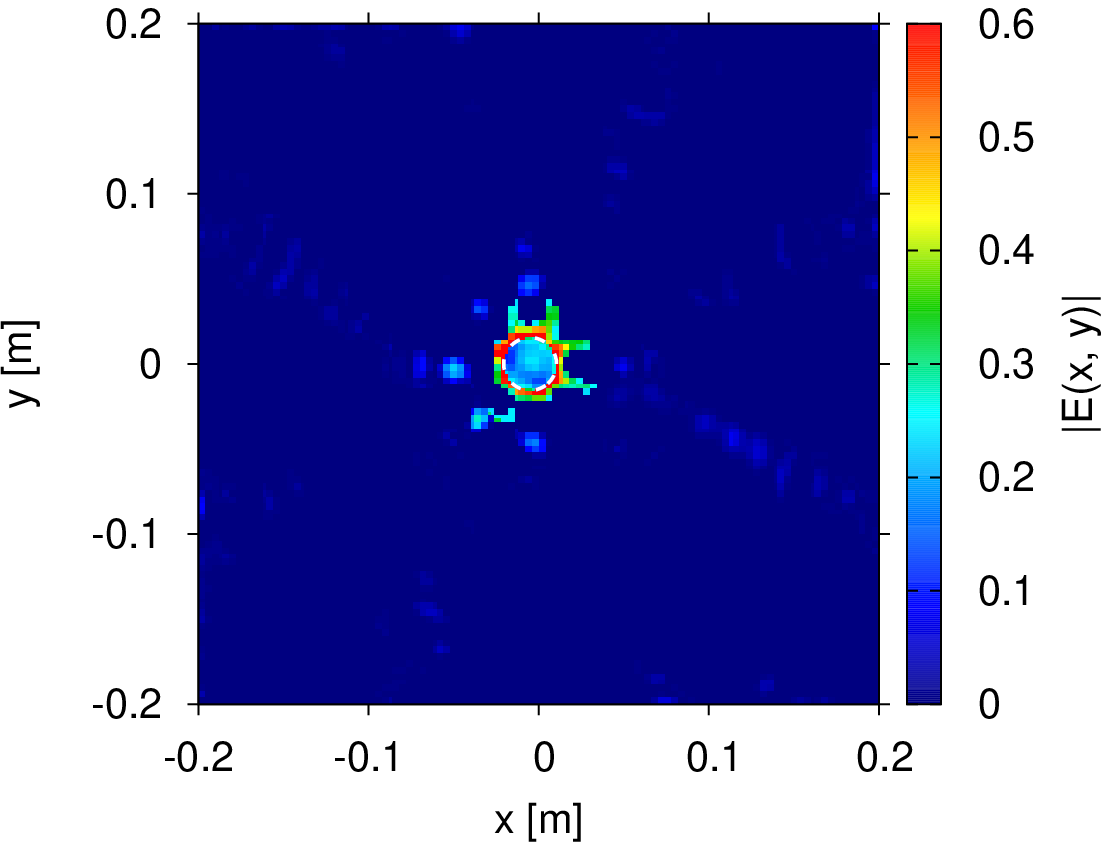}\tabularnewline
(\emph{a})&
(\emph{b})\tabularnewline
\emph{MS-DLSIE}&
\emph{DLSIE}\tabularnewline
\includegraphics[%
  width=0.48\textwidth]{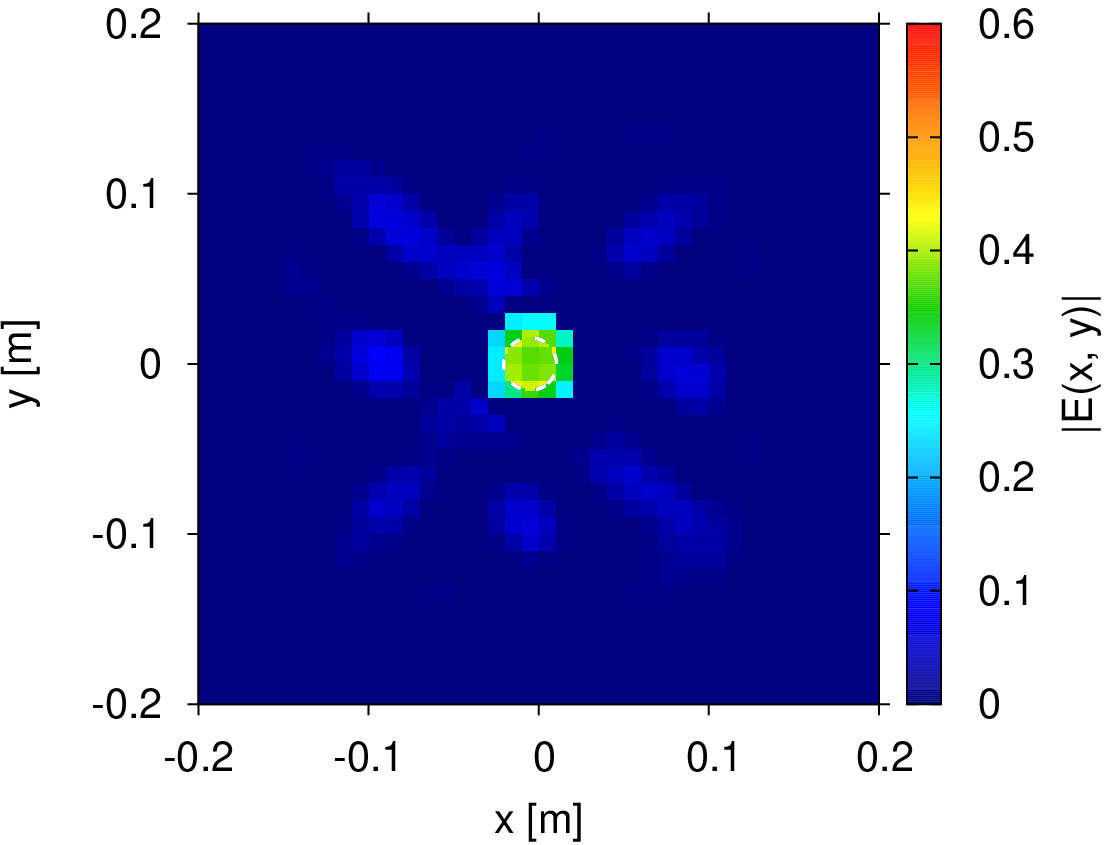}&
\includegraphics[%
  width=0.48\textwidth]{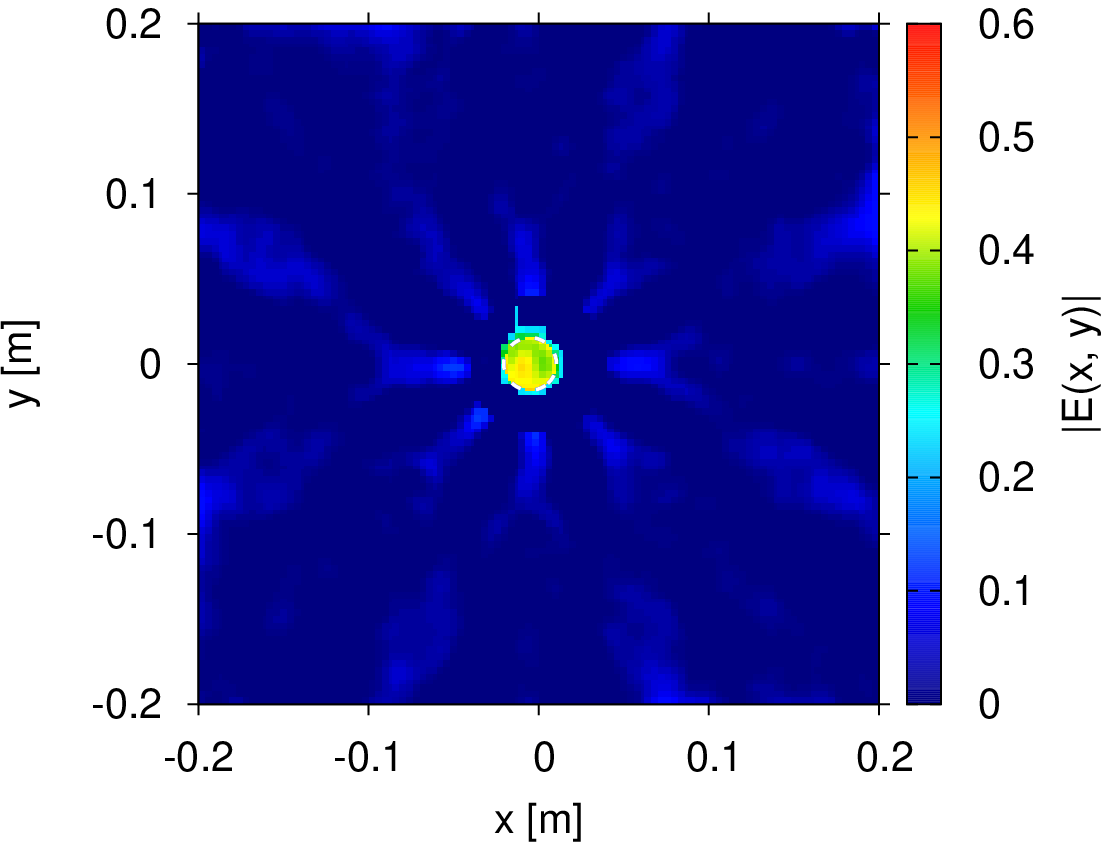}\tabularnewline
(\emph{c})&
(\emph{d})\tabularnewline
\end{tabular}\end{center}

\begin{center}~\vfill\end{center}

\begin{center}\textbf{Fig. 16 - Zhong et} \textbf{\emph{al.}}\textbf{,}
\textbf{\emph{{}``}}Multi-Scaling Differential Contraction ...''\end{center}

\newpage
\begin{center}~\vfill\end{center}

\begin{center}\begin{tabular}{|c||c|c|c|c|}
\hline 
$f$~{[}GHz{]}&
\multicolumn{4}{c|}{$\Xi_{tot}$~~~~{[}$\times10^{-3}${]} }\tabularnewline
\cline{2-5} 
&
\emph{MS-DCIE}&
\emph{~~DCIE~~}&
\multicolumn{1}{c|}{\emph{MS-DLSIE}}&
\emph{~~DLSIE~~}\tabularnewline
\hline
\hline 
$6.0$&
$4.57$&
$6.39$&
$8.28$&
$12.47$\tabularnewline
\hline 
$7.0$&
$5.01$&
$6.01$&
$8.31$&
$12.93$\tabularnewline
\hline 
$8.0$&
$4.69$&
$8.31$&
$10.91$&
$12.32$\tabularnewline
\hline 
$9.0$&
$4.39$&
$6.70$&
$8.84$&
$11.34$\tabularnewline
\hline
\end{tabular}\end{center}

\begin{center}~\vfill\end{center}

\begin{center}\textbf{Tab. I - Zhong et} \textbf{\emph{al.}}\textbf{,}
\textbf{\emph{{}``}}Multi-Scaling Differential Contraction ...''\end{center}
\end{document}